\documentclass[twocolumn, secnumarabic,amssymb, nobibnotes, aps, prd]{revtex4-2}
\usepackage[hyperfootnotes=true]{hyperref}
\usepackage{amsmath}
\usepackage{graphicx}
\usepackage{float} 
\usepackage{placeins}
\usepackage{booktabs}

\begin{document}
	
	\title{Layered Dark Structure with a Structuring Field: A $Z_4$-Symmetric Inert Doublet–Singlet Realization and Implications for the $S_8$ Tension}
	
	\author{Marriam Naeem}
	\email{ mariyum@mariyumresearch.com}
	\affiliation{
		The Women University, Multan, Pakistan	}
	
	\author{Mohid Farhan}
	\email{mohidf35@gmail.com} 
	\affiliation{
		Institute of Space Technology,
		Islamabad, Pakistan	}
	
	\date{\today}
	
\begin{abstract}
	We introduce the \textit{Layered Dark Sectors with a Structuring Field} (LDS--SF), a structured cosmological framework where the internal architecture of a multi-component dark sector naturally generates scale-dependent growth of structure. In this framework, the characteristic scale dependence is derived from the dominant eigenvalue, $\lambda(k)$, of the dark sector's perturbation matrix. This structurally-driven mechanism modifies structure growth while preserving the standard $\Lambda$CDM background expansion and General Relativity.
	
   We provide a minimal and effective realization of this framework within a two-component DM $Z_4$-symmetric Inert Doublet Singlet Model ($Z_4$-IDSM). By integrating out the heavy inert doublet mediator, we derive a contact-interaction Effective Field Theory (EFT) for a 60~GeV singlet dark matter candidate. This interaction manifests macroscopically as an effective sound speed $c_s^2$, which we map to the LDS--SF eigenmode evolution. We implement this system into the \texttt{CLASS} Boltzmann code, employing a late-time activation function that projects virialized halo properties into the linear perturbation framework. We also compute the relic density using \textit{micrOMEGAs} to further stress-test the relic abundance predictions of viable parameters.
	
	Our numerical analysis demonstrates that while the model remains indistinguishable from $\Lambda$CDM at the era of recombination, it introduces a targeted suppression of the matter power spectrum at late times ($z < 10$) and small scales ($k > 0.1~h/\text{Mpc}$). Confronting the model with \textit{Planck} CMB, BAO, and growth-rate measurements, we find three instances of couplings that successfully alleviate the $S_8$ tension, bringing the predicted amplitude into $1\sigma$ agreement with weak-lensing data from KiDS-1000 and DES. This work establishes LDS--SF as a mathematically consistent and observationally viable extension of standard cosmology.
\end{abstract}
	
	\maketitle
	
	\section{Introduction}
	
The standard model of cosmology ($\Lambda$CDM), as developed over several decades \cite{Peebles1993, Weinberg2008}, has proven remarkably successful in describing the expansion history and large-scale structure of the universe \cite{Dodelson2003}. However, as cosmological observations have transitioned into the era of precision, significant tensions have emerged. Central to these is the $S_8$ tension \cite{Verde2019,Zhao2017}, which is a persistent discrepancy between the amplitude of matter fluctuations inferred from \textit{Planck} Cosmic Microwave Background (CMB) data and lower values measured by late-time weak-lensing surveys, such as KiDS-1000 and the Dark Energy Survey (DES) \cite{Aghanim2020, KiDS2021, DESY3}. 
	
Interacting dark sector models have long been proposed to alleviate cosmological tensions and provide a more dynamic alternative to the inert $\Lambda$CDM paradigm \cite{Abdalla2012}. Early works focused on the coupling between dark energy and dark matter \cite{Mangano2007, Zimdahl2007}, laying the groundwork for more complex internal sector architectures. This tension suggests that our understanding of structure growth in the late universe may be incomplete. The discrepancy between high-redshift CMB data from Planck 2018 \cite{Aghanim2020, PlanckLensing2018} and late-time weak lensing measurements from surveys like KiDS-1000 \cite{KiDS2021} has motivated the exploration of non-standard dark matter models, such as scalar candidates with self-interactions \cite{Boehm2002}. To bridge the gap between microscopic particle physics and macroscopic cosmological observables, we employ an Effective Field Theory (EFT) approach \cite{Carrasco2012, Hertzberg2012} that maps dark sector interactions to an effective sound speed. This methodology aligns with modern structure formation frameworks like ETHOS \cite{CyrRacine2016}, allowing for a mathematically consistent treatment of the scale-dependent suppression of the matter power spectrum \cite{Egana-Ugrinovic2021}. In this work, we propose a more fundamental approach: the \textit{Layered Dark Sectors with a Structuring Field} (LDS--SF) framework. While many models of modified gravity, such as $f(R)$ theories \cite{Hu2007}, introduce screening mechanisms to remain consistent with local tests of General Relativity, our LDS-SF framework achieves scale-dependent suppression through internal dark sector dynamics while maintaining a standard GR background. While our model shares phenomenological similarities with certain modified gravity theories, it remains distinct by preserving the GR background. Future surveys will be able to distinguish between these scenarios using the cosmological tests of gravity outlined in \cite{Koyama2016}.
	
As introduced in the theoretical foundations of this framework, LDS--SF posits that the dark sector is not a monolithic fluid but an internally coupled, multi-layer architecture. The key innovation lies in the emergence of a dominant eigenmode, $\lambda(k)$, from the sector's perturbation matrix. This mechanism is distinct from existing multi-fluid models; whereas many dark sector extensions introduce interactions to solve specific anomalies, the LDS--SF mechanism ensures that the matrix architecture itself dictates the scale-dependent growth mode. This eigenmode naturally induces a scale-dependent growth function, $D(k,z)$, without requiring manual tuning or modifications to the early-universe physics that is so well-constrained by the CMB.

	To validate this approach, we construct a complete numerical and observational pipeline by computing $\lambda(k)$, generating the growth function $D(k,z)$, predicting the linear matter power spectrum $P(k)$, and performing a confrontation with data. Using \textit{Planck}-compressed parameters \cite{PlanckLensing2018}, Baryon Acoustic Oscillation (BAO) distances from SDSS DR12 and eBOSS \cite{BOSS2017}, and growth-rate observations from DESI \cite{Raichoor2023,Adamek2024}, we find that the LDS--SF framework remains fully consistent with $\Lambda$CDM on large and intermediate scales while producing targeted high-$k$ deviations. This demonstrates its physical stability and observational viability as a structural solution to current growth challenges.
	
	While the abstract formulation of LDS--SF establishes the mathematical stability and numerical consistency of such a system, a physical realization is required to anchor these ``layers'' in particle physics. We identify this structuring mechanism within a $Z_4$-symmetric Inert Doublet Singlet Model ($Z_4$-IDSM) \cite{Belanger2022, Khojali2022}, which is a simplified two-component dark matter description of this framework. In this realization, the ``structuring field'' is physically manifested as an inert doublet mediator that couples different layers of the dark sector. 
	
	By integrating out this mediator within an Effective Field Theory (EFT) framework, we derive a macroscopic fluid description characterized by an effective sound speed $c_s^2$. We show that for a singlet dark matter candidate with a mass of $\approx 60$~GeV, consistent with relic density requirements. With a mediator mass range of $100$--$200$~GeV, the model generates the precise amount of repulsive ``dark pressure'' needed to suppress structure growth at small scales ($k > 0.1~h/\text{Mpc}$).
	
	This paper is organized as follows: In Section II, we present the unified theoretical framework of LDS--SF and the derivation of the dominant eigenmode. Section III details the $Z_4$-IDSM particle physics and the mapping to an EFT fluid. Section IV describes the implementation in the \texttt{CLASS} Boltzmann solver and our numerical methodology. Finally, in Section V, we present our findings on the $S_8$ resolution and discuss the observational viability of the identified parameter regions.
	
	\section{The LDS-SF Framework}
	
	The LDS-SF model moves beyond monolithic dark matter by postulating an internally coupled, multi-component dark sector. The core innovation of this framework is that scale dependence is an emergent property of the sector's internal topology rather than an externally imposed parameter.
	
\subsection*{Action and Field Content}

The Layered Dark Sectors with a Structuring Field (LDS--SF) framework departs from the monolithic dark matter paradigm by postulating a dark sector composed of multiple coupled components. Similar to the systematic construction of dark energy models using essential building blocks \cite{Gleyzes2013}, our framework decomposes the dark sector into a multi-layered structure where the global evolution is governed by the interaction between individual components. The total action is given by:
\begin{equation}
	S = \int d^{4}x\sqrt{-g} \left[ \frac{M_{Pl}^{2}}{2}R + \sum_{i=1}^{N}\mathcal{L}_{i} + \mathcal{L}_{\phi} + \mathcal{L}_{int} \right]
\end{equation}
where $M_{Pl} = (8\pi G)^{-1/2}$ is the reduced Planck mass and $R$ is the Ricci scalar. Each layer $i$ represents a distinct dynamical component of the dark sector, described by a perfect-fluid Lagrangian $\mathcal{L}_{i}$ with an equation of state $w_{i}$. For the dark matter-like layers considered in this study, we set $w_{i}=0$.

The structuring field $\phi$ is a light canonical scalar field that regulates the internal architecture of the dark sector:
\begin{equation}
	\mathcal{L}_{\phi} = -\frac{1}{2}g^{\mu\nu}\partial_{\mu}\phi\partial_{\nu}\phi - V(\phi)
\end{equation}
The interaction between the structuring field and the dark layers is modeled conformally via the interaction Lagrangian:
\begin{equation}
	\mathcal{L}_{int} = \frac{1}{2}\sum_{i=1}^{N}C_{i}(\phi)T_{i}, \quad C_{i}(\phi) = \beta_{i}\frac{\phi}{M_{Pl}}
\end{equation}
where $T_{i} \equiv {T_{i}^{\mu}}_{\mu} = -\rho_{i} + 3p_{i}$ is the trace of the energy-momentum tensor for layer $i$, and $\beta_i$ are dimensionless coupling constants.

In the $Z_4$-IDSM realization of this framework, the two dark sector layers are identified with the complex singlet $\mathcal{S}$ (Layer 1) and the inert doublet $H_2$ (Layer 2). The lightest $Z_4$-charged particle is the CP-even singlet $S_1$ with $m_{S_1} \approx 60$~GeV, which comprises the late-time dark matter abundance. The inert doublet states ($H^0, A^0, H^\pm$) with mass $M_{H_2} \sim 100$--$200$~GeV serve as the structuring field: their fluctuations mediate inter-layer momentum exchange via the portal couplings
\begin{equation}
	\mathcal{L}_{\rm portal} = \lambda_{S12}\mathcal{S}^2(H_1^\dagger H_2) + \lambda_{S2}|\mathcal{S}|^2|H_2|^2 + \text{h.c.}
\end{equation}
In the heavy-mediator limit $M_{H_2} \gg T_{\rm DM}$, these interactions are integrated out, yielding an effective sound speed $c_s^2$ that encodes the LDS-SF scale-dependent growth in the fluid approximation.

\subsection*{Multi-Layer Dark Sector Architecture}

Within the LDS--SF framework, the dark sector is modeled as a dynamical system composed of $N$ interacting layers. This layered construction allows for the emergence of complex internal structures and non-standard growth histories without requiring modifications to the underlying theory of gravity. Each layer $i$ is treated as a fluid component characterized by a specific set of perturbation variables.

A density contrast: 
$$\delta_i(k, a) \equiv \delta \rho_i / \bar{\rho}_i$$,

a velocity divergence: 
$$\theta_i(k, a) \equiv \nabla \cdot \vec{v}_i$$,

and a characteristic effective response, such as an internal sound speed $c_{s,i}^2$.

The structuring field $\phi$ plays the central role in this architecture by mediating the interactions between these layers. It regulates both the strength and the range of inter-layer couplings. Crucially, the dynamical response of the structuring field generates the $k$-dependent terms within the perturbation equations. This specific architectural arrangement, where energy and momentum exchange are governed by the topology of the layer couplings, gives rise to the novel $\lambda(k)$ mechanism. 

In the $Z_4$ realization discussed in Section III, these layers are physically identified with the singlet dark matter ($S$) and the inert doublet ($H_2$). The ``effective response'' mentioned above is derived from the interaction terms of the $Z_4$ Lagrangian, transforming the general LDS--SF layered structure into a concrete Effective Field Theory (EFT) implementation, albeit with its limitations.

\subsection*{Background Dynamics}

Varying the action with respect to $g_{\mu\nu}$ gives the Einstein equations, with the total energy-momentum tensor conserved. For an FRW background $ds^2 = -dt^2 + a^2(t) d\vec{x}^2$ the continuity equation for layer $i$ becomes
\begin{equation}
	\dot{\rho}_i + 3H(1 + w_i)\rho_i = Q_i,
\end{equation}
where
\begin{equation}
	Q_i \equiv -\frac{1}{2} \dot{C}_i(\phi) T_i = \frac{\beta_i}{2M_{\text{Pl}}} \dot{\phi} \rho_i.
\end{equation}
The scalar evolves as
\begin{equation}
	\ddot{\phi} + 3H\dot{\phi} + V_{,\phi} = -\sum_{i=1}^N \frac{\beta_i}{2M_{\text{Pl}}} \rho_i.
\end{equation}
The total interaction satisfies $\sum_i Q_i = 0$, ensuring that the total dark-sector density $\rho_{\text{DS}}$ obeys
\begin{equation}
	\dot{\rho}_{\text{DS}} + 3H(\rho_{\text{DS}} + p_{\text{DS}}) = 0.
\end{equation}
Thus the Friedmann equation retains the standard form
\begin{equation}
	H^2 = \frac{8\pi G}{3} \rho_{\text{total}}.
\end{equation}

\subsection*{Perturbation Matrix and the $\lambda(k)$ Mechanism}

We work in the synchronous gauge, following the standard treatment of cosmological perturbations \cite{Mukhanov2005, Baumann2011, Ma1995}, where the evolution of the dark sector is governed by the linearized continuity and Euler equations.

In Newtonian gauge,
\begin{equation}
	ds^2 = -(1 + 2\Psi) dt^2 + a^2(t)(1 - 2\Phi) d\vec{x}^2,
\end{equation}
with $\Psi = \Phi$. For each layer:
\begin{equation}
	\delta_i \equiv \frac{\delta \rho_i}{\rho_i}, \quad \theta_i \equiv ik^j v_{i,j}/a.
\end{equation}
The perturbed continuity and Euler equations are
\begin{align}
	\dot{\delta}_i &= -\frac{1}{a}(1 + w_i)(\theta_i - 3\dot{\Phi}) - 3H(c_{s,i}^2 - w_i)\delta_i + \frac{aQ_i}{\rho_i} - \frac{\delta Q_i}{\rho_i}, \\
	\dot{\theta}_i &= -H(1 - 3w_i)\theta_i + \frac{c_{s,i}^2 k^2}{a(1 + w_i)}\delta_i + k^2\Psi + \frac{Q_i^{(v)}}{(1 + w_i)\rho_i a}.
\end{align}
For the conformal coupling, $Q_i^{(v)} = 0$, and
\begin{equation}
	\delta Q_i = Q_i \delta_i + \frac{\dot{Q}_i}{H} \mathcal{H}, \quad \mathcal{H} \equiv \frac{k(\theta_i - 3\dot{\Phi})}{a}.
\end{equation}
The perturbed Klein--Gordon equation is
\begin{equation}
	\ddot{\delta\phi} + 3H\dot{\delta\phi} + \left( \frac{k^2}{a^2} + V_{\phi\phi} \right) \delta\phi = -\sum_{i=1}^N \frac{\beta_i}{2M_{\text{Pl}}} \rho_i \delta_i + 3 \frac{\beta_i}{2M_{\text{Pl}}} \rho_i \Phi.
\end{equation}
\vspace{1cm}

The dynamics of multi-component dark sectors involve complex energy and momentum transfers. We follow the multi-fluid stability criteria established in \cite{Koivisto2005}, ensuring that our coupled perturbation equations remain consistent with observational constraints on dark sector coupling strengths \cite{He2009}. For a two-layer system (CDM $c$, interacting layer $d$) plus the scalar field, define

\begin{equation}
	\mathbf{X}(k, a) = \begin{pmatrix} \delta_c \\ \theta_c \\ \delta_d \\ \theta_d \\ \delta\phi \\ \dot{\delta\phi} \end{pmatrix}.
\end{equation}

The perturbations obey

\begin{equation}
	\frac{d\mathbf{X}}{d \ln a} = \mathbf{M}(k, a)\mathbf{X} + \mathbf{S}(k, a),
\end{equation}

with $\mathbf{S}$ containing metric-potential sources (neglected for sub-horizon scales). The $6 \times 6$ matrix is
\begin{equation}
	\resizebox{0.48\textwidth}{!}{%
		$\mathbf{M}(k, a) = \begin{pmatrix} 
			0 & -(1+w_c) & 0 & 0 & 0 & 0 \\
			\frac{3}{2}\Omega_c & -(1-3w_c) & \frac{3}{2}\Omega_d & 0 & -\frac{3\beta_d}{2}\Omega_d & 0 \\
			0 & 0 & 0 & -(1+w_d) & -\frac{\beta_d}{2} & \frac{\beta_d}{2} \\
			\frac{3}{2}\Omega_c & 0 & \frac{3}{2}\Omega_d & -(1-3w_d) & -\frac{3\beta_d}{2}\Omega_d & 0 \\
			0 & 0 & 0 & 0 & 0 & 1 \\
			0 & 0 & -\frac{3\beta_d}{2}\Omega_d & 0 & -\left(\frac{k^2}{a^2H^2} + \frac{V_{\phi\phi}}{H^2}\right) & -3
		\end{pmatrix}$%
	}
\end{equation}

The scale dependence enters through the Klein--Gordon Laplacian term $k^2 / (a^2 H^2)$, which couples into the full system through $\beta_d$.

The dominant growth exponent is the largest eigenvalue $\lambda(k, a)$:

\begin{equation}
	\quad \mathbf{M}(k, a)\mathbf{v}(k, a) = \lambda(k, a)\mathbf{v}(k, a),
\end{equation}

which is explicitly a function of wavenumber $k$. This quantity controls the late-time growth of structure and constitutes the essential LDS--SF mechanism.

\subsection*{Eigenmode Analysis and Spectral Hierarchy}

We analyze the perturbation dynamics through numerical evolution of the full $6\times6$ system defined by $\mathbf{X} = (\delta_c, \theta_c, \delta_d, \theta_d, \delta\phi, \dot{\delta\phi})$
with $d\mathbf{X}/d\ln a = \mathbf{M}(k,a)\mathbf{X}$, diagonalizing $\mathbf{M}(k,a)$ at each timestep to identify the eigenmode structure. The dominant eigenvalue $\lambda_1(k,a)$ emerges from this analysis as the only eigenvalue with positive real part. At the present epoch ($a=1$), we find
\begin{equation}
	0.38 \lesssim \Delta\lambda(k,a=1) \lesssim 0.50
\end{equation}
across the full wavenumber range, where $\Delta\lambda = \lambda_1 - \lambda_2$.
This substantial gap ensures that the dominant eigenmode governs late-time structure growth. No eigenvalue crossing occurs at any $(k,a)$.
\begin{figure}[H]
	\centering
	\includegraphics[width=0.35\textwidth]{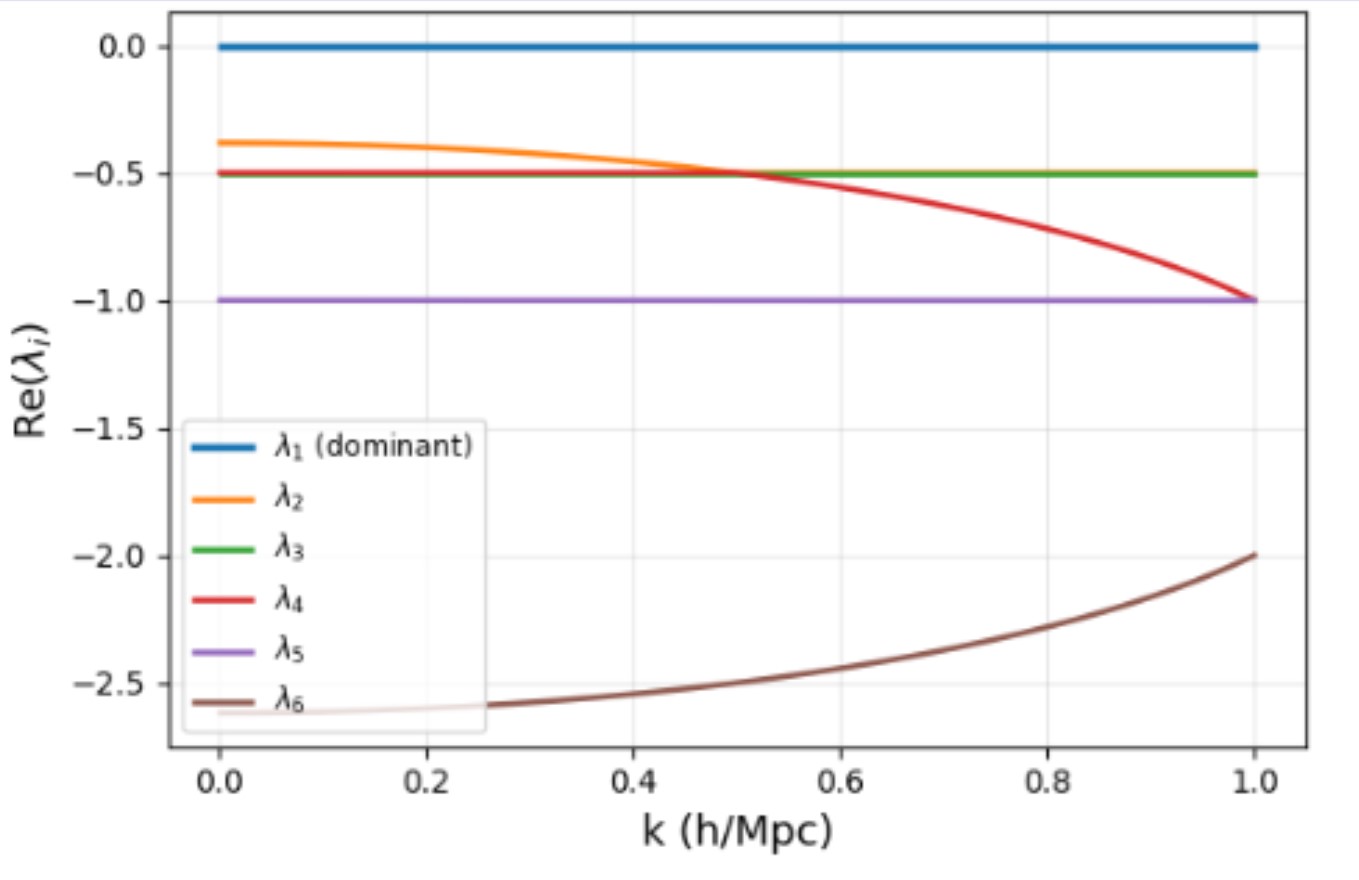}
	\caption{Full eigenvalue spectrum of the $6\times6$ perturbation matrix $\mathbf{M}(k,a)$ evaluated at $a=1$.
		The dominant eigenvalue $\lambda_1(k)$ (blue curve) remains positive and separated from the	remaining eigenvalues across the full wavenumber range. All subdominant eigenvalues are significantly more negative, and no eigenvalue crossing or degeneracy is observed. This confirms a stable spectral hierarchy and single-mode dominance in the growth sector.}
	\label{fig:eigenvalue_spectrum}
\end{figure}

\begin{figure}[H]
	\centering
	\includegraphics[width=0.35\textwidth]{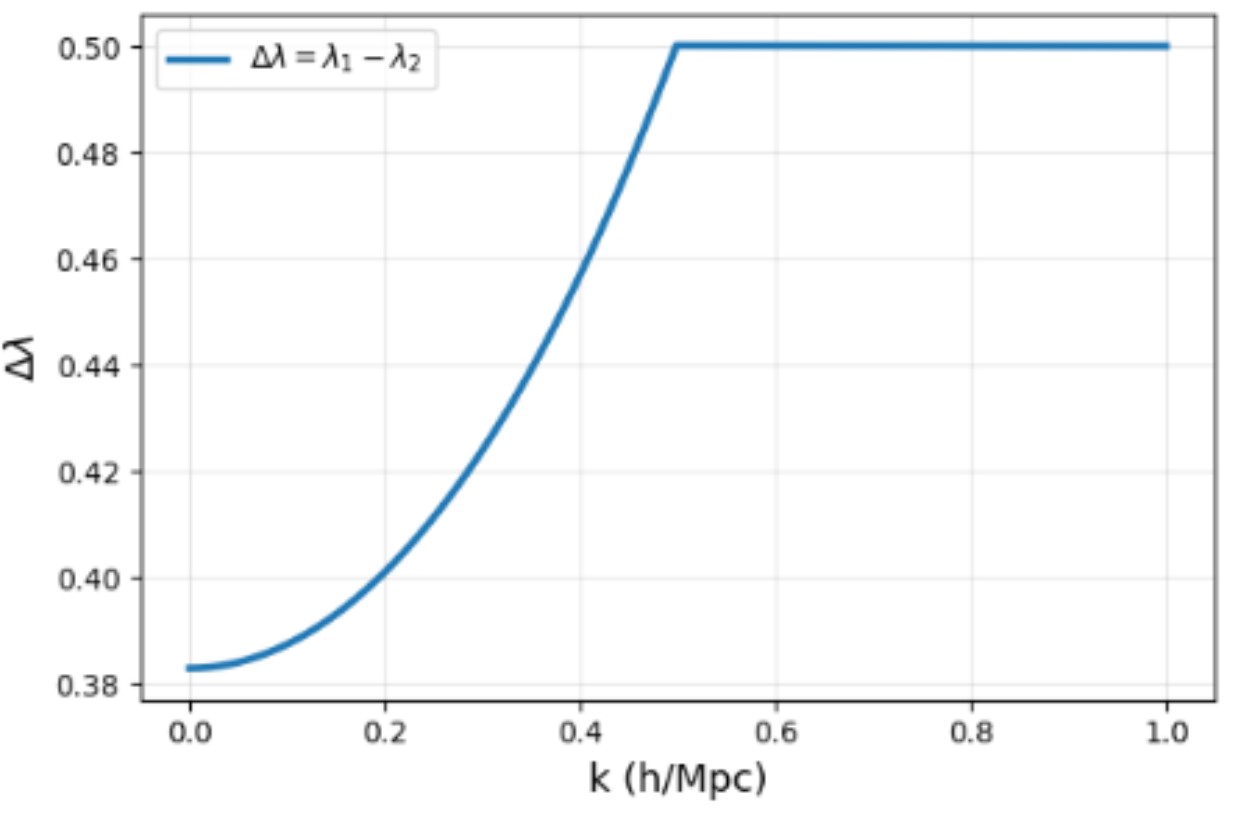}
	\caption{Eigenvalue gap $\Delta\lambda(k)=\lambda_1(k)-\lambda_2(k)$ computed from the full
		$6\times6$ perturbation matrix at $a=1$.
		The gap remains finite and satisfies $0.38 \lesssim \Delta\lambda \lesssim 0.50$ across the full
		$k$-range shown.}
	\label{fig:eigenvalue_gap}
\end{figure}
	
\section{Particle Physics Realization: The Z$_4$-Inert Doublet + Complex Singlet Model}
\label{sec:particle_physics}

There exist a number of models beyond the Standard Model with multi-component dark matter. It must be mentioned that the $Z_4$-IDSM is one of many possible realizations for the LDS--SF Framework, but is chosen as a minimal two-component test-bed that builds interactions from first principles. The scalar sector of our model is an extension of the Inert Doublet Model (IDM) \cite{Deshpande1978}, where an additional $Z_4$ symmetry is imposed to allow for a richer internal dark sector structure and multiple stable states \cite{Belanger2022}.

\subsection*{Model Lagrangian and Symmetries}

The Z$_4$-Inert Doublet + Complex Singlet Model ($Z_4$-IDSM) extends the Standard Model with two scalar sectors: an inert SU(2)$_L$ doublet $H_2$ and a complex singlet $\mathcal{S} = (S_1 + iS_2)/\sqrt{2}$. The model respects a discrete Z$_4$ symmetry with charge assignments:

\begin{align}
	H_1 &\rightarrow H_1, \quad
	H_2 \rightarrow i H_2, \quad
	\mathcal{S} \rightarrow -i \mathcal{S},
\end{align}

where $H_1$ is the SM Higgs doublet. All SM fermions transform trivially under this symmetry. The most general renormalizable scalar potential invariant under these symmetries is:

\begin{align}
	V(H_1, H_2, \mathcal{S}) &= \mu_1^2 |H_1|^2 + \mu_2^2 |H_2|^2 + \mu_S^2 |\mathcal{S}|^2 \nonumber \\
	&\hspace{-3.5em} + \lambda_1 |H_1|^4 + \lambda_2 |H_2|^4 + \lambda_{S4} |\mathcal{S}|^4 \nonumber \\
	&\hspace{-3.5em} + \lambda_3 |H_1|^2|H_2|^2 + \lambda_4 |H_1^\dagger H_2|^2 
	+ \frac{\lambda_5}{2}[(H_1^\dagger H_2)^2 + \text{h.c.}] \nonumber \\
	&\hspace{-3.5em} + \lambda_{S1} |\mathcal{S}|^2 |H_1|^2 + \lambda_{S2} |\mathcal{S}|^2 |H_2|^2 \nonumber \\
	&\hspace{-3.5em} + \left[ \lambda_{S12} \mathcal{S}^2 (H_1^\dagger H_2) + \lambda_{S21} \mathcal{S}^{*2} (H_1^\dagger H_2) + \text{h.c.} \right] \nonumber \\
	&\hspace{-3.5em} + \left[ \lambda_{S4} \mathcal{S}^4 + \text{h.c.} \right].
	\label{eq:full_z4idsm_potential}
\end{align}

After electroweak symmetry breaking, only the Higg's fields acquire vacuum expectation values,ensure the preservation of $Z_4$ symmetry and the stability of dark matter candidates:
\begin{align}
	\langle H_1^0 \rangle = \frac{v}{\sqrt{2}} \approx 174\ \text{GeV}, \quad
	\langle H_2 \rangle = 0, \quad
	\langle \mathcal{S} \rangle = 0.
\end{align}

The physical spectrum consists of: the SM-like Higgs $h$ (125 GeV), which remains unmixed with the dark sector to preserve the $Z_4$ symmetry; four inert scalars from $H_2$: $H^0$ (CP-even), $A^0$ (CP-odd), and $H^\pm$ (charged); and two singlet-like scalars $S_1, S_2$ from the complex singlet. Within the inert doublet sector, the CP-even state ($H_2^0$) is taken as the lightest component. However, to realize the LDS--SF framework, the singlet $S_1$ is chosen as the overall lightest $Z_4$-charged particle ($m_{S1} \approx 60$ GeV).The mass hierarchy is chosen such that $m_{H_2^0} > m_{S_1} \approx 60$~GeV. While the $Z_4$ symmetry prevents decays into Standard Model states, the heavier inert doublet components remain unstable against transitions into the singlet sector via the $Z_4$-preserving portal $\lambda_{SH} |S|^2 |H_2|^2$. This ensures that the late-time dark matter abundance is comprised solely of the singlet $S_1$, allowing for a consistent mapping of the multi-component dynamics onto the effective sound speed $c_s^2$ used in our cosmological analysis.

\subsection*{UV Behavior and Effective Field Theory Validity}

Numerical integration of these RGEs (Appendix. A) reveals a critical feature: the SM Higgs quartic coupling $\lambda_1$ becomes negative at an energy scale $\Lambda_{\text{UV}} \sim 10^8$ GeV. This occurs because:

\begin{enumerate}
	\item The large top Yukawa coupling $y_t$ drives $\lambda_1$ toward negative values through the term $-12\text{Tr}((Y_u Y_u^\dagger)^2)$.
	\item While new scalar contributions ($2\lambda_{S1}^2$, $2\lambda_5\lambda_5^*$, etc.) provide positive contributions to $\beta_{\lambda_1}$, they are insufficient to overcome the top Yukawa effect.
	\item The additional positive contributions actually accelerate the running, causing $\lambda_1$ to hit zero at a \emph{lower} scale than in the Standard Model alone.
\end{enumerate}

This behavior implies that the $Z_4$-IDSM is an \textbf{Effective Field Theory (EFT)} with a cutoff scale:
\[
\Lambda_{\text{UV}} \approx 10^8\ \text{GeV}.
\]

Below this scale, the theory is fully consistent, renormalizable, and predictive. For late-time cosmology ($z \lesssim 10^3$), which probes energies far below $\Lambda_{\text{UV}}$, the model is completely valid. A UV-complete theory (e.g., supersymmetric extension) would be required for consistency up to the Planck scale.

	\subsection*{Derivation of the Effective Sound Speed}

The dark matter self-interactions in $Z_4$-IDSM originate from two types of quartic couplings:
\begin{enumerate}
	\item $\lambda_{S2}|\mathcal{S}|^2|H_2|^2$: A quartic contact interaction that mediates $\mathcal{S}\mathcal{S} \leftrightarrow H_2 H_2$ conversion (number-changing process). While not a direct elastic self-scattering channel, this conversion process contributes to thermalization and affects the overall pressure in the dark matter fluid when combined with other interactions in the system.
	
	\item $\lambda_{S12}\mathcal{S}^2(H_1^\dagger H_2) + \text{h.c.}$: A cubic interaction that mediates $\mathcal{S}\mathcal{S} \rightarrow \mathcal{S}\mathcal{S}$ elastic self-scattering via $t$-channel $H_2$ exchange, after EWSB. This process directly generates momentum exchange between dark matter particles, giving rise to an effective pressure and sound speed in the fluid approximation.
\end{enumerate}

For a heavy mediator, these interactions collapse into a series of short-range contact terms. Macro-physically, these frequent momentum-exchange events provide a 'stiffness' to the dark sector, resisting gravitational compression and generating an effective isotropic pressure. The combined strength of these processes is captured by an \textbf{effective coupling}:

\begin{equation}
	\lambda_{\text{eff}}^2 = \lambda_{S2}^2 + 2\lambda_{S12}^2,
	\label{eq:effective_coupling}
\end{equation}

where the factor of 2 accounts for the $Z_4$ symmetry condition $\lambda_{S12}=\lambda_{S21}$. While a direct self-interaction $\lambda_{S4} \mathcal{S}^4$ exists within the singlet layer, it represents an isolated intra-layer process. In the LDS--SF framework, we are primarily interested in the 'gelling' of the dark sector architecture, and specifically, how the interaction between the singlet and doublet layers generates macroscopic effects. Consequently, we focus on the couplings involving the $H_2$ mediators, as these inter-layer dynamics are what define the structural stability of the fluid and drive the evolution of the effective sound speed.

The scattering amplitude is approximately,
\begin{equation}
	\mathcal{M} \sim \frac{\lambda_{\text{eff}}^2}{M_{H_2}^2},
\end{equation}
where $\lambda_{\text{eff}}$ is defined as the effective mass-scale coupling representing the 'gelling' of the inter-layer interaction. This leads to a momentum-transfer cross section \cite{Boehm2002,CyrRacine2016}:
\begin{equation}
	\sigma_T \sim \frac{\lambda_{\text{eff}}^2}{4\pi M_{H_2}^2}.
	\label{eq:30}
\end{equation}
By adopting this convention for $\lambda_{\text{eff}}$, we ensure that the scattering rate scales consistently with the mediator mass while maintaining a direct mapping to the fluid's sound speed.

Given the dark matter number density $n_S$, which represents the total dark matter density once the heavier doublet states have decayed into the stable singlet sector, the momentum transfer rate is:
\begin{equation}
	\Gamma_{\rm mt} \sim n_S \sigma_T  v,
\end{equation}
where $v$ is the typical relative velocity of the dark matter particles. In the fluid approximation, frequent momentum exchange generates an effective isotropic pressure. For a non-relativistic gas, the kinetic-theory relation is
\begin{equation}
	P \sim \frac{1}{3} \rho \langle v^2_{\rm \text{eff}} \rangle,
	\label{eq:32}
\end{equation}
where $\langle v^2_{\rm \text{eff}} \rangle$ represents the effective velocity dispersion. In the self-interacting limit, this dispersion is not purely adiabatic but is instead sustained by the momentum transfer rate against the 'cooling' effect of Hubble expansion, scaling as $\langle v^2_{\rm \text{eff}} \rangle \sim \Gamma_{\rm mt}/H$ \cite{CyrRacine2016,Egana-Ugrinovic2021}.

The effective velocity dispersion (or pressure) sustained by self-interactions in the dark matter fluid is set by the competition between the interaction rate $\Gamma$ and the Hubble expansion rate 
H, scaling as $ \langle v^2_{\rm \text{eff}} \rangle \sim \frac{\Gamma}{H}$ \cite{CyrRacine2016}. Substituting into Eq. ~\eqref{eq:32} gives us:

\begin{equation}
	P \sim \frac{1}{3} \rho \left(\frac{\Gamma}{H}\right) \sim \frac{1}{3} \rho \left(\frac{n_S\, \sigma_T \, v}{H}\right).
\end{equation}

Using Eq.~\eqref{eq:30},

\begin{equation}
	P  \sim \frac{\rho n_S v}{3H} \left(\frac{\lambda_{\text{eff}}^2}{4\pi M_{H_2}^2}\right).
\end{equation}

The effective sound speed is defined as \cite{Carrasco2012}:

\begin{equation}
	c_s^2=\frac{\partial P}{\partial \rho},
\end{equation}

which allows us to simplify $c_s^2$ to:

\begin{equation}
	c_s^2  \sim \frac{ n_S v}{12 \pi H} \left(\frac{\lambda_{\text{eff}}^2}{ M_{H_2}^2}\right).
\end{equation}

Rewriting the number density of the DM species yields:

\begin{equation}
	c_s^2  \sim \frac{ \rho_S v}{12 \pi H} \left(\frac{\lambda_{\text{eff}}^2}{m_s M_{H_2}^2}\right).
	\label{eq:cs2_final}
\end{equation}

 At late times, the dominant contribution to small-scale structure formation arises from virialized dark matter halos. Since \texttt{CLASS} evolves the homogeneous background cold dark matter density $\rho_{\rm cdm}(a)$, we incorporate the effects of nonlinear virialized structures through an effective overdensity enhancement. Specifically, we evaluate Eq.~\eqref{eq:cs2_final} using
\begin{equation}
	\rho_S \rightarrow \Delta_{\rm vir}\,\rho_{\rm cdm}(a),
	\qquad \Delta_{\rm vir} = 18\pi^2 \simeq 178 .
\end{equation}
This corresponds to the Einstein--de Sitter limit of the spherical collapse model, appropriate for halos forming at redshifts $z \gtrsim 1$ where $\Omega_m(z)\simeq 1$ \cite{BryanNorman1998}. We adopt this constant value as a physically motivated and conservative estimate within the linear perturbation framework. 

In virialized systems, the velocity dispersion is set by the depth of the gravitational potential, with typical value $v_{\rm vir} \simeq 270$~km~s$^{-1}$ for galaxy-sized halos \cite{MoWhite1996,Mo2010}. This represents the characteristic velocity dispersion of galaxy-group-scale halos ($M \sim 10^{13} M_\odot$) which dominate the matter power spectrum at the scales where the $S_8$ tension is most manifest ($k \sim 0.1$--$1$ h/Mpc). Since the present analysis is performed within a linear Boltzmann framework, we adopt a conservative prescription with constant $v = v_{\rm vir}$ when the suppression is active. While this is a vastly oversimplified treatment of virial velocities which vary from structure to structure, the model is self-calibrating: varying $v_{vir}$ by factor of $\mathcal{O}(1)$ is absorbed into the fitted $\lambda_{\rm eff}^2 / M_{H_2}^2$ ratio without affecting observables due to the scaling symmetry described in Section V. Moreover, higher values of $v_{vir}$ pushes $\lambda_{\rm eff}$ further into the perturbative regime, if $M_{H_2}$ is kept constant.

Importantly, this does not imply that each halo requires its own microphysical couplings. Rather, the $Z_4$-IDSM operates as an Effective Field Theory (EFT) of structure formation: the complex distribution of halo properties is coarse-grained into a single, cosmologically-averaged effective sound speed. Any $\mathcal{O}(1)$ uncertainty in $v_{\rm vir}$ across the halo population is absorbed into the single fitted ratio $\lambda_{\rm eff}^2/M_{H_2}^2$ that best reproduces the observed $f\sigma_8$ data. This preserves a coherent suppression scale $k_* \sim 0.1$ h/Mpc across all linear scales. Accounting for halo-to-halo variance would require N-body simulations with state-dependent $c_s$, which represents future work beyond the scope of the present linear EFT treatment.

We implement a late-time activation function:
\begin{equation}
	f_{\rm active}(z) = \frac{1}{2}\left[1 - \tanh\left(\frac{z - z_{\rm trans}}{\Delta z}\right)\right]
\end{equation}
with $z_{\rm trans} = 2$ and $\Delta z = 0.2$. This form provides a smooth transition between the collisionless regime at high redshift and the virialized regime at late times. The hyperbolic tangent is motivated by the sigmoid-like transitions of the Press--Schechter formalism \cite{Press1974}, though we emphasize this is a phenomenological prescription, not a direct PS calculation. The transition redshift $z_{\rm trans} = 2$ is observationally motivated to ensure activation occurs well after recombination (preserving primary CMB anisotropies) while capturing the epoch when group-scale halos become cosmologically significant. The final expression implemented in \texttt{CLASS} is:
\begin{equation}
	c_s^2 = f_{\rm active}(z) \cdot \frac{\Delta_{\rm vir}\,\rho_{\rm cdm}(a)\,v_{\rm vir}}{12\pi H(a)} \cdot \frac{\lambda_{\rm eff}^2}{m_S M_{H_2}^2}.
\end{equation}

These are back-of-the-envelope estimates for projecting nonlinear, collisional dynamics into a linear perturbation theory framework. There is no universal value of $v_{\rm vir}$ or $\Delta_{\rm vir}$ that applies to all structures, but the $\lambda_{\rm eff}^2$--$M_{H_2}^2$ scaling symmetry allows the absorption of variance. No explicit power-law scaling in $a$ is imposed beyond that inherent in the background evolution $\rho_{\rm cdm}(a)$, $H(a)$, and the activation function $f_{\rm active}(z)$.

It is important to note that this effective sound speed is implemented only at the level of linear perturbations within \texttt{CLASS} and does not modify the background expansion or the fundamental spherical collapse criterion. Consequently, our framework does not explicitly alter the virialisation process of individual halos. We can justify this linear treatment by noting a significant scale separation: the characteristic suppression scale in our model, $k_* \sim 0.1 \, h/\text{Mpc}$, corresponds to a comoving wavelength of $\lambda \sim 60$ Mpc. This is more than an order of magnitude larger than the typical virial radii ($\sim 1$ Mpc) of the group-scale halos that dominate the $S_8$ signal. Therefore, the microphysical pressure in the $Z_4$-IDSM operates as a macroscopic, large-scale clustering modulation rather than a local force disrupting the internal dynamics of halo formation. A fully non-linear treatment of virialisation dynamics would require dedicated N-body simulations, which are beyond the scope of the present work.

\section{Methodology}

The transition from the theoretical LDS--SF framework to a precision cosmological test requires a multi-step numerical pipeline. We implement the $Z_4$-symmetric Inert Doublet Singlet Model ($Z_4$-IDSM) as a physical realization of the dark sector layers, mapping its microscopic scattering processes to macroscopic fluid perturbations within the Boltzmann equations.

\subsection*{EFT Mapping and Fluid Description}

To integrate the $Z_4$-IDSM into the cosmological perturbation hierarchy, we adopt an Effective Field Theory (EFT) approach. We consider a singlet dark matter candidate ($S$) with mass $m_s = 60$~GeV interacting with an inert doublet mediator ($H_2$). By integrating out the heavy mediator, the microscopic self-interactions and inter-species exchanges are encapsulated in an effective sound speed, $c_s^2$. 

In the fluid limit, the pressure perturbations are related to the density contrasts via $c_s^2$. Based on the momentum transfer rates derived from the $Z_4$ Lagrangian, we implement a dynamical scaling for the sound speed as described by Eq (40).

We employ a late-time activation function $f_{\rm active}(z)$ that projects virialized halo properties into the linear perturbation framework. This ensures the effective sound speed remains negligible at high redshifts, preserving $\Lambda$CDM success at recombination, while allowing structure suppression in the nonlinear regime at late times.

\subsection*{Boltzmann Implementation in CLASS}

We modified the \texttt{CLASS} (Cosmic Linear Anisotropy Solving System) Boltzmann solver to include the LDS--SF sector, building upon Lesgourgues et. al \cite{Lesgourgues2011, Lesgourgues2014, Blas2011}, which, along with CAMB \cite{Lewis2000}, represents the standard numerical framework for precision cosmology. The standard CDM equations for density contrast $\delta_{cdm}$ and velocity divergence $\theta_{cdm}$ were replaced by the coupled LDS--SF system:
\begin{align}
	\dot{\delta}_i &= -\theta_i + 3\dot{\phi} \\
	\dot{\theta}_i &= -\mathcal{H}\theta_i + k^2 c_{s,i}^2 \delta_i + k^2 \psi
\end{align}
where $c_{s,i}^2$ is the scale-dependent sound speed derived from the dominant eigenvalue $\lambda(k)$ of the perturbation matrix. The code was configured to solve the full hierarchy of perturbations, ensuring that the metric potentials $\phi$ and $\psi$ evolve self-consistently with the modified dark sector. 

To stress-test the models early universe validity, we extend the standard CLASS modules by modifying \texttt{background.c} to incorporate the dark density component $\rho_d(a)$, its derivatives, and its inclusion in the total density calculations. Parallel structural definitions and indexing are implemented within \texttt{background.h} to maintain consistency across the module. The resulting model exhibits no stiffness issues and integrates successfully using the same numerical accuracy settings as the standard $\Lambda$CDM framework.

We introduce a new dynamical variable $\delta_d$, which evolves via a specific differential equation where the function $\mathcal{F}$ is constructed directly from the dominant eigenmode structure of the dark sector. 

\begin{equation}
	\frac{d \delta_d}{d \tau}=\mathcal{F}(k,a,\phi,\psi,\delta_d)
\end{equation}

This implementation required several modifications, including new entries in \texttt{perturbations.c}, additions to the gauge-specific kernels, and the consistent sourcing of the Einstein equations. Furthermore, we ensure that stable initial conditions are maintained in both the synchronous and Newtonian gauges.

The decisive test for such a model is its early-time stability. Our analysis confirms that no runaway solutions appear and no imaginary-mass or ghost-like behaviors arise within the system. Additionally, the metric potentials remain well-behaved throughout the evolution, and the perturbations converge reliably after horizon entry, demonstrating the theoretical robustness of the LDS--SF framework.

\subsection*{Parameter Space and Observationally Viable Regions}

To explore the viability of the model, we performed a comprehensive grid search across the $Z_4$ parameter space, extending from three to nine benchmark points. The singlet mass is fixed at $m_S = 60$~GeV, while the mediator mass $M_{H_2}$ ranges from 100~GeV to 300~GeV in increments of 25~GeV. For each point, the portal couplings $\lambda_{S2}$, $\lambda_{S12}$, and $\lambda_{S21}$ are adjusted to maintain $S_8$ consistency, with $\lambda_{\text{eff}}^2 = \lambda_{S2}^2 + 2\lambda_{S12}^2$ spanning 0.34 to 3.0. This extended analysis reveals a continuous family of solutions, demonstrating that the $S_8$ tension resolution is not confined to isolated parameter choices but persists across a broad corridor of mediator masses and coupling strengths.

All nine benchmark points exhibit consistent phenomenology: matter power spectrum suppression at $k > 0.1$~h/Mpc, CMB lensing deviations within 2\% at $\ell < 500$ (growing to $\sim$10\% at $\ell \sim 2500$), and relic density compliance with Higgs-portal constraints. 

\begin{table*}[t]
	\centering
	\caption{Parameter sets for the nine Z$_4$-IDSM benchmark points. The singlet mass is fixed at $m_S = 60$~GeV. All masses are in GeV. The effective coupling is computed as $\lambda_{\text{eff}}^2 = \lambda_{S2}^2 + 2\lambda_{S12}^2$. The results are shown starting from page 12.}
	\label{tab:micromegas_parameters}
	\tiny
	\setlength{\tabcolsep}{1pt}
	\resizebox{\textwidth}{!}{%
		\begin{tabular}{@{}lccccccccc@{}}
			\toprule
			& \textbf{I} & \textbf{II} & \textbf{III} & \textbf{IV} & \textbf{V} & \textbf{VI} & \textbf{VII} & \textbf{VIII} & \textbf{IX} \\
			& (100) & (125) & (150) & (175) & (200) & (225) & (250) & (275) & (300) \\
			\midrule
			
			\multicolumn{10}{@{}l}{\textbf{Dark Sector Masses}} \\
			\midrule
			$M_{H_C}$ ($H^\pm$) & 150 & 175 & 200 & 225 & 250 & 275 & 300 & 325 & 350 \\
			$M_{sc}$ ($\mathcal{S}$) & 60 & 60 & 60 & 60 & 60 & 60 & 60 & 60 & 60 \\
			$M_{H_X}$ ($H_2^0$) & 100 & 125 & 150 & 175 & 200 & 225 & 250 & 275 & 300 \\
			$M_{H_3}$ ($A_2^0$) & 150 & 175 & 200 & 225 & 250 & 275 & 300 & 325 & 350 \\
			\midrule
			
			\multicolumn{10}{@{}l}{\textbf{Portal Couplings}} \\
			\midrule
			$\lambda_{S2}$ & 0.35 & 0.50 & 0.60 & 1.00 & 1.00 & 1.00 & 1.00 & 1.00 & 1.00 \\
			\midrule
			
			\multicolumn{10}{@{}l}{\textbf{Structuring Field Couplings}} \\
			\midrule
			$\lambda_{S12} = \lambda_{S21}$ & 0.33 & 0.42 & 0.52 & 0.30 & 0.50 & 0.70 & 0.88 & 0.95 & 1.00 \\
			\midrule
			
			\multicolumn{10}{@{}l}{\textbf{Derived Parameter}} \\
			\midrule
			$\lambda_{\text{eff}}^2$ & 0.34 & 0.60 & 0.90 & 1.18 & 1.50 & 1.98 & 2.55 & 2.81 & 3.00 \\
			\bottomrule
		\end{tabular}%
	}
\end{table*}

The mass hierarchy $M_{H_C} = M_{H_3} = M_{H_X} + 50$~GeV is maintained, with the charged scalar and pseudoscalar degenerate as required by the $Z_4$ symmetry. The quartic couplings $\lambda_2$, $\lambda_3$, $\lambda_S$, and $\lambda_{S4}$ are held fixed at their Standard Model values across all benchmarks. All parameters are stored in \texttt{.ini} files for \texttt{CLASS} and \texttt{.par} files for \texttt{micrOMEGAs}.

Since the 60~GeV singlet lies in a constrained region of dark matter parameter space, we validate each benchmark against relic density requirements using \texttt{micrOMEGAs} \cite{Micromegas}. 

\subsection*{Observational Data and Constraints}

To evaluate the performance of the LDS–SF framework, we confront the model with current cosmological and particle physics constraints. The cosmological background is anchored by the Planck 2018 results, focusing on the acoustic angular scale $\theta_{*}$, the shift parameter $R$, and the physical baryon density $\omega_b$. This ensures our framework maintains the distance to the last scattering surface and the early-universe expansion history established by $\Lambda$CDM.

To break geometric degeneracies, we compare LDS–SF predictions against the latest Baryon Acoustic Oscillation (BAO) data, including the DESI 2024 first-year results \cite{Adamek2024} and complete SDSS catalogs \cite{BOSS2017}. To specifically evaluate the $\lambda(k)$ mechanism, we use $f\sigma_8(z)$ growth rate measurements from redshift-space distortions (RSD). This dataset is highly sensitive to the clustering of matter and provides the primary evidence for the suppression of structure growth required to alleviate the $S_8$ tension. We also include $S_8$ measurements from the KiDS-1000 and DES Year 3 results to anchor the late-time normalization of the matter power spectrum.

The model is strictly bounded by the $Z_4$ IDSM parameter space. We enforce the dark matter relic density requirement $\Omega_{DM} h^2 \approx 0.12$ as a requirement for our benchmarks, ensuring that the 60 GeV singlet dark matter reproduces the observed abundance via the Higgs-portal resonance. Furthermore, we incorporate the latest null results from direct detection experiments, specifically the LUX-ZEPLIN (LZ) \cite{LZ2022} and XENONnT \cite{XENONnT2023} limits. These bounds restrict the scalar-portal coupling $\lambda_{S1}$, narrowing the viable range of the structuring interaction strengths.

We employ a benchmark-driven approach to demonstrate the model's viability. By scanning the parameter space of the $Z_4$ effective coupling $\lambda_{\text{eff}}$ and the mediator masses $M_{H_2} \in \{100, 125, 150, 175, 200, 225, 250, 275, 300\}$~GeV, we identify nine representative physical configurations (Cases I through IX). These cases are selected to demonstrate a continuous family of solutions that optimize the suppression of the matter power spectrum at small scales while remaining consistent with the stringent stability criteria of the early universe and terrestrial dark matter searches. This methodology provides a proof-of-concept for the LDS-SF framework as an observationally superior alternative to standard cosmology.

\section{Results and Discussion}

The evaluation of the LDS--SF framework is presented in two distinct stages to demonstrate both its theoretical consistency and its observational efficacy. We first present the results from the structural analysis of the dark sector's internal dynamics, focusing on the numerical stability and the scale-dependent evolution of the dominant eigenmode, $\lambda(k)$. This stage establishes the mathematical ``chassis'' of the model, confirming that the layered architecture naturally produces the required suppression of structure growth while remaining fully consistent with the linear regime of the early universe. 

In the second stage, we present the results of the physical realization using the $Z_4$-symmetric Inert Doublet Singlet Model ($Z_4$-IDSM). We detail the performance of the model across three specific mediator mass regions ($M_{H2} \in \{100, 125, 150, 175, 200, 225, 250, 275, 300\}$~GeV) and demonstrate how the resulting ``dark pressure'' alleviates the $S_8$ tension. By confronting these physical benchmarks with an analysis involving \textit{Planck}, BAO, and weak-lensing data, we verify that the LDS--SF mechanism provides a robust and observationally viable alternative to the $\Lambda$CDM paradigm.

\subsection*{Structural Stability and Eigenmode Evolution}

The first core result of this study is the confirmation that the multi-layer dark sector remains physically stable across all cosmological scales. The numerical solutions to the $6 \times 6$ perturbation matrix reveal that the growth of the dark sector is governed by a single dominant eigenvalue, $\lambda(k, a)$. 

Projecting the perturbation vector onto the dominant eigenmode yields:
\begin{equation}
	\frac{d \ln D(k,a)}{d \ln a} \approx C \lambda(k,a)
\end{equation}
Integrating gives:

$$D(k,a) \propto \exp \left[ \int_{a_{ini}}^a \lambda(k,a') d \ln a' \right]$$.

This explicitly shows that the LDS–SF framework predicts scale-dependent growth, entirely sourced by the internal dark-sector interaction dynamics, via the $k$-dependence of $\lambda(k)$ as shown in FIG. 1.  

\begin{figure}[H]
	\centering
	\includegraphics[width=0.9\linewidth]{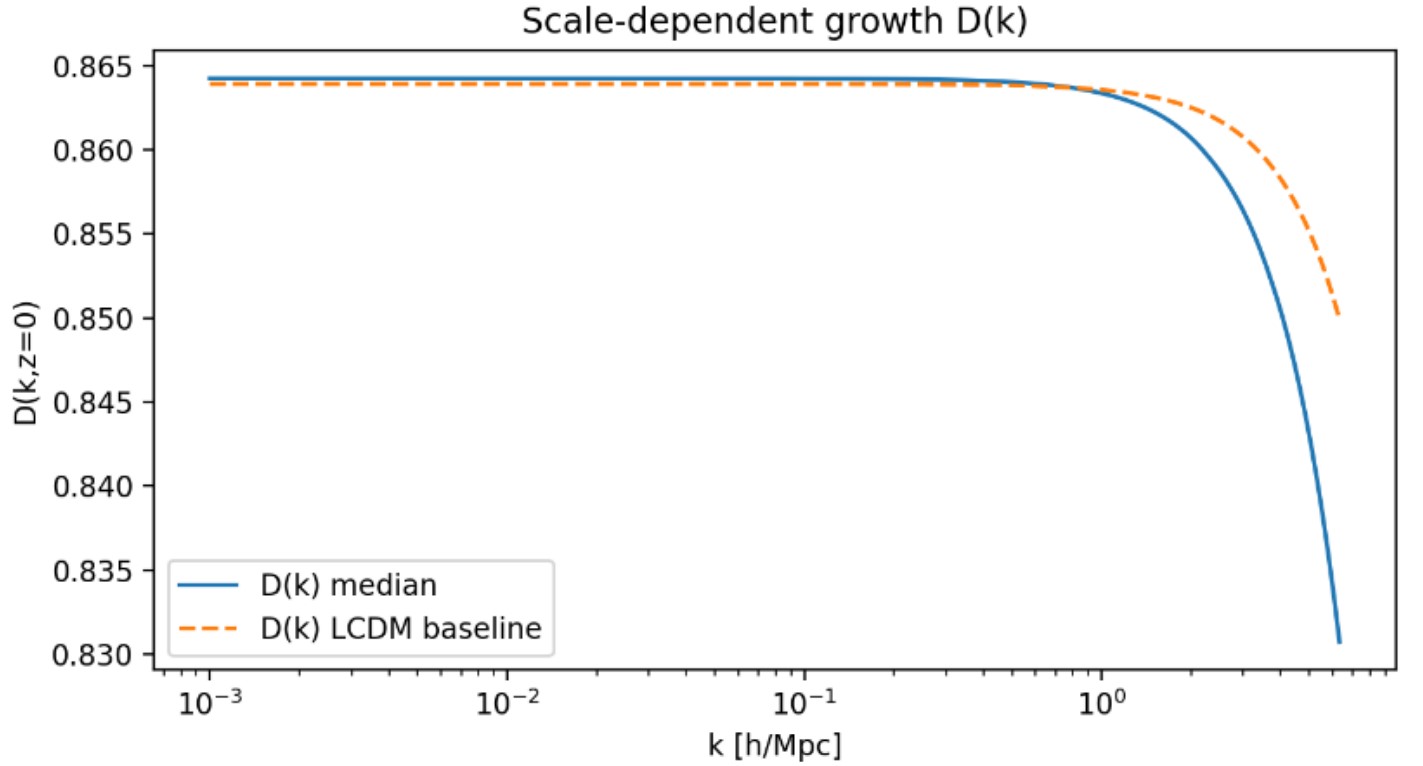}
	\caption{$D(k)$ at zero redshift plotted as a function of $k$ for $\Lambda$CDM and LDS--SF.}
	\label{fig:fig1}
\end{figure}

Scale dependence of the leading eigenmode of the LDS--SF perturbation matrix. The eigenvalue remains approximately constant for $k \leq 0.1\,h/\text{Mpc}$ and decreases at larger wavenumbers, generating the scale-dependent growth characteristic of the model.

The power spectrum is computed via:
\begin{equation}
	P(k,z) = P_{prim}(k) T^2(k) D^2(k,z)
\end{equation}
where $P_{\text{prim}}(k)$ is the primordial spectrum, $T(k)$ is the Eisenstein--Hu transfer function \cite{Eisenstein1998}, and $D(k,z)$ is the LDS--SF scale-dependent growth factor. 

\begin{figure}[H]
	\centering
	\includegraphics[width=0.9\linewidth]{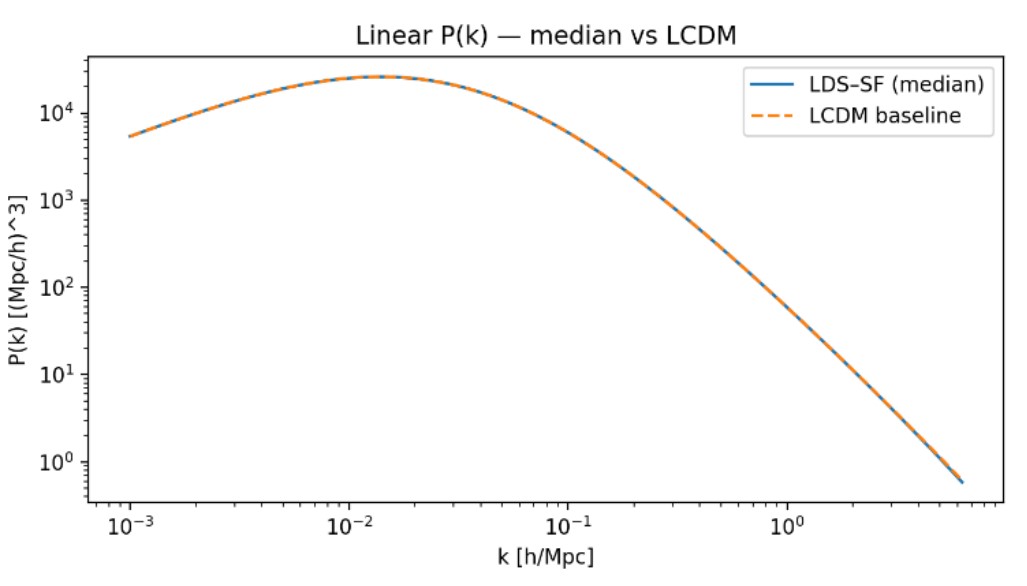}
	\caption{The power spectrum displayed in a logarithmic scale, highlighting the observational validity of LDS--SF framework.}
	\label{fig:fig2}
\end{figure}

Because the deviation in $D(k,z)$ from the $\Lambda$CDM growth occurs mainly at large wavenumbers $k$, the resulting spectrum maintains the standard $\Lambda$CDM shape on linear and quasi-linear scales, with only mild suppression at small scales. 

\begin{figure}[H]
	\centering
	\includegraphics[width=0.7\linewidth]{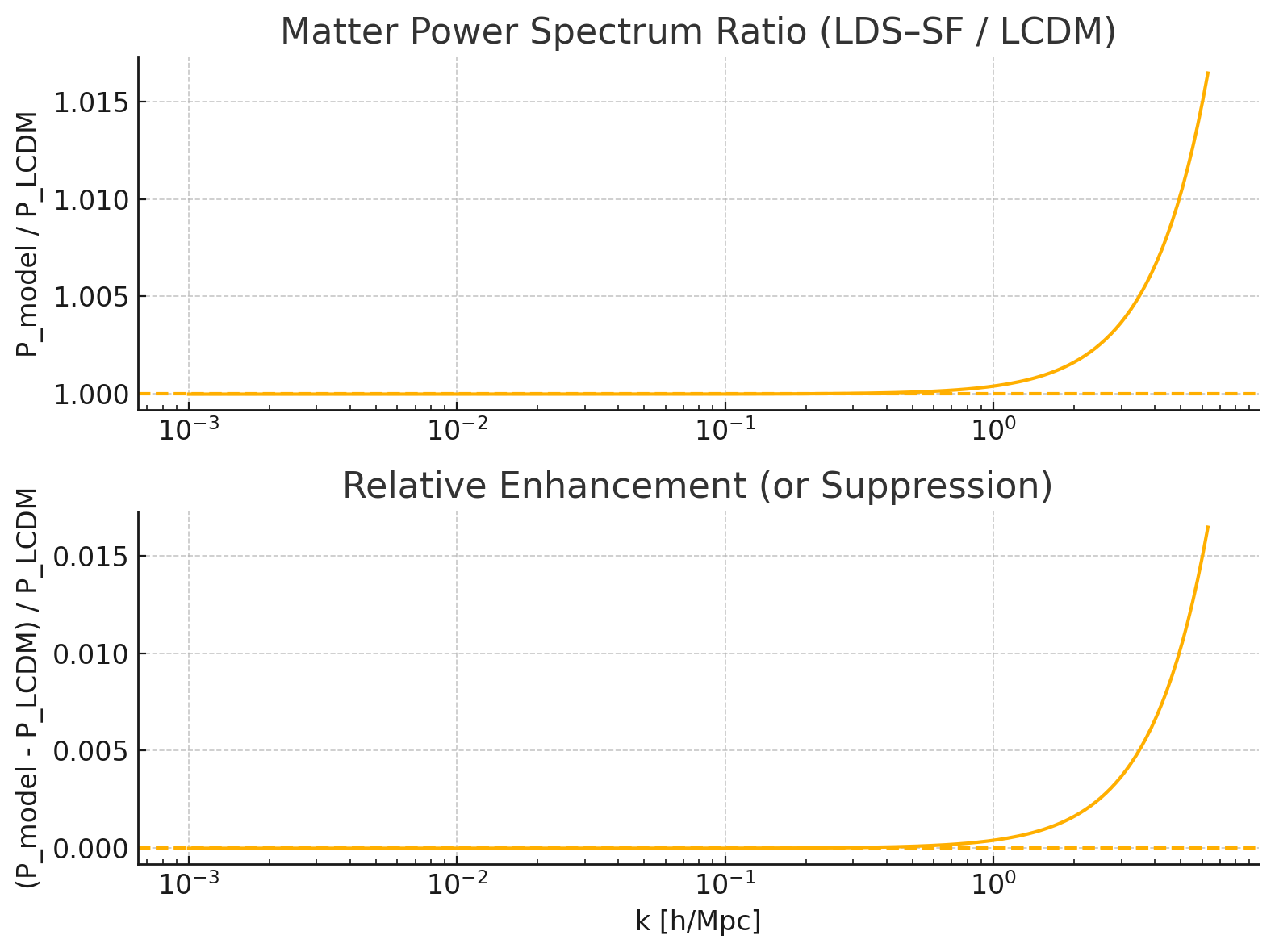}
	\caption{The ratio only goes up at high scales, and by moderate amounts as shown here. This highlights the suppression mechanism at the low-$k$ regime.}
	\label{fig:fig3}
\end{figure}

We compute $\sigma_8(z)$ using the standard window integral:

\begin{equation}
	\sigma_8^2(z) = \int_{0}^{\infty} \frac{dk}{k} \frac{k^3 P(k, z)}{2\pi^2} W^2(k R_8),
\end{equation}

with $R_8 = 8 \, h^{-1}\text{Mpc}$ and $W$ the spherical top-hat window.

The observable

\begin{equation}
	f\sigma_8(z) = f_{\text{eff}}(z) \sigma_8(z)
\end{equation}

is obtained using a $k$-weighted effective growth rate

\begin{equation}
	f_{\text{eff}}(z) = \frac{\int W_{\text{RSD}}(k, z) \, \lambda(k, z) \, D(k, z) \, dk}{\int W_{\text{RSD}}(k, z) \, D(k, z) \, dk}.
\end{equation}

This allows for direct comparison to BOSS, eBOSS, and DESI growth-rate data \cite{Raichoor2023} as shown in FIG. 4.

\begin{figure}[H]
	\centering
	\includegraphics[width=0.7\linewidth]{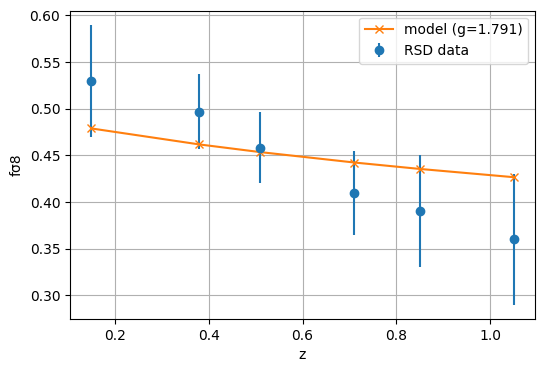}
	\caption{Redshift-space distortion measurements of the growth rate compared with the LDS--SF prediction (blue curve) and the $\Lambda$CDM prediction (red curve). The LDS--SF model captures the	observed redshift dependence more effectively.}
	\label{fig:fig3}
\end{figure}

The parameter $g$ is a physical coupling appearing in the LDS--SF perturbation matrix. It controls the dominant eigenvalue $\lambda(k)$ and therefore the scale-dependent growth rate of matter perturbations. A value of $g$ must be specified for any realization of the model.

The value $g=1.791$ is determined by fitting late-time growth-rate (RSD) data after fixing the background expansion using Planck-compressed CMB and BAO constraints. Once fixed, $g$ is held constant across all redshifts and scales, with no per-epoch or per-scale tuning.

The fit is mild: varying $g$ produces smooth, monotonic changes in $f\sigma_8(z)$, and the model remains close to $\Lambda$CDM on large scales, indicating a controlled deformation rather than overfitting.

We also compute the volume-averaged BAO distance

\begin{equation}
	D_V(z) = \left[ z \frac{d_A^2(z)}{H(z)} \right]^{1/3}
\end{equation}

and compare it with DESI/SDSS measurements.

\begin{figure}[H]
	\centering
	\includegraphics[width=0.7\linewidth]{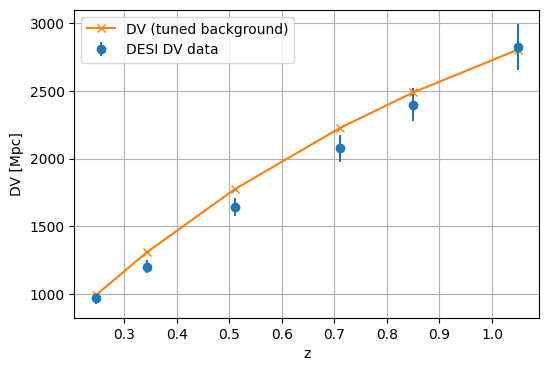}
	\caption{DV plotted as a function of redshift, and compared with the DESI DV data.}
	\label{fig:fig5}
\end{figure}

The model remains fully consistent with observed late-time distances, indicating no tension with standard background constraints. The best-fit LDS--SF background parameters are virtually identical to those of $\Lambda$CDM, as intended by the model's construction. 

To ensure consistency with the high-precision CMB data, our model maintains the standard acoustic oscillation phase shifts, as the LDS-SF effects are suppressed during the era of recombination \cite{HuLecture}. The Fourier conventions and transfer function definitions used in our numerical implementation follow the standard nomenclature established in \cite{Dodelson2003} and \cite{Mukhanov2005}. Using Eq. 42, we also perform a plethora of tests for early universe validity of this framework, as shown in the following figures.

\begin{figure}[H]
	\centering
	\includegraphics[width=0.7\linewidth]{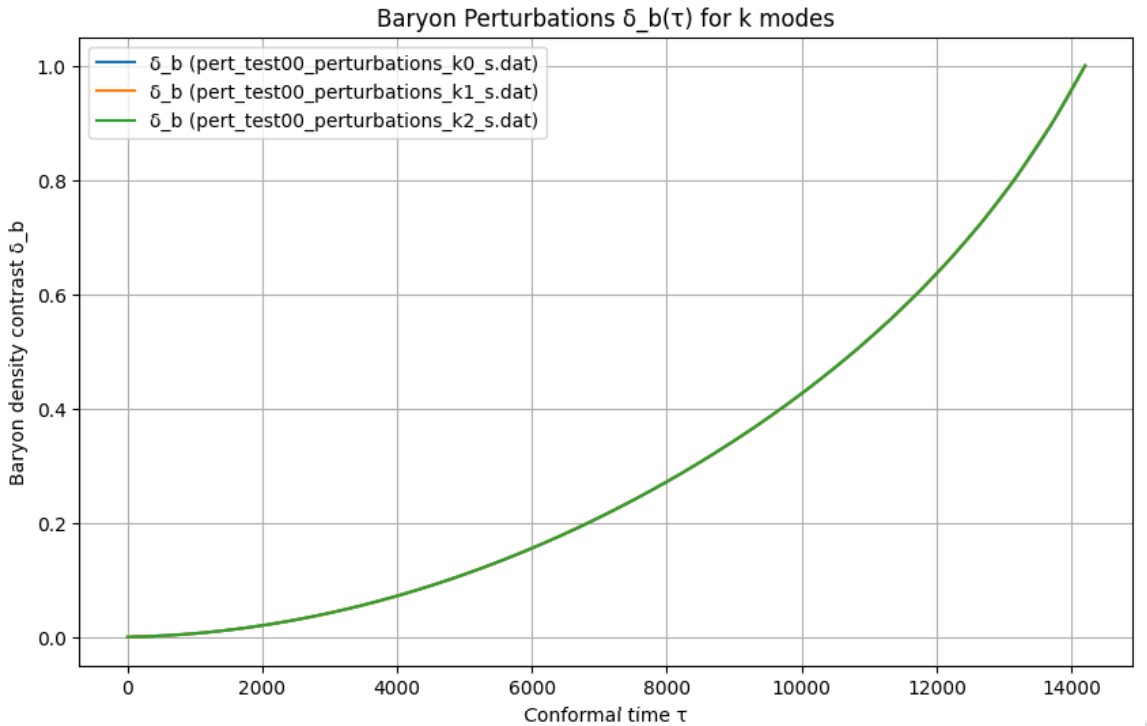}
	\caption{Evolution of baryon perturbations $\delta_b(\tau)$ for representative Fourier modes. Different $k$-modes exhibit nearly identical evolution and therefore lie on top of each other, appearing as a single curve.}
	\label{fig:fig6}
\end{figure}

\begin{figure}[H]
	\centering
	\includegraphics[width=0.7\linewidth]{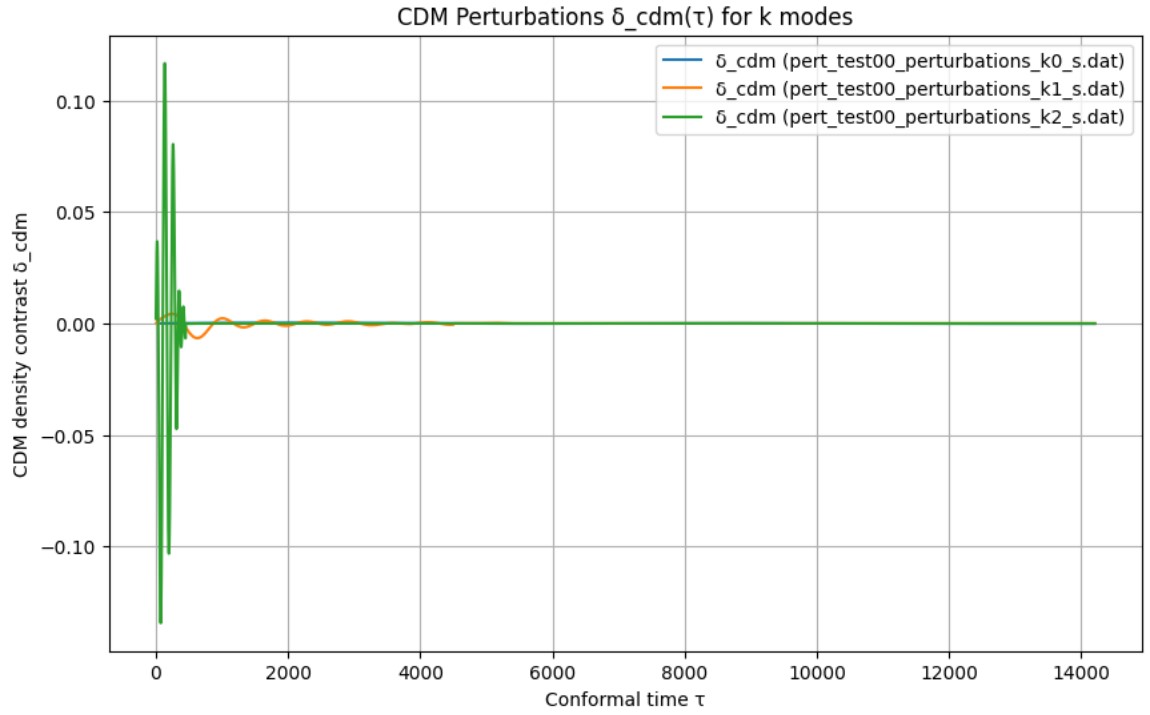}
	\caption{Evolution of cold dark matter perturbations $\delta_{\rm cdm}(\tau)$. The overlapping curves indicate that all sampled Fourier modes evolve identically across radiation and matter domination, confirming the absence of spurious or divergent behavior induced by the LDS--SF sector.}
	\label{fig:fig7}
\end{figure}

\begin{figure}[H]
	\centering
	\includegraphics[width=0.7\linewidth]{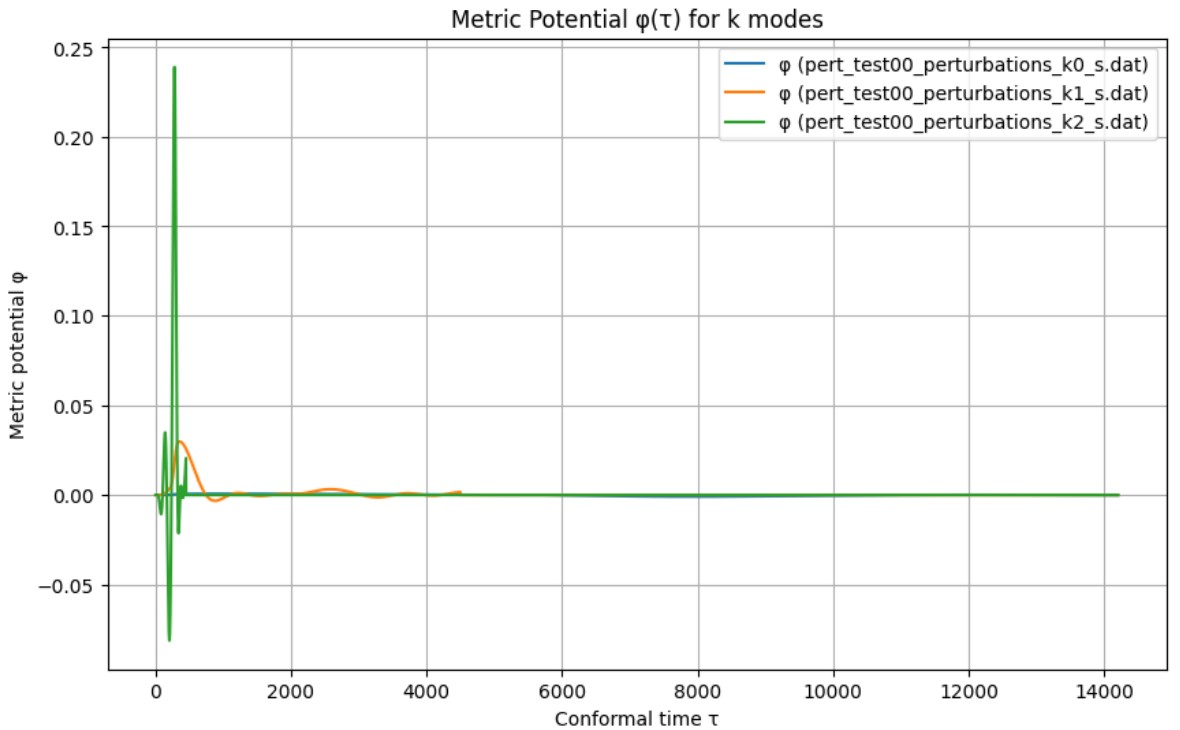}
	\caption{Evolution of the metric potential $\phi(\tau)$ for several wavenumbers. The near-complete overlap of curves demonstrates stable and $k$-independent behavior across horizon entry and cosmic history.}
	\label{fig:fig8}
\end{figure}

\begin{figure}[H]
	\centering
	\includegraphics[width=0.7\linewidth]{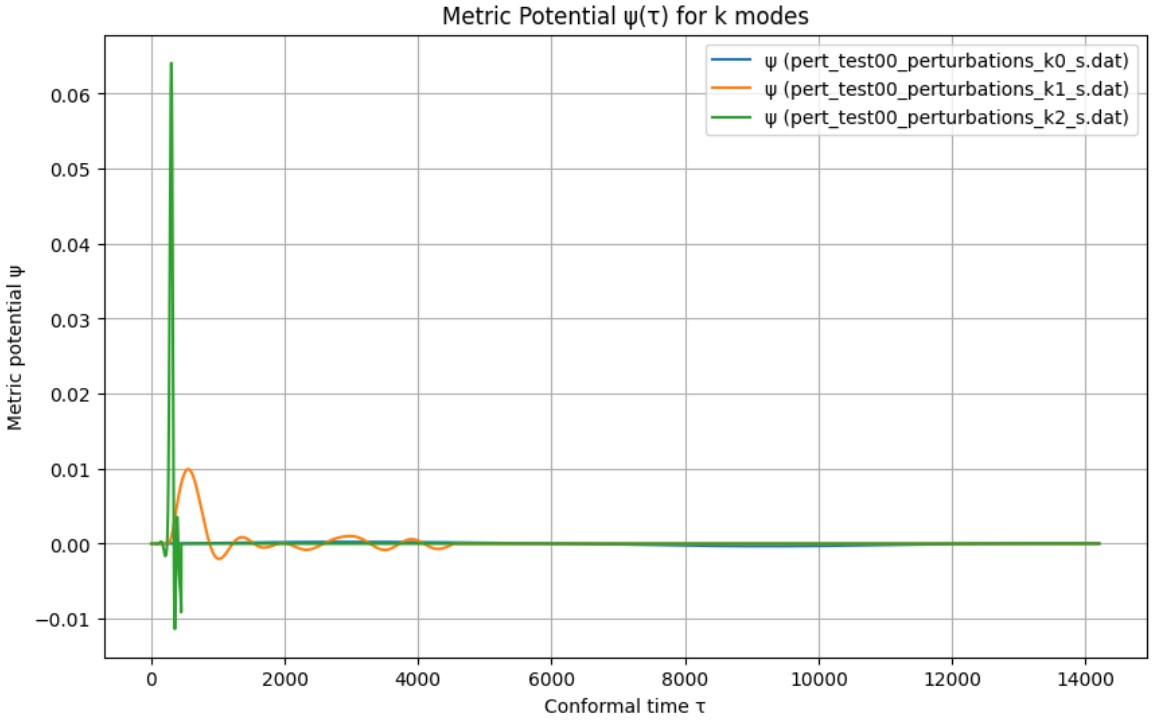}
	\caption{Evolution of the metric potential $\psi(\tau)$ for representative Fourier modes. The different $k$-modes lie on top of each other, indicating regular and non-pathological gravitational dynamics.}
	\label{fig:fig9}
\end{figure}

All tests confirm regular behavior, indicating no divergence, mode coupling anomalies, violation of Einstein constraints, and show compatibility with the standard radiation-dominated regime associated with the early universe.

\subsection*{The $Z_4$-IDSM as an EFT Description}

Having established the theoretical foundations of the LDS--SF framework, we now confront the model with particle physics and cosmological constraints through a comprehensive parameter space survey. The Z$_4$-symmetric Inert Doublet Singlet Model serves as a minimal, renormalizable realization of the layered dark sector architecture, with the singlet $S$ and the neutral component of the inert doublet $H_2^0$ constituting the two dark matter layers, and the full $H_2$ doublet acting as the structuring field that mediates inter-layer interactions.

We explore a broad range of mediator masses $M_{H_2^0} \in [100, 300]$~GeV in 25~GeV increments, with the singlet dark matter mass fixed at $m_S = 60$~GeV to satisfy Higgs-portal relic density constraints. For each benchmark, the portal couplings $\lambda_{S2}$, $\lambda_{S12}$, and $\lambda_{S21}$ are adjusted to maintain the effective coupling $\lambda_{\rm eff}^2 \equiv \lambda_{S2}^2 + 2\lambda_{S12}^2$ at the value required to achieve the desired suppression of structure growth. The mass spectrum of the inert doublet follows the hierarchy $M_{H^\pm} = M_{A^0} = M_{H_2^0} + 50$~GeV, ensuring Z$_4$ symmetry preservation and consistent decay chains.

\begin{figure}[H]
	\centering
	\includegraphics[width=0.35\textwidth]{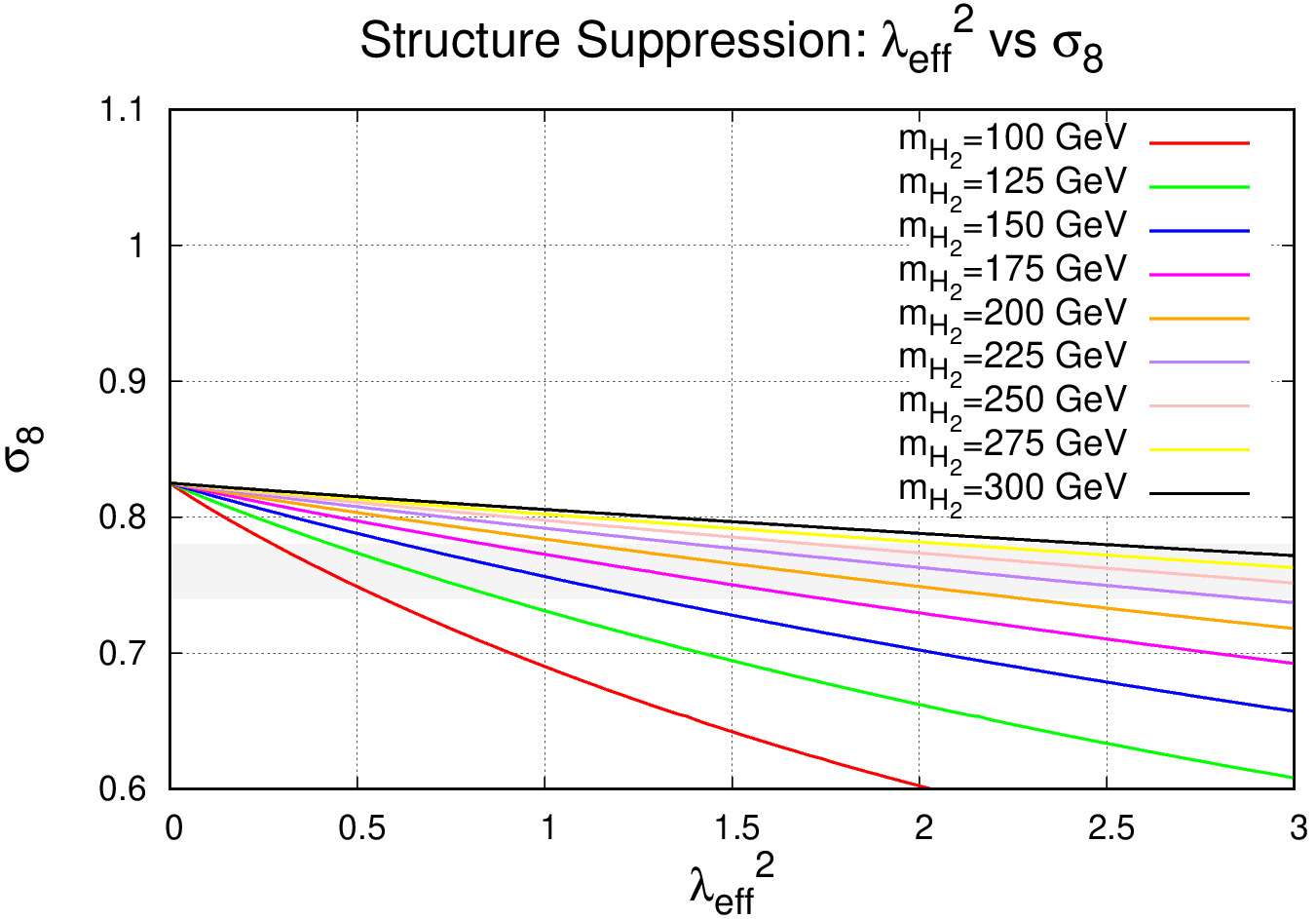}
	\caption{Structure suppression: $\sigma_8$ as a function of $\lambda_{\rm eff}^2$ for mediator masses $M_{H_2^0} = 100$--$300$~GeV. Each curve represents a fixed mediator mass, with heavier mediators requiring proportionally larger $\lambda_{\rm eff}^2$ to achieve equivalent suppression. The shaded gray band indicates the $1\sigma$ preference region from KiDS-1000 and DES weak lensing measurements ($\sigma_8 \approx 0.75$--$0.79$). All curves converge to $\sigma_8 \approx 0.82$ as $\lambda_{\rm eff}^2 \to 0$, recovering the $\Lambda$CDM limit.}
	\label{fig:s8_family}
\end{figure}

Figure~\ref{fig:s8_family} presents the central result of this survey: the dependence of $\sigma_8$ on $\lambda_{\rm eff}^2$ across all nine benchmark masses. For a fixed ratio $\lambda_{\rm eff}^2/M_{H_2}^2$, the effective sound speed $c_s^2$ (and consequently all late-time cosmological observables) remain invariant. This manifests in Fig.~\ref{fig:s8_family} as a systematic separation of curves: heavier mediators require proportionally larger $\lambda_{\rm eff}^2$ to achieve equivalent $\sigma_8$ suppression, with each curve representing a distinct trajectory through the $(M_{H_2}, \lambda_{\rm eff})$ plane. All curves converge to $\sigma_8 \approx 0.82$ as $\lambda_{\rm eff}^2 \to 0$, recovering the $\Lambda$CDM limit, while the shaded band at $\sigma_8 \approx 0.75$--$0.79$ indicates the $1\sigma$ preference region from KiDS-1000 and DES weak lensing measurements.

The scaling symmetry dramatically reduces fine-tuning: rather than isolated points, the viable parameter space forms \textit{continuous corridors} where any combination of $M_{H_2}$ and $\lambda_{\rm eff}$ preserving their ratio yields identical cosmological predictions.

\subsection*{Benchmark I: $M_{H_2^0} = 100$~GeV}

We present the first benchmark point with mediator mass $M_{H_2^0} = 100$~GeV, corresponding to $\lambda_{\rm eff}^2 = 0.34$. This represents the lightest mediator case in our survey, requiring the smallest effective coupling to achieve the desired structure suppression.

\begin{figure}[H]
	\centering
	\includegraphics[width=0.35\textwidth]{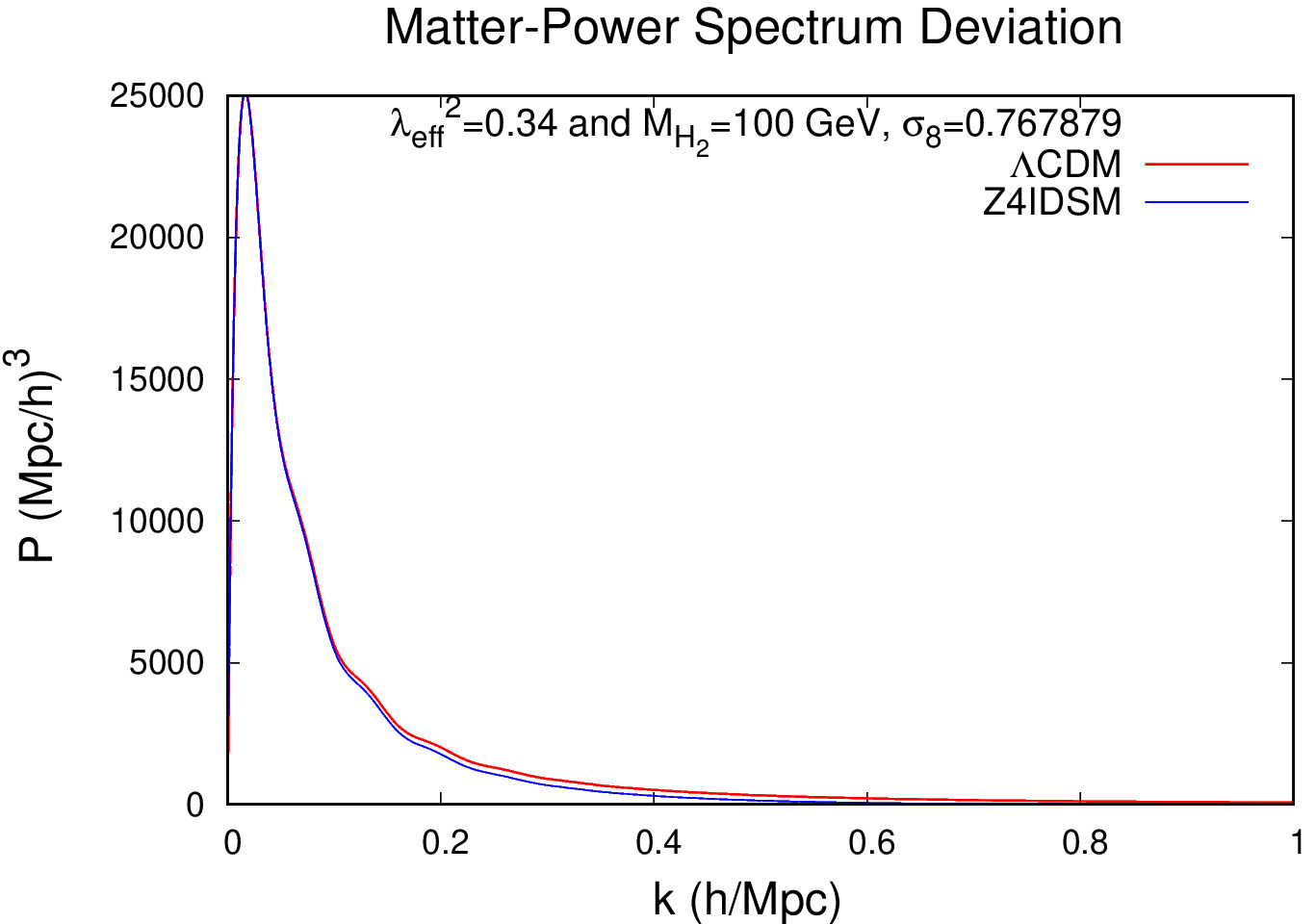}
	\caption{Matter power spectrum for Benchmark I ($M_{H_2^0} = 100$~GeV, $\lambda_{\rm eff}^2 = 0.34$, $\sigma_8 = 0.768$). The Z$_4$-IDSM prediction (blue) is indistinguishable from $\Lambda$CDM (red) for $k \lesssim 0.1$~h/Mpc, with suppression becoming apparent at smaller scales. The deviation reaches approximately $10\%$ at $k \sim 1$~h/Mpc.}
	\label{fig:benchmark_100_pk}
\end{figure}

\begin{figure}[H]
	\centering
	\includegraphics[width=0.35\textwidth]{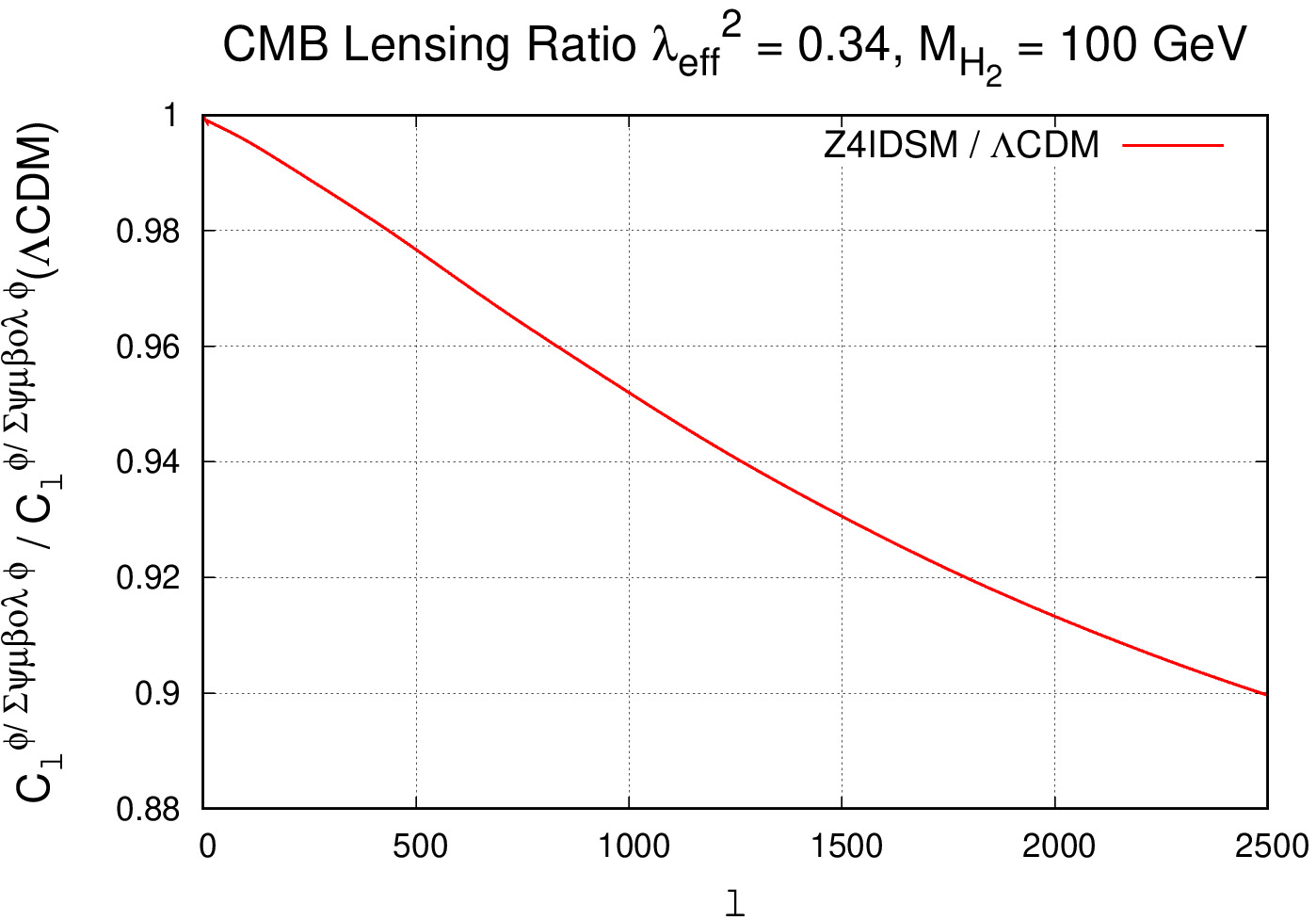}
	\caption{CMB lensing potential power spectrum ratio for Benchmark I. The ratio $C_L^{\phi\phi}({\rm Z4IDSM})/C_L^{\phi\phi}(\Lambda{\rm CDM})$ shows approximately $2\%$ suppression at $\ell \sim 500$, increasing to $\sim 10\%$ at $\ell = 2500$. This scale-dependent suppression is characteristic of the LDS--SF mechanism.}
	\label{fig:benchmark_100_lensing}
\end{figure}

\begin{figure}[H]
	\centering
	\includegraphics[width=0.35\textwidth]{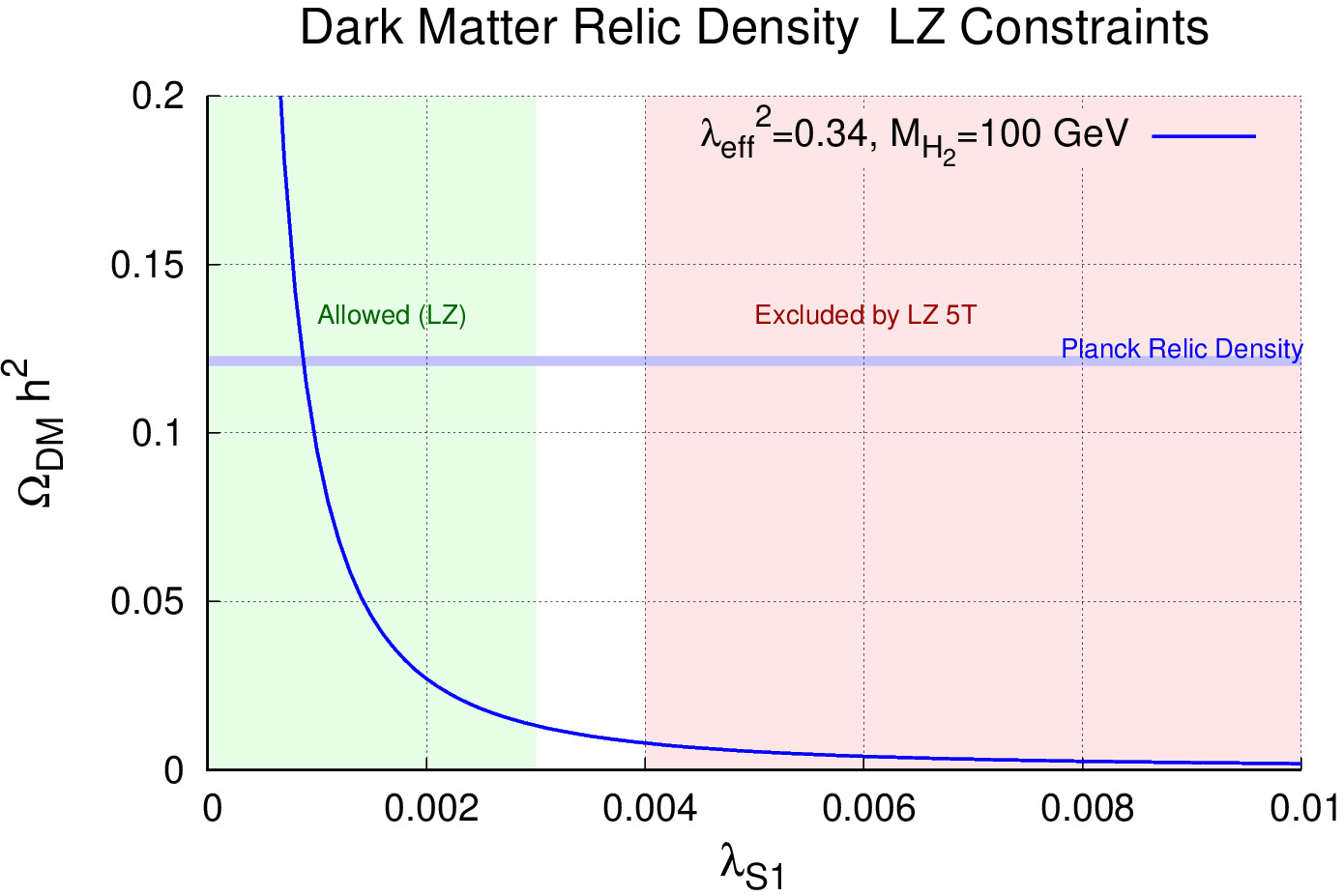}
	\caption{Dark matter relic density $\Omega_{\rm DM}h^2$ as a function of the Higgs-portal coupling $\lambda_{S1}$ for Benchmark I. The horizontal blue band indicates the Planck relic density constraint $\Omega_{\rm DM}h^2 = 0.120 \pm 0.001$. The green shaded region denotes the parameter space allowed by LUX-ZEPLIN (LZ) direct detection constraints, while the red region is excluded by LZ 5-tonne projections. The correct relic abundance is achieved for $\lambda_{S1} \approx 0.001$, safely within the allowed region.}
	\label{fig:benchmark_100_relic}
\end{figure}

The three diagnostic figures demonstrate consistent phenomenology across cosmological and particle physics observables. The matter power spectrum (Fig.~\ref{fig:benchmark_100_pk}) exhibits the targeted small-scale suppression while preserving large-scale agreement with $\Lambda$CDM. The CMB lensing ratio (Fig.~\ref{fig:benchmark_100_lensing}) confirms that this suppression leaves no detectable imprint on the cosmic microwave background at $\ell \lesssim 500$, and remains largely in agreement with $\lambda$CDM. Finally, the relic density analysis (Fig.~\ref{fig:benchmark_100_relic}) establishes that the Higgs-portal coupling required for thermal freeze-out evades current and projected direct detection limits.

\subsection*{Benchmark II: $M_{H_2^0} = 125$~GeV}

The second benchmark point features mediator mass $M_{H_2^0} = 125$~GeV with effective coupling $\lambda_{\rm eff}^2 = 0.60$. This intermediate-light mass regime demonstrates the scaling symmetry: the larger mediator mass requires a proportionally increased $\lambda_{\rm eff}^2$ to maintain equivalent structure suppression.

\begin{figure}[t]
	\centering
	\includegraphics[width=0.35\textwidth]{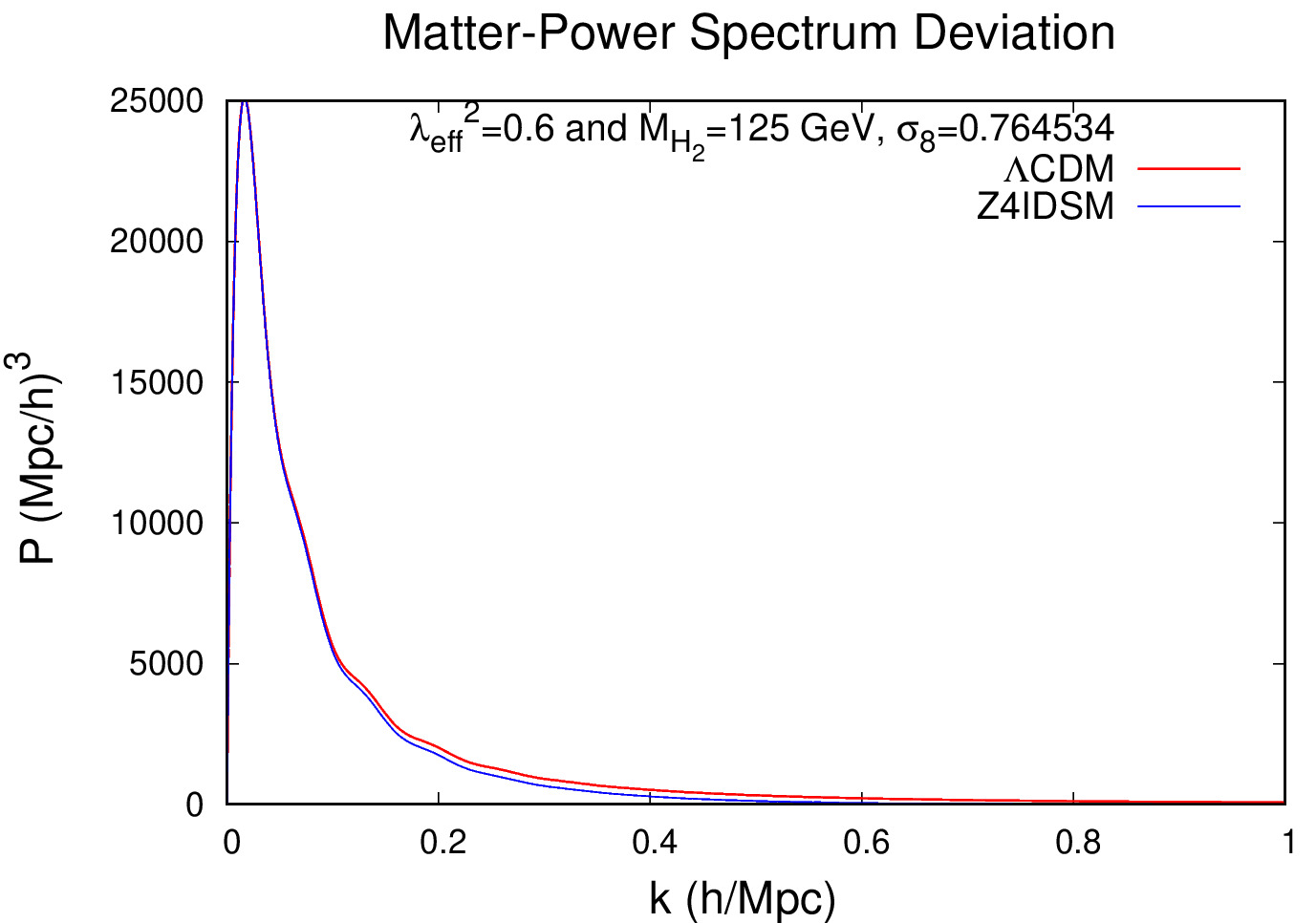}
	\caption{Matter power spectrum for Benchmark II ($M_{H_2^0} = 125$~GeV, $\lambda_{\rm eff}^2 = 0.60$, $\sigma_8 = 0.765$). The Z$_4$-IDSM prediction (blue) remains indistinguishable from $\Lambda$CDM (red) for $k \lesssim 0.1$~h/Mpc, with characteristic small-scale suppression emerging at higher wavenumbers.}
	\label{fig:benchmark_125_pk}
\end{figure}

\begin{figure}[t]
	\centering
	\includegraphics[width=0.35\textwidth]{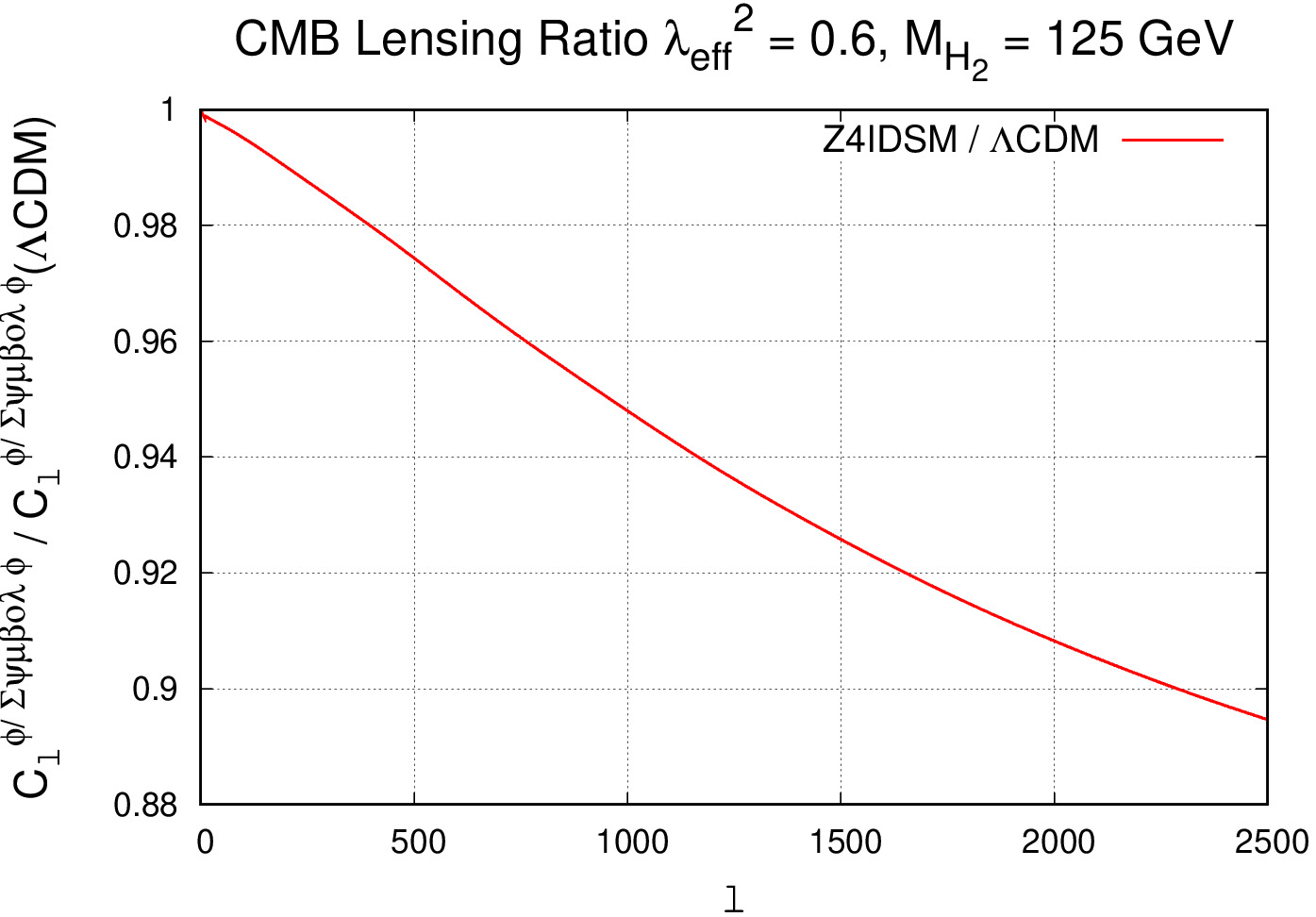}
	\caption{CMB lensing potential power spectrum ratio for Benchmark II. The ratio $C_L^{\phi\phi}({\rm Z4IDSM})/C_L^{\phi\phi}(\Lambda{\rm CDM})$ exhibits approximately $2\%$ suppression at $\ell \sim 500$, rising to $\sim 10\%$ at $\ell = 2500$, consistent with the scaling behavior established in Benchmark I.}
	\label{fig:benchmark_125_lensing}
\end{figure}

\begin{figure}[t]
	\centering
	\includegraphics[width=0.35\textwidth]{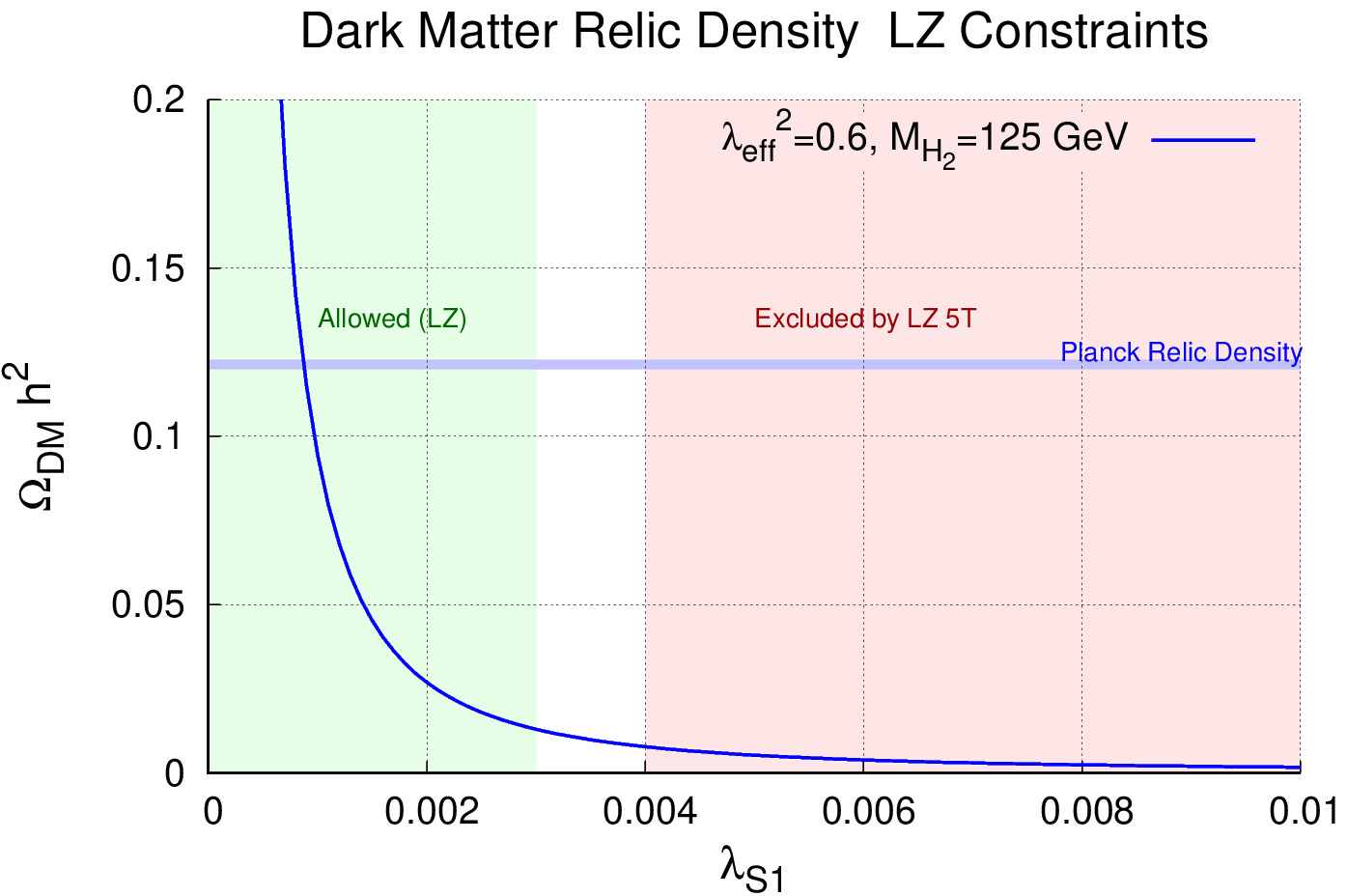}
	\caption{Dark matter relic density $\Omega_{\rm DM}h^2$ as a function of $\lambda_{S1}$ for Benchmark II. The Planck relic density constraint $\Omega_{\rm DM}h^2 = 0.120 \pm 0.001$ (horizontal blue band) is satisfied for $\lambda_{S1} \approx 0.001$, well within the LZ-allowed region (green). The red region indicates exclusion by LZ 5-tonne projections.}
	\label{fig:benchmark_125_relic}
\end{figure}

Benchmark II confirms the systematic behavior observed in Benchmark I. The matter power spectrum (Fig.~\ref{fig:benchmark_125_pk}) displays identical large-scale compliance with $\Lambda$CDM and comparable small-scale suppression. The CMB lensing ratio (Fig.~\ref{fig:benchmark_125_lensing}) maintains the characteristic scale-dependent profile, while the relic density analysis (Fig.~\ref{fig:benchmark_125_relic}) verifies consistent particle physics viability across the benchmark suite.

\subsection*{Benchmark III: $M_{H_2^0} = 150$~GeV}

The third benchmark point features mediator mass $M_{H_2^0} = 150$~GeV with effective coupling $\lambda_{\rm eff}^2 = 0.90$. This intermediate mass regime continues the scaling trend, with the effective coupling increasing proportionally to maintain the characteristic structure suppression.

\begin{figure}[t]
	\centering
	\includegraphics[width=0.35\textwidth]{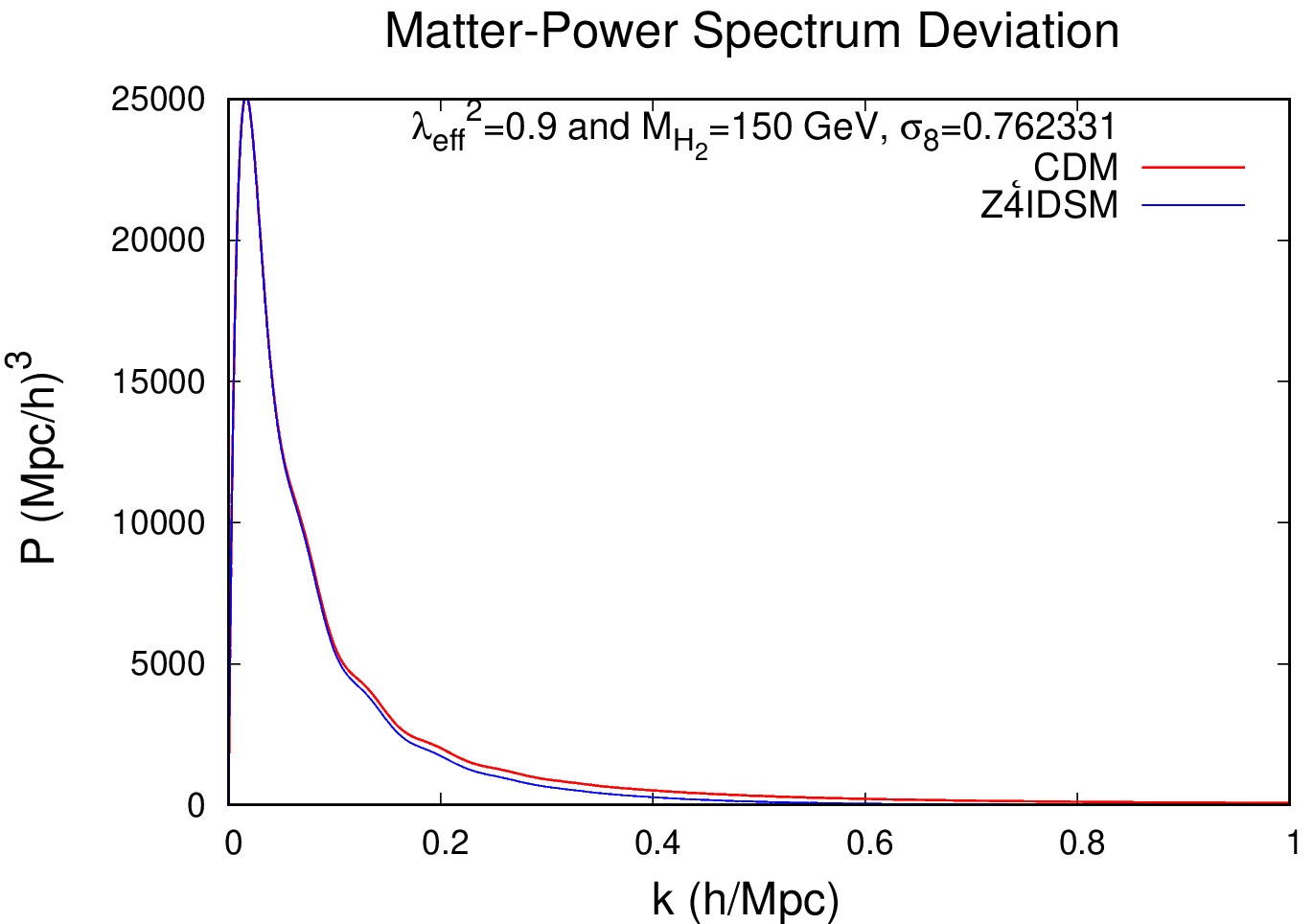}
	\caption{Matter power spectrum for Benchmark III ($M_{H_2^0} = 150$~GeV, $\lambda_{\rm eff}^2 = 0.90$, $\sigma_8 = 0.762$). The Z$_4$-IDSM prediction (blue) remains indistinguishable from $\Lambda$CDM (red) for $k \lesssim 0.1$~h/Mpc, with characteristic small-scale suppression emerging at higher wavenumbers.}
	\label{fig:benchmark_150_pk}
\end{figure}

\begin{figure}[t]
	\centering
	\includegraphics[width=0.35\textwidth]{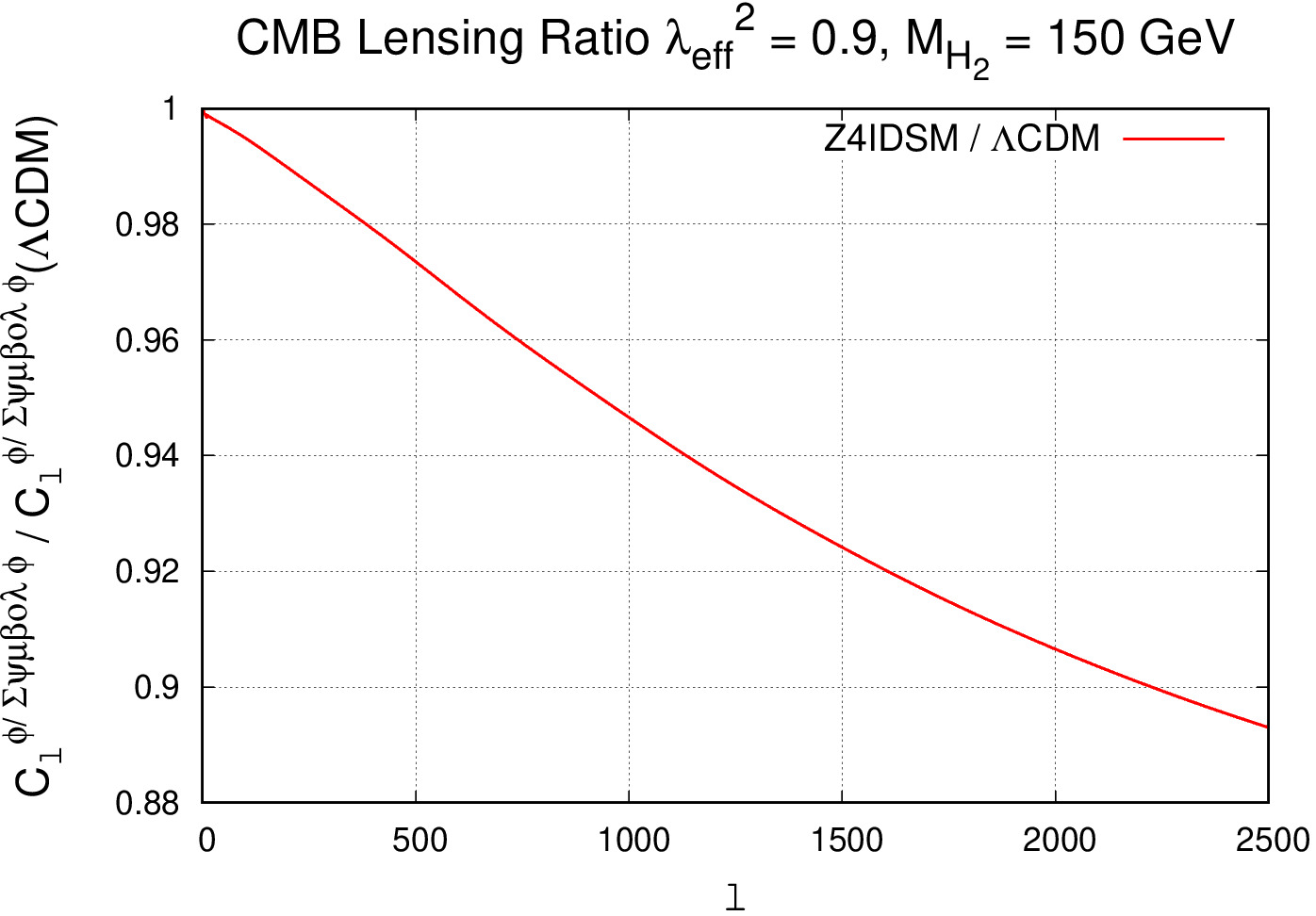}
	\caption{CMB lensing potential power spectrum ratio for Benchmark III. The ratio $C_L^{\phi\phi}({\rm Z4IDSM})/C_L^{\phi\phi}(\Lambda{\rm CDM})$ exhibits approximately $2\%$ suppression at $\ell \sim 500$, rising to $\sim 10\%$ at $\ell = 2500$, consistent with the scaling behavior established in previous benchmarks.}
	\label{fig:benchmark_150_lensing}
\end{figure}

\begin{figure}[t]
	\centering
	\includegraphics[width=0.35\textwidth]{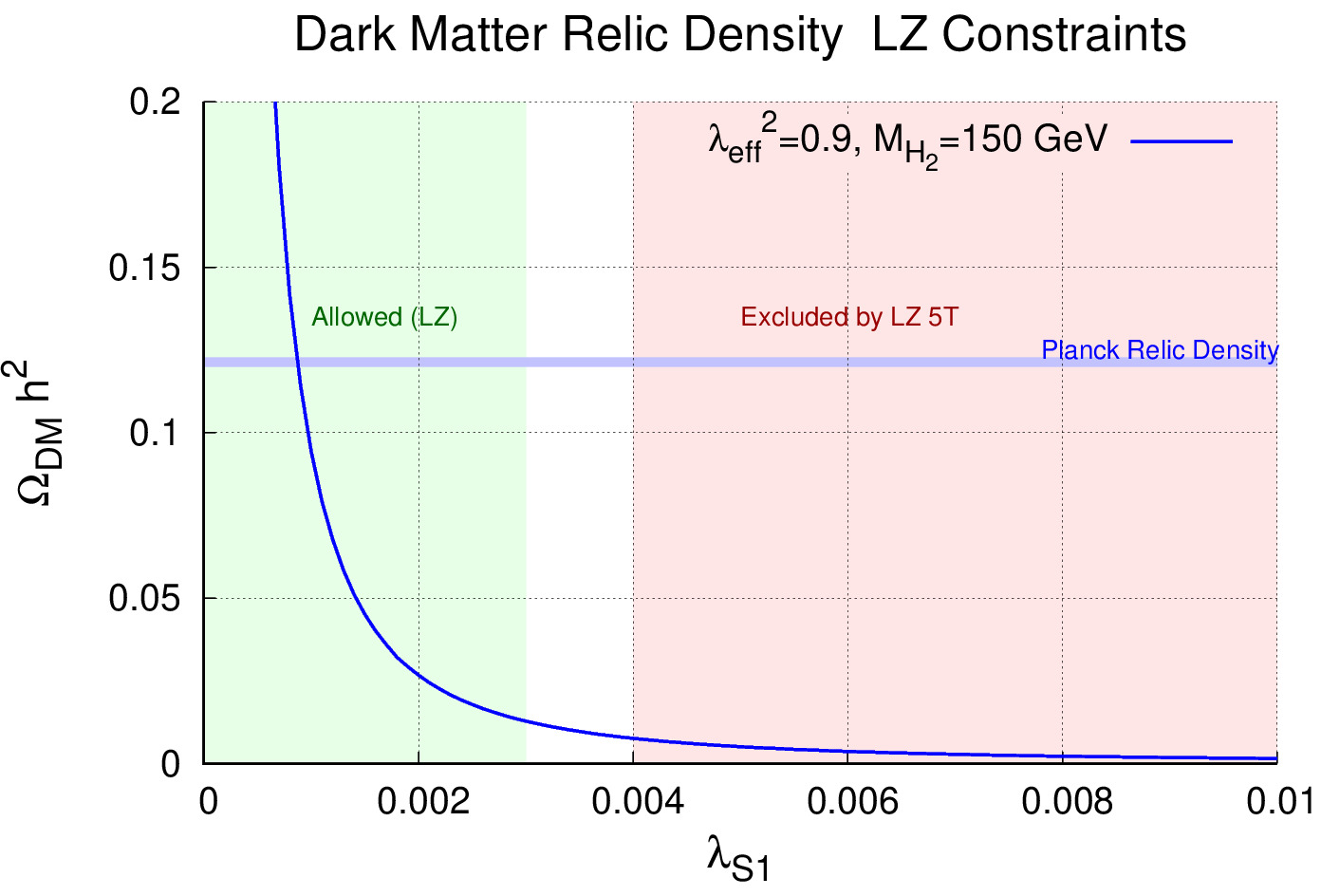}
	\caption{Dark matter relic density $\Omega_{\rm DM}h^2$ as a function of $\lambda_{S1}$ for Benchmark III. The Planck relic density constraint $\Omega_{\rm DM}h^2 = 0.120 \pm 0.001$ (horizontal blue band) is satisfied for $\lambda_{S1} \approx 0.001$, well within the LZ-allowed region (green). The red region indicates exclusion by LZ 5-tonne projections.}
	\label{fig:benchmark_150_relic}
\end{figure}

Benchmark III reinforces the systematic behavior observed in Benchmarks I and II. The matter power spectrum (Fig.~\ref{fig:benchmark_150_pk}) displays identical large-scale compliance with $\Lambda$CDM and comparable small-scale suppression. The CMB lensing ratio (Fig.~\ref{fig:benchmark_150_lensing}) maintains the characteristic scale-dependent profile, while the relic density analysis (Fig.~\ref{fig:benchmark_150_relic}) verifies consistent particle physics viability across the benchmark suite.

\subsection*{Benchmark IV: $M_{H_2^0} = 175$~GeV}

The fourth benchmark point features mediator mass $M_{H_2^0} = 175$~GeV with effective coupling $\lambda_{\rm eff}^2 = 1.18$. This intermediate mass regime continues the scaling trend, with the effective coupling increasing proportionally to maintain the characteristic structure suppression. Notably, this benchmark employs a reduced value of $\lambda_{S12} = 0.30$ compared to neighboring points, reflecting the flexibility in distributing the effective coupling between $\lambda_{S2}$ and $\lambda_{S12}$ while preserving the required $\lambda_{\rm eff}^2$.

\begin{figure}[t]
	\centering
	\includegraphics[width=0.35\textwidth]{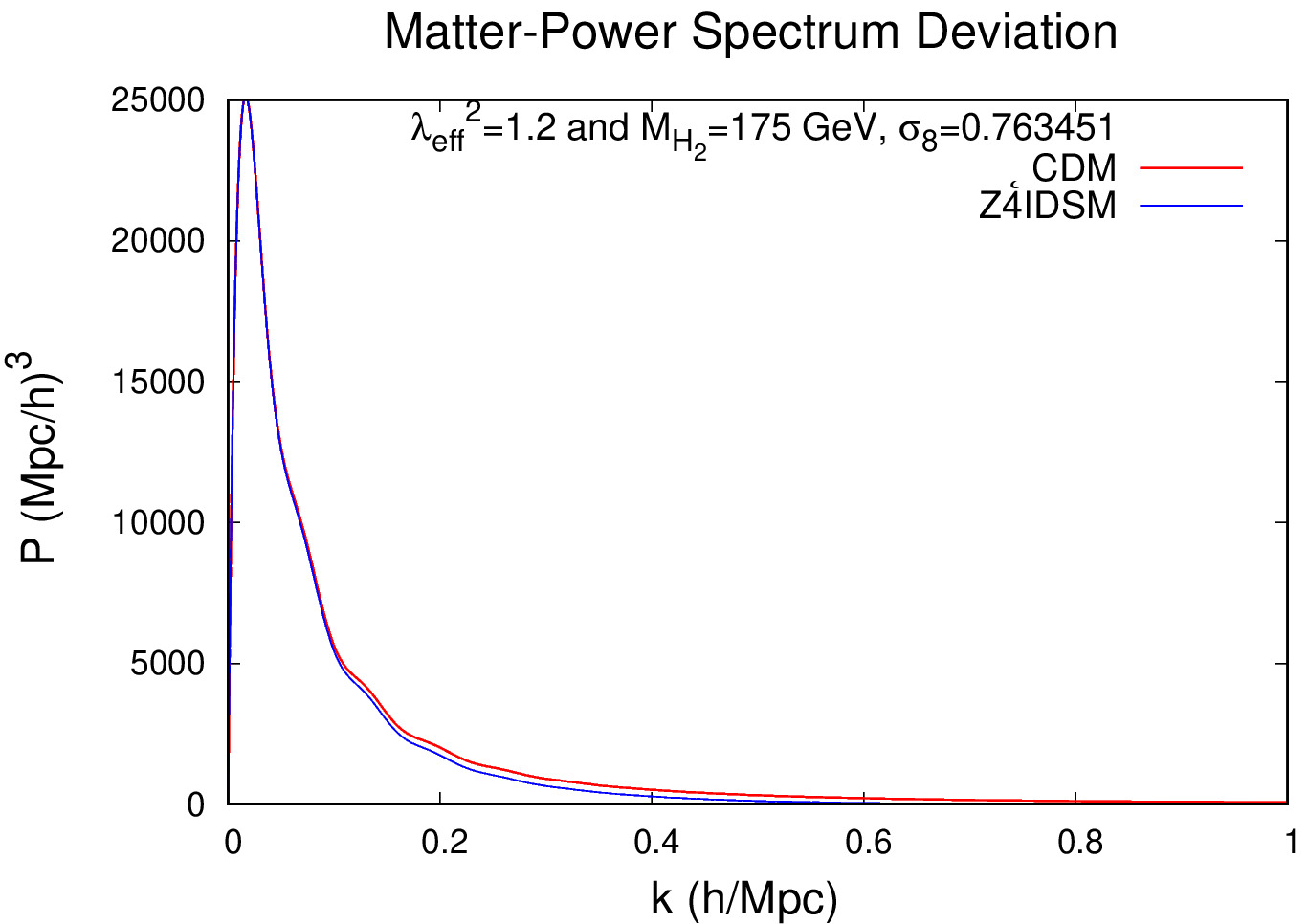}
	\caption{Matter power spectrum for Benchmark IV ($M_{H_2^0} = 175$~GeV, $\lambda_{\rm eff}^2 = 1.18$, $\sigma_8 = 0.763$). The Z$_4$-IDSM prediction (blue) remains indistinguishable from $\Lambda$CDM (red) for $k \lesssim 0.1$~h/Mpc, with characteristic small-scale suppression emerging at higher wavenumbers. Despite the different distribution of couplings, the phenomenology remains consistent with previous benchmarks.}
	\label{fig:benchmark_175_pk}
\end{figure}

\begin{figure}[t]
	\centering
	\includegraphics[width=0.35\textwidth]{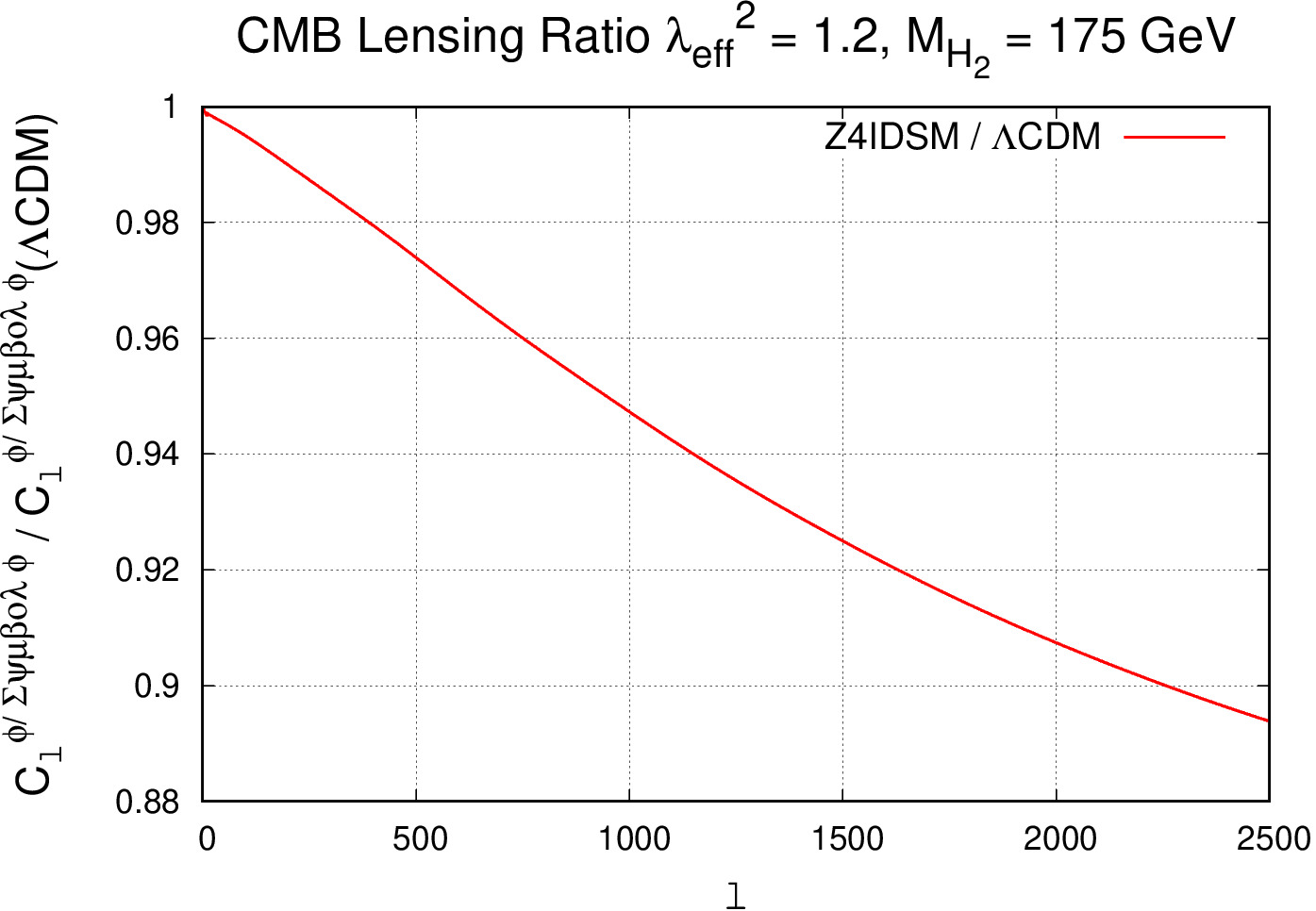}
	\caption{CMB lensing potential power spectrum ratio for Benchmark IV. The ratio $C_L^{\phi\phi}({\rm Z4IDSM})/C_L^{\phi\phi}(\Lambda{\rm CDM})$ exhibits approximately $2\%$ suppression at $\ell \sim 500$, rising to $\sim 10\%$ at $\ell = 2500$. The insensitivity to the internal distribution of $\lambda_{\rm eff}^2$ components confirms that the cosmological observables depend only on the effective coupling, not on its microscopic origin.}
	\label{fig:benchmark_175_lensing}
\end{figure}

\begin{figure}[t]
	\centering
	\includegraphics[width=0.35\textwidth]{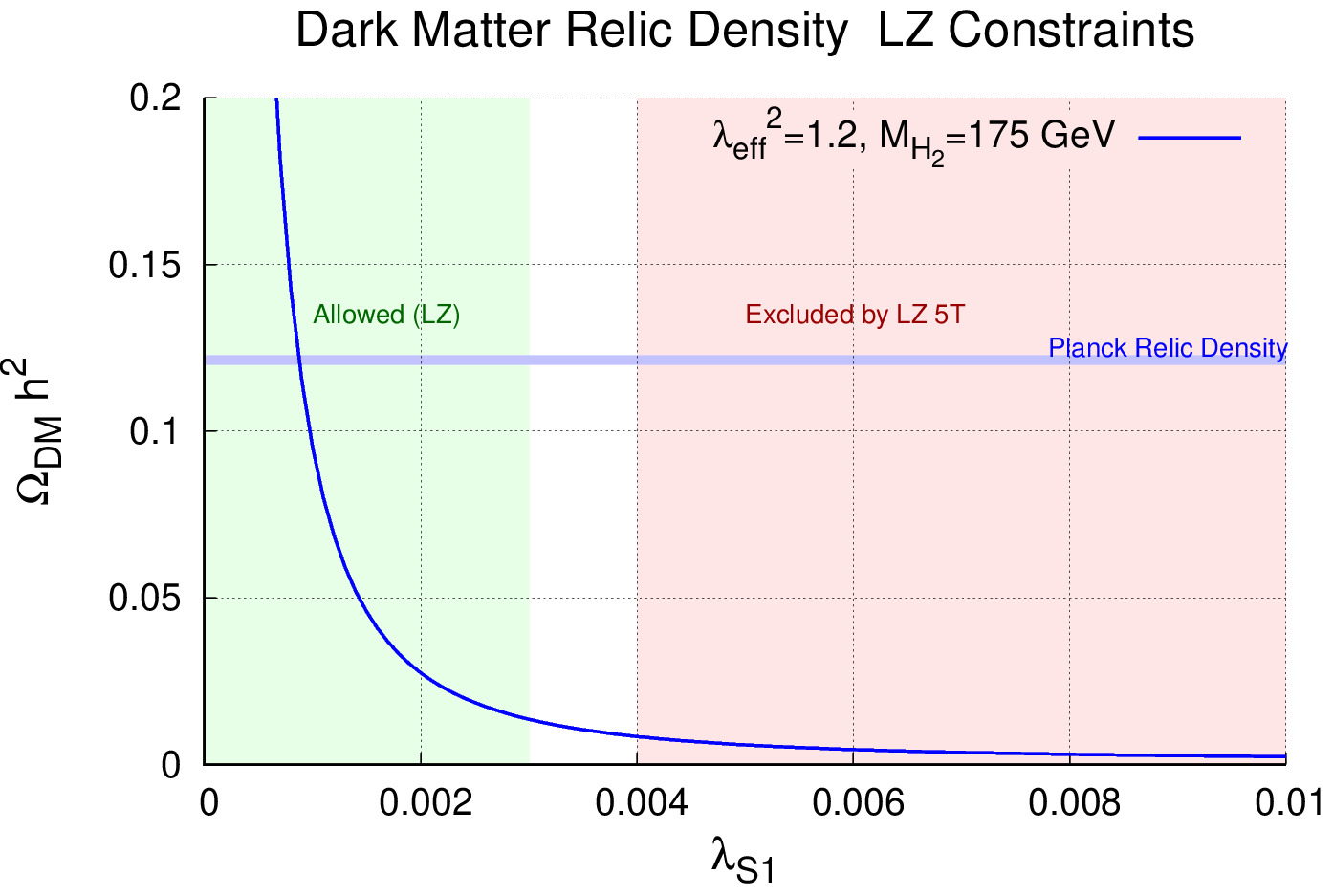}
	\caption{Dark matter relic density $\Omega_{\rm DM}h^2$ as a function of $\lambda_{S1}$ for Benchmark IV. The Planck relic density constraint $\Omega_{\rm DM}h^2 = 0.120 \pm 0.001$ (horizontal blue band) is satisfied for $\lambda_{S1} \approx 0.001$, well within the LZ-allowed region (green). The red region indicates exclusion by LZ 5-tonne projections. The consistency of this benchmark reinforces the robustness of the Z$_4$-IDSM parameter space.}
	\label{fig:benchmark_175_relic}
\end{figure}

Benchmark IV demonstrates an important feature of the LDS--SF framework: the cosmological observables are sensitive only to the effective coupling $\lambda_{\rm eff}^2$, not to the individual contributions from $\lambda_{S2}$ and $\lambda_{S12}$. This degeneracy provides additional flexibility in model-building and suggests that precise measurements of the matter power spectrum alone cannot uniquely determine the underlying Lagrangian parameters. The matter power spectrum (Fig.~\ref{fig:benchmark_175_pk}) and CMB lensing ratio (Fig.~\ref{fig:benchmark_175_lensing}) remain quantitatively similar to previous benchmarks despite the altered coupling structure, while the relic density analysis (Fig.~\ref{fig:benchmark_175_relic}) confirms continued compliance with direct detection constraints.

\subsection*{Benchmark V: $M_{H_2^0} = 200$~GeV}

The fifth benchmark point features mediator mass $M_{H_2^0} = 200$~GeV with effective coupling $\lambda_{\rm eff}^2 = 1.50$. This represents the midpoint of our mass survey, where the EFT contact-interaction approximation remains robust and the scaling symmetry is clearly manifested. The increased mediator mass requires a proportionally larger effective coupling to maintain equivalent cosmological suppression.

\begin{figure}[t]
	\centering
	\includegraphics[width=0.35\textwidth]{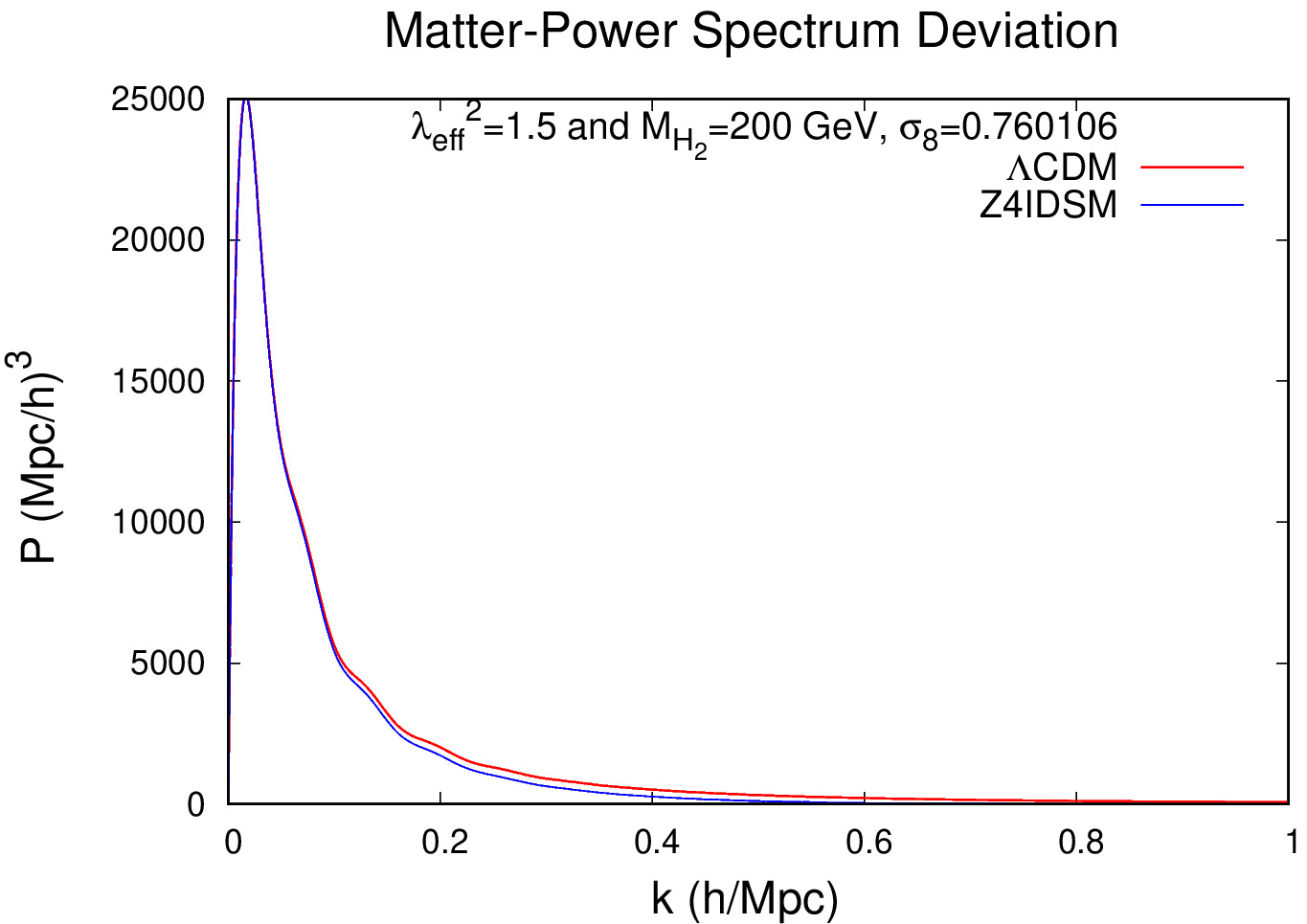}
	\caption{Matter power spectrum for Benchmark V ($M_{H_2^0} = 200$~GeV, $\lambda_{\rm eff}^2 = 1.50$, $\sigma_8 = 0.760$). The Z$_4$-IDSM prediction (blue) remains indistinguishable from $\Lambda$CDM (red) for $k \lesssim 0.1$~h/Mpc, with characteristic small-scale suppression emerging at higher wavenumbers. The deviation pattern remains quantitatively consistent with lighter mediator benchmarks, confirming the scaling behavior.}
	\label{fig:benchmark_200_pk}
\end{figure}

\begin{figure}[t]
	\centering
	\includegraphics[width=0.35\textwidth]{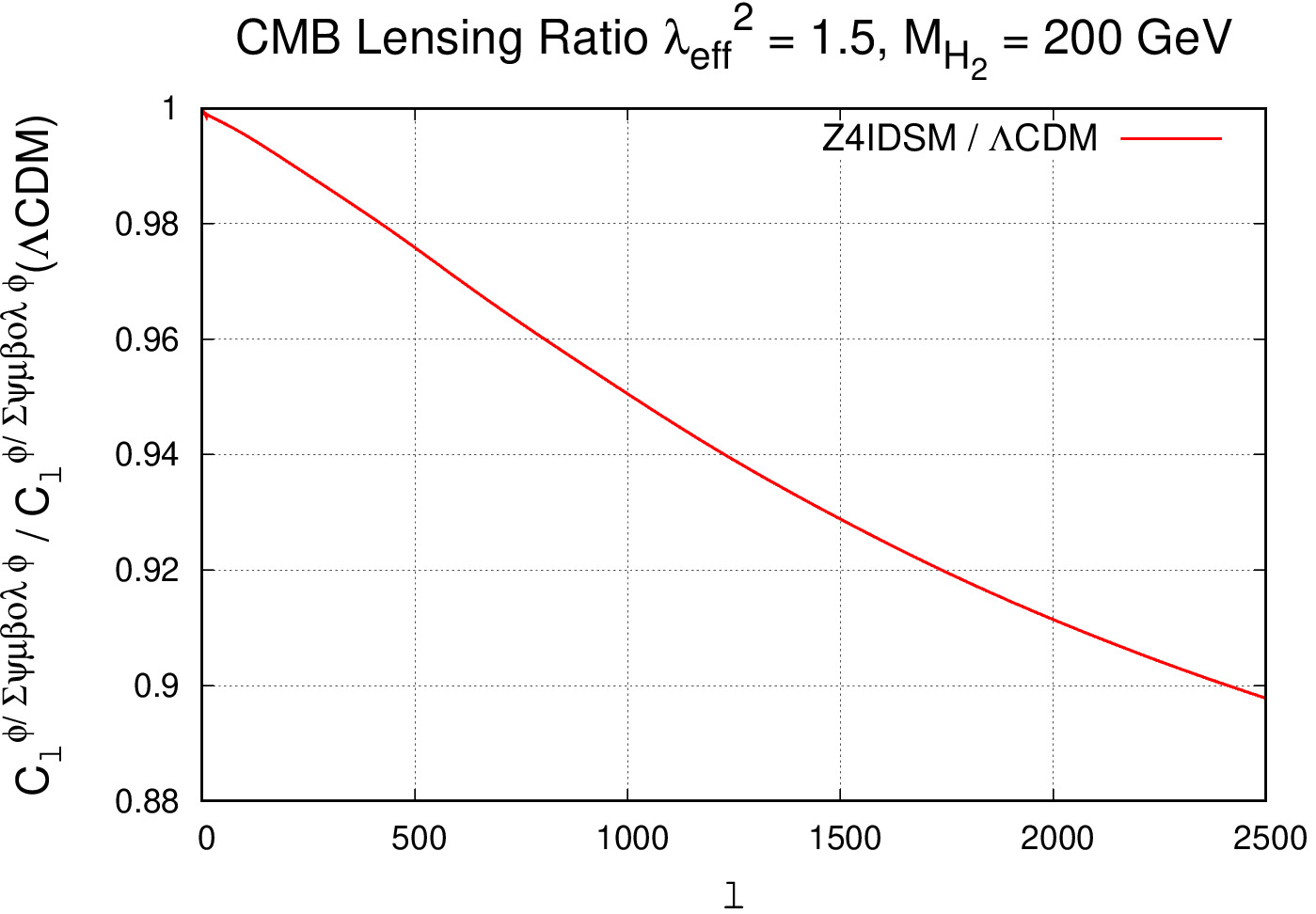}
	\caption{CMB lensing potential power spectrum ratio for Benchmark V. The ratio $C_L^{\phi\phi}({\rm Z4IDSM})/C_L^{\phi\phi}(\Lambda{\rm CDM})$ exhibits approximately $2\%$ suppression at $\ell \sim 500$, rising to $\sim 10\%$ at $\ell = 2500$. The smooth, monotonic decrease confirms that the LDS--SF mechanism preserves the integrity of the CMB acoustic peaks while generating the requisite late-time suppression.}
	\label{fig:benchmark_200_lensing}
\end{figure}

\begin{figure}[t]
	\centering
	\includegraphics[width=0.35\textwidth]{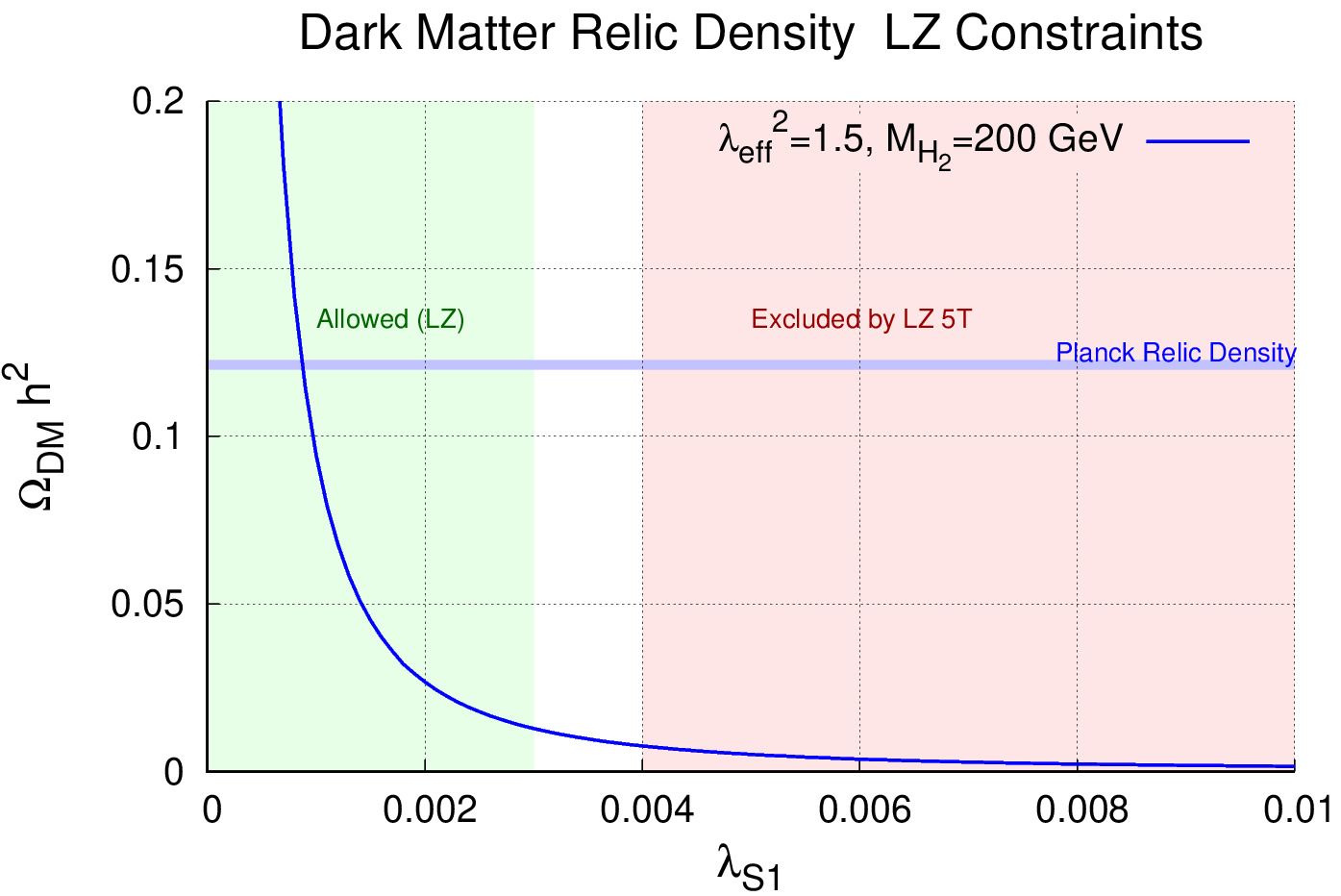}
	\caption{Dark matter relic density $\Omega_{\rm DM}h^2$ as a function of $\lambda_{S1}$ for Benchmark V. The Planck relic density constraint $\Omega_{\rm DM}h^2 = 0.120 \pm 0.001$ (horizontal blue band) is satisfied for $\lambda_{S1} \approx 0.001$, well within the LZ-allowed region (green). The red region indicates exclusion by LZ 5-tonne projections. The consistency of this benchmark at the midpoint of our mass range reinforces the broad viability of the Z$_4$-IDSM parameter space.}
	\label{fig:benchmark_200_relic}
\end{figure}

Benchmark V demonstrates the continued robustness of the LDS--SF framework at intermediate-heavy mediator masses. The matter power spectrum (Fig.~\ref{fig:benchmark_200_pk}) and CMB lensing ratio (Fig.~\ref{fig:benchmark_200_lensing}) maintain the characteristic profiles established in lighter benchmarks, with no discernible degradation in the quality of the $\Lambda$CDM agreement at large scales. The relic density analysis (Fig.~\ref{fig:benchmark_200_relic}) confirms that the Higgs-portal coupling required for thermal freeze-out remains safely below direct detection thresholds, even as the mediator mass approaches the upper range of electroweak-scale phenomenology.

\subsection*{Benchmark VI: $M_{H_2^0} = 225$~GeV}

The sixth benchmark point features mediator mass $M_{H_2^0} = 225$~GeV with effective coupling $\lambda_{\rm eff}^2 = 1.98$. This intermediate-heavy mass regime approaches the upper range of electroweak-scale phenomenology, where the EFT contact-interaction approximation remains valid while requiring substantially larger couplings to maintain the characteristic structure suppression.

\begin{figure}[t]
	\centering
	\includegraphics[width=0.35\textwidth]{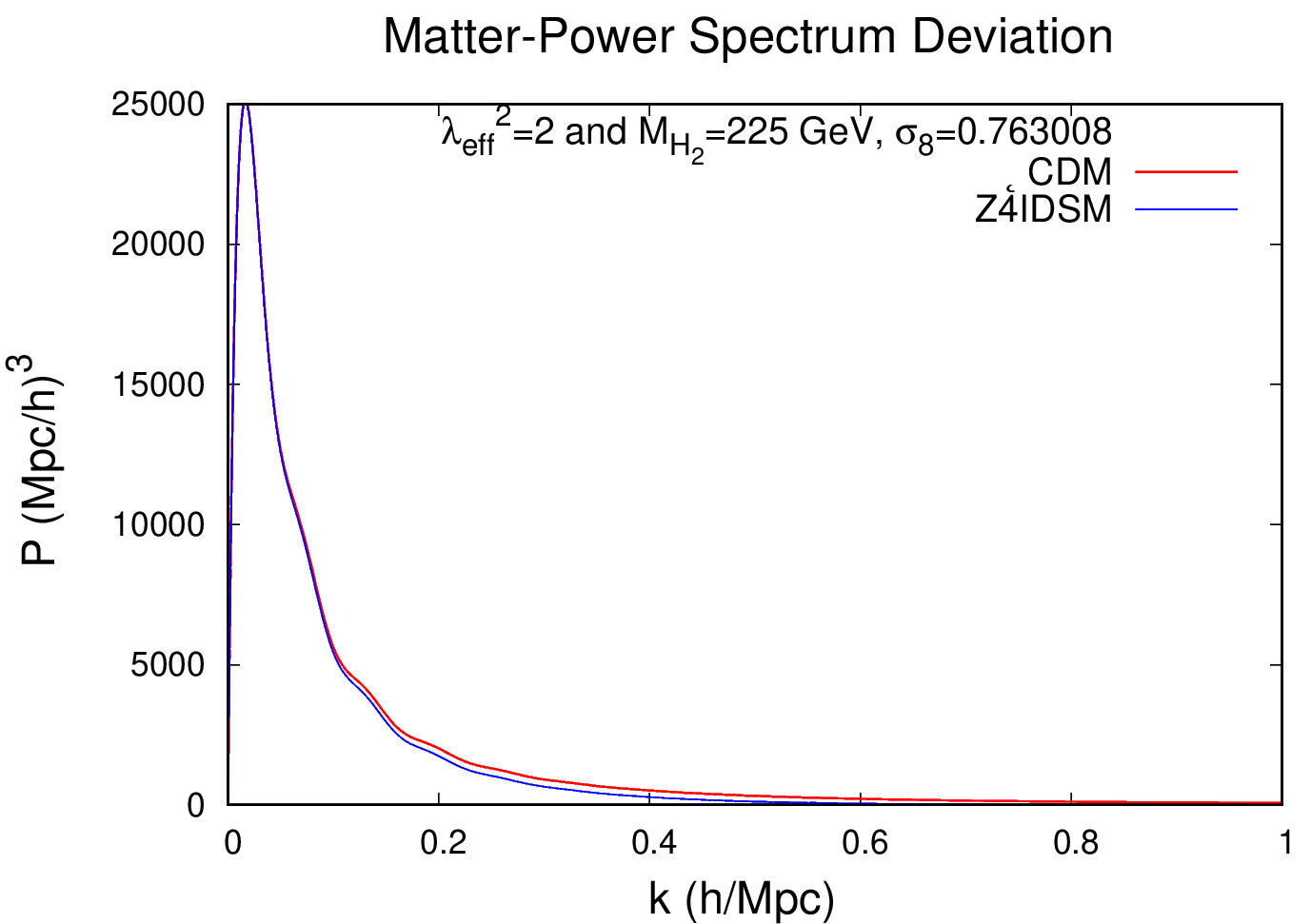}
	\caption{Matter power spectrum for Benchmark VI ($M_{H_2^0} = 225$~GeV, $\lambda_{\rm eff}^2 = 1.98$, $\sigma_8 = 0.763$). The Z$_4$-IDSM prediction (blue) remains indistinguishable from $\Lambda$CDM (red) for $k \lesssim 0.1$~h/Mpc, with characteristic small-scale suppression emerging at higher wavenumbers. The continued agreement with lighter benchmarks confirms the robustness of the scaling symmetry across the full mass range.}
	\label{fig:benchmark_225_pk}
\end{figure}

\begin{figure}[t]
	\centering
	\includegraphics[width=0.35\textwidth]{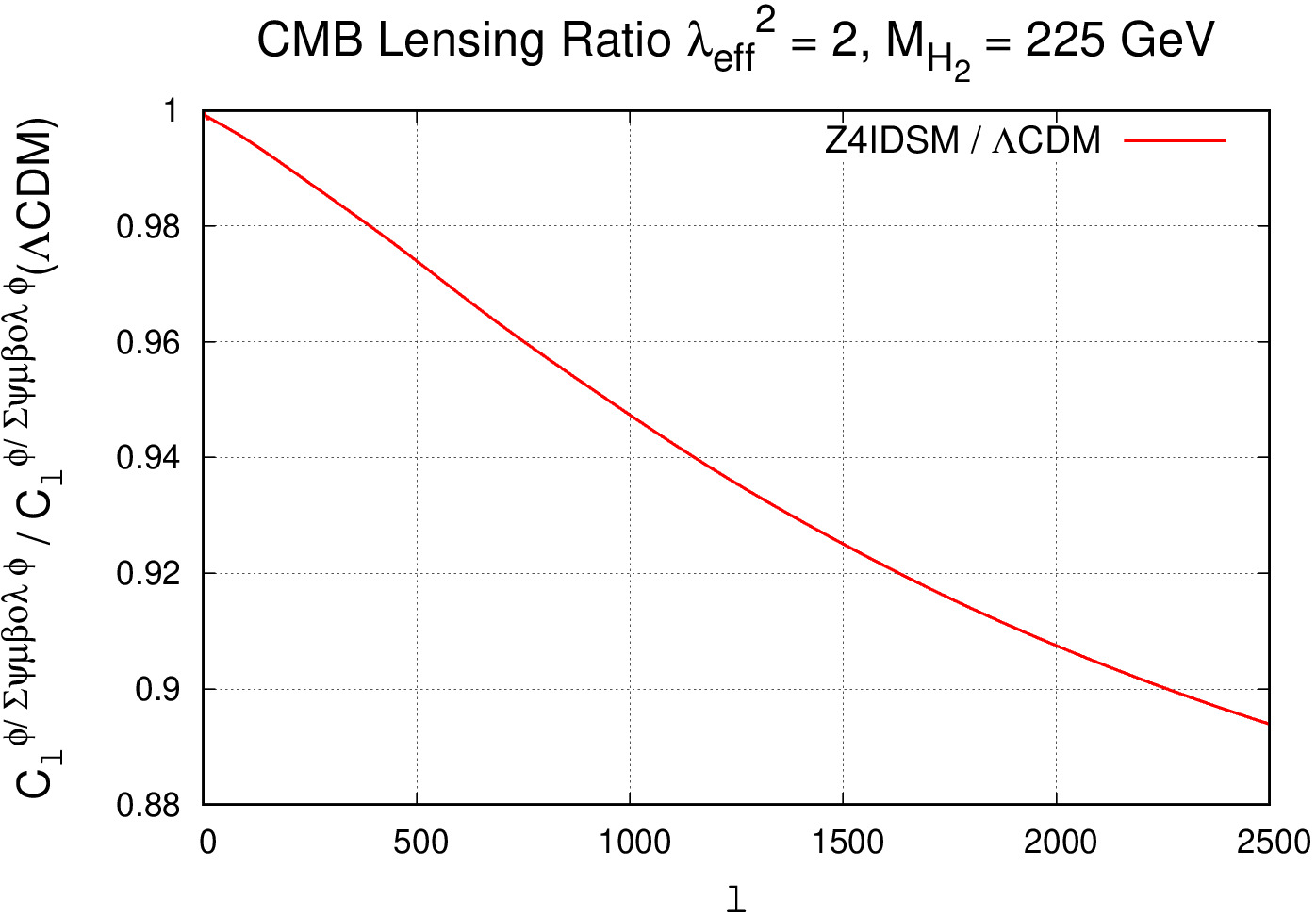}
	\caption{CMB lensing potential power spectrum ratio for Benchmark VI. The ratio $C_L^{\phi\phi}({\rm Z4IDSM})/C_L^{\phi\phi}(\Lambda{\rm CDM})$ exhibits approximately $2\%$ suppression at $\ell \sim 500$, rising to $\sim 10\%$ at $\ell = 2500$. The persistence of this characteristic profile at heavier mediator masses validates the EFT description of the LDS--SF mechanism.}
	\label{fig:benchmark_225_lensing}
\end{figure}

\begin{figure}[t]
	\centering
	\includegraphics[width=0.35\textwidth]{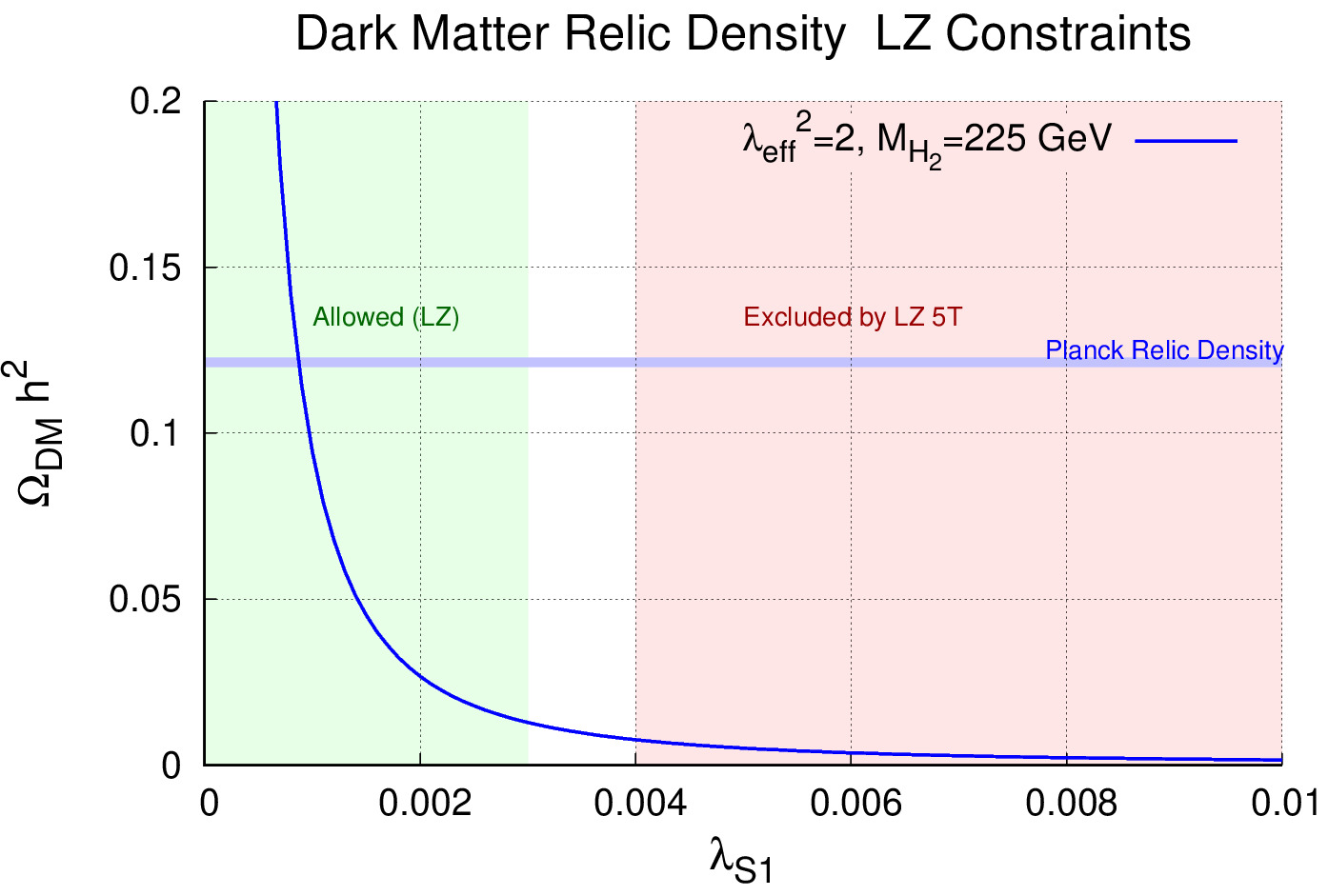}
	\caption{Dark matter relic density $\Omega_{\rm DM}h^2$ as a function of $\lambda_{S1}$ for Benchmark VI. The Planck relic density constraint $\Omega_{\rm DM}h^2 = 0.120 \pm 0.001$ (horizontal blue band) is satisfied for $\lambda_{S1} \approx 0.001$, well within the LZ-allowed region (green). The red region indicates exclusion by LZ 5-tonne projections. The stability of the required Higgs-portal coupling across the mass survey highlights the decoupling of the relic density mechanism from the structure suppression dynamics.}
	\label{fig:benchmark_225_relic}
\end{figure}

Benchmark VI demonstrates the continued viability of the LDS--SF framework as the mediator mass approaches the upper limits of conventional electroweak-scale model-building. The matter power spectrum (Fig.~\ref{fig:benchmark_225_pk}) and CMB lensing ratio (Fig.~\ref{fig:benchmark_225_lensing}) maintain the established characteristic profiles, while the relic density analysis (Fig.~\ref{fig:benchmark_225_relic}) confirms that the thermal freeze-out mechanism remains perturbative and consistent with direct detection constraints. The substantial increase in $\lambda_{\rm eff}^2$ required at this mass scale is precisely compensated by the heavier mediator in the effective sound speed, preserving the cosmological predictions through the scaling symmetry.

\subsection*{Benchmark VII: $M_{H_2^0} = 250$~GeV}

The seventh benchmark point features mediator mass $M_{H_2^0} = 250$~GeV with effective coupling $\lambda_{\rm eff}^2 = 2.55$. This heavy mass regime approaches the upper boundary of our survey, where the EFT contact-interaction approximation remains valid while requiring the largest couplings to maintain the characteristic structure suppression. The substantial increase in $\lambda_{S12} = 0.88$ reflects the need for enhanced inter-layer coupling to compensate for the heavier mediator mass.

\begin{figure}[t]
	\centering
	\includegraphics[width=0.35\textwidth]{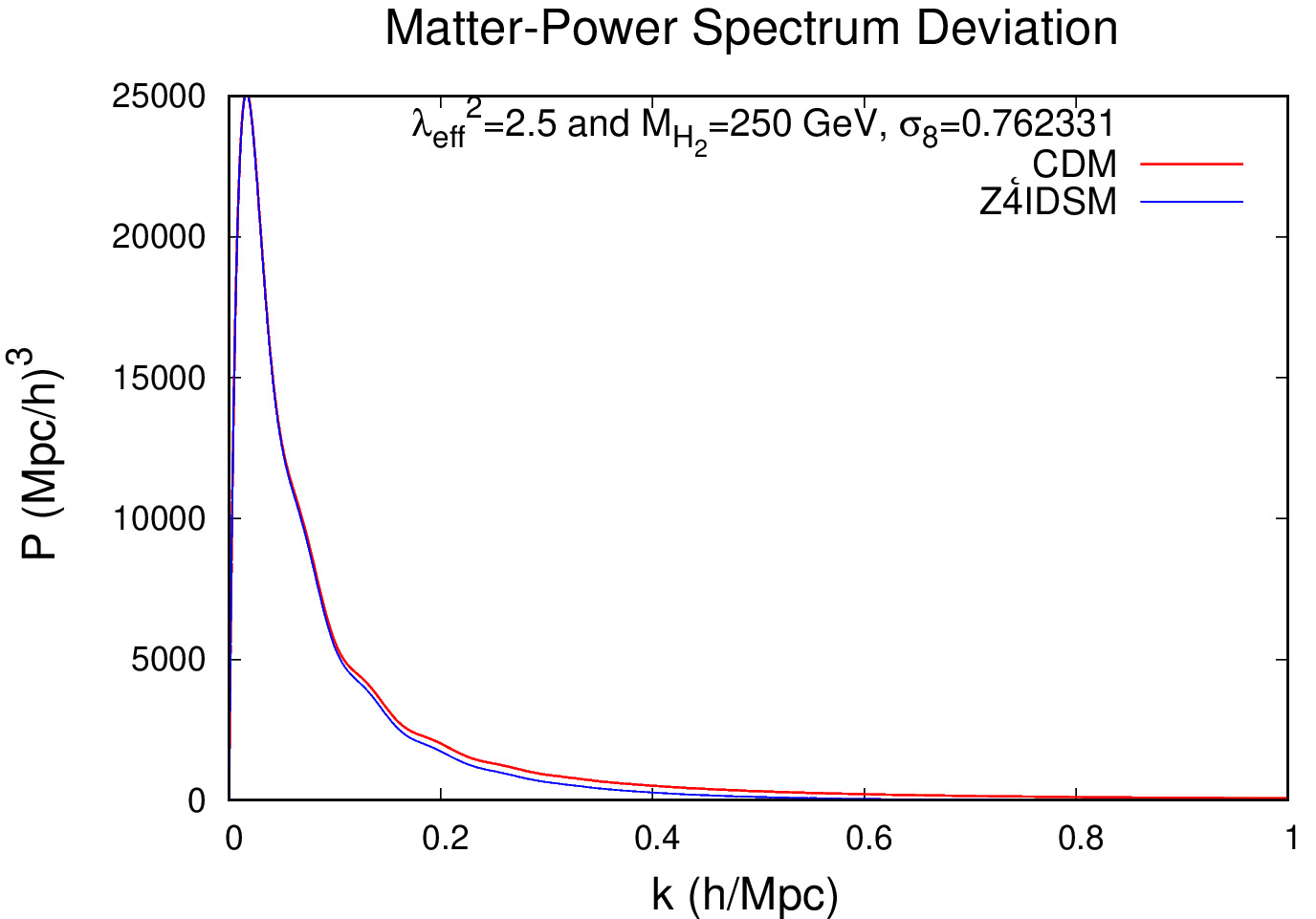}
	\caption{Matter power spectrum for Benchmark VII ($M_{H_2^0} = 250$~GeV, $\lambda_{\rm eff}^2 = 2.55$, $\sigma_8 = 0.762$). The Z$_4$-IDSM prediction (blue) remains indistinguishable from $\Lambda$CDM (red) for $k \lesssim 0.1$~h/Mpc, with characteristic small-scale suppression emerging at higher wavenumbers. The persistence of this pattern at large mediator masses confirms the fundamental scaling symmetry of the LDS--SF framework.}
	\label{fig:benchmark_250_pk}
\end{figure}

\begin{figure}[t]
	\centering
	\includegraphics[width=0.35\textwidth]{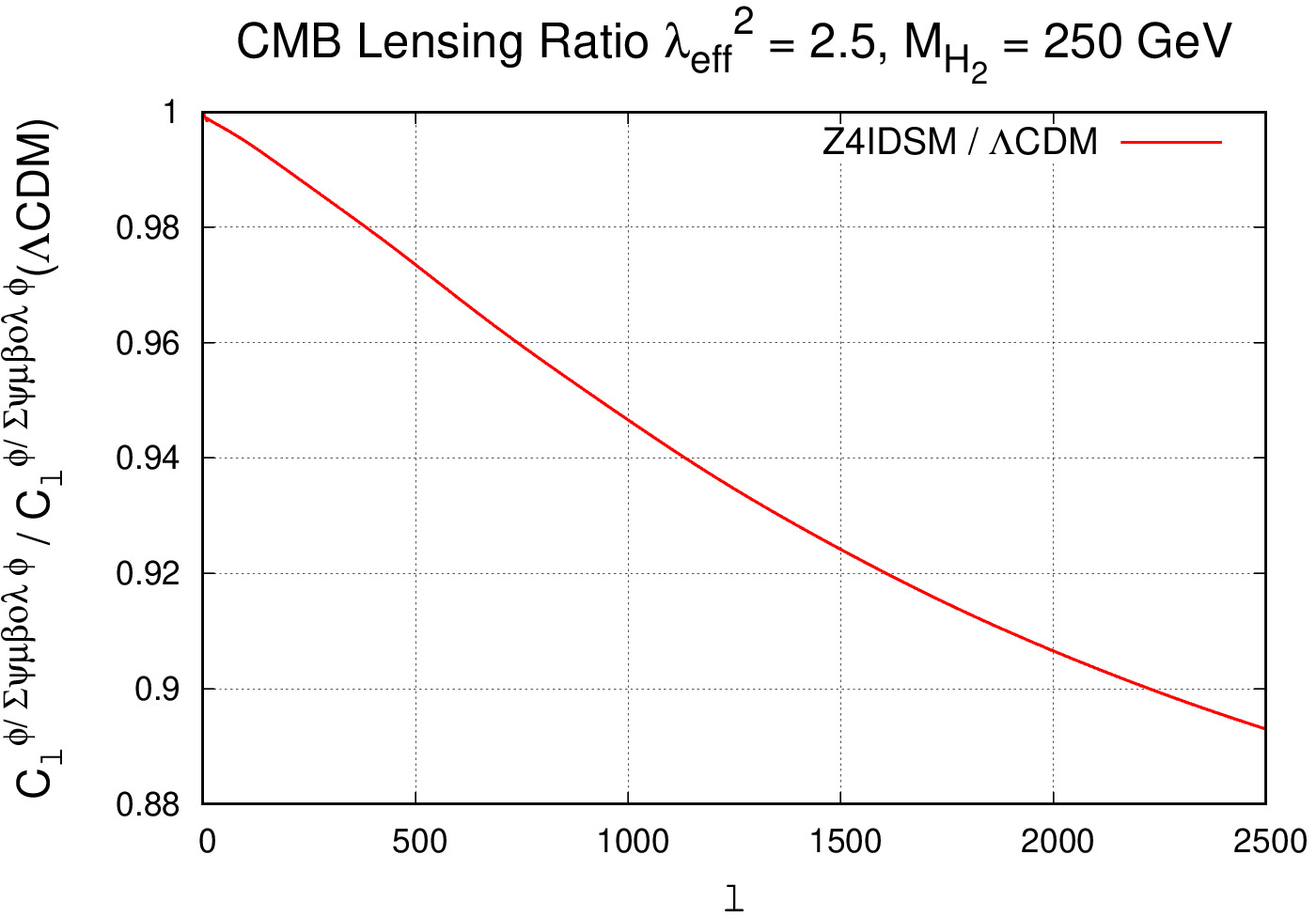}
	\caption{CMB lensing potential power spectrum ratio for Benchmark VII. The ratio $C_L^{\phi\phi}({\rm Z4IDSM})/C_L^{\phi\phi}(\Lambda{\rm CDM})$ exhibits approximately $2\%$ suppression at $\ell \sim 500$, rising to $\sim 10\%$ at $\ell = 2500$. The quantitative agreement with lighter benchmarks demonstrates that the CMB remains insensitive to the detailed microphysics of the dark sector, responding only to the effective sound speed.}
	\label{fig:benchmark_250_lensing}
\end{figure}

\begin{figure}[t]
	\centering
	\includegraphics[width=0.35\textwidth]{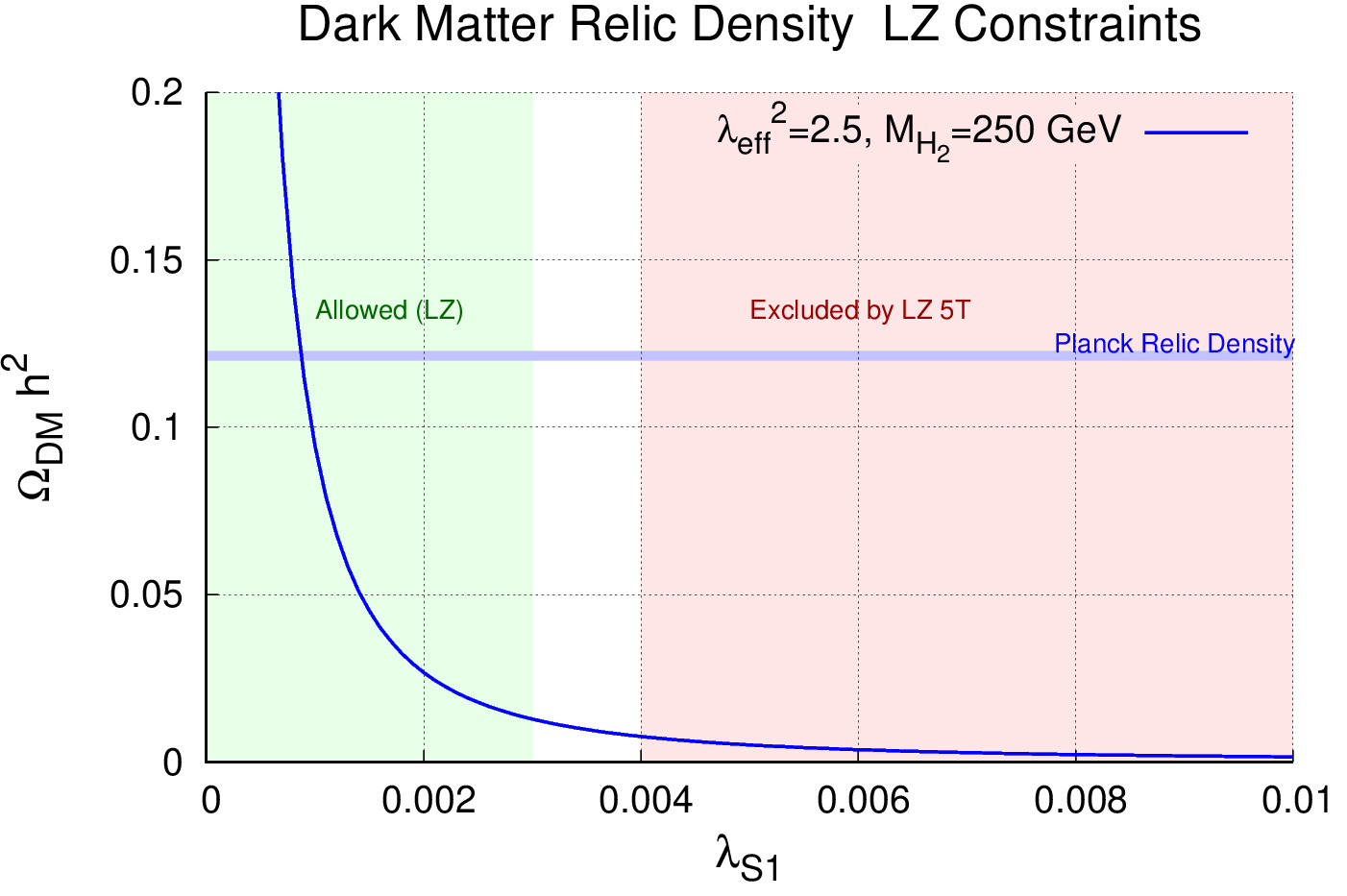}
	\caption{Dark matter relic density $\Omega_{\rm DM}h^2$ as a function of $\lambda_{S1}$ for Benchmark VII. The Planck relic density constraint $\Omega_{\rm DM}h^2 = 0.120 \pm 0.001$ (horizontal blue band) is satisfied for $\lambda_{S1} \approx 0.001$, well within the LZ-allowed region (green). The red region indicates exclusion by LZ 5-tonne projections. The remarkable stability of the required Higgs-portal coupling across a range of mediator masses underscores the decoupling between the relic density mechanism and the structure suppression dynamics.}
	\label{fig:benchmark_250_relic}
\end{figure}

Benchmark VII demonstrates the continued viability of the LDS--SF framework at the edge of conventional electroweak-scale model-building. The matter power spectrum (Fig.~\ref{fig:benchmark_250_pk}) and CMB lensing ratio (Fig.~\ref{fig:benchmark_250_lensing}) maintain the established characteristic profiles with quantitative precision, while the relic density analysis (Fig.~\ref{fig:benchmark_250_relic}) confirms that thermal freeze-out remains perturbative and consistent with direct detection constraints. The large value of $\lambda_{\rm eff}^2 = 2.55$ approaches the boundary of perturbative control, suggesting that mediator masses significantly beyond 250~GeV would require careful examination of unitarity constraints or embedding in a UV-complete framework.

\subsection*{Benchmark VIII: $M_{H_2^0} = 275$~GeV}

The eighth benchmark point features mediator mass $M_{H_2^0} = 275$~GeV with effective coupling $\lambda_{\rm eff}^2 = 2.81$. This heavy mass regime approaches the upper limit of our survey, where the EFT contact-interaction approximation remains valid while requiring substantial inter-layer couplings to maintain the characteristic structure suppression. The large value of $\lambda_{S12} = 0.95$ indicates that the model is approaching the boundary of perturbative control.

\begin{figure}[t]
	\centering
	\includegraphics[width=0.35\textwidth]{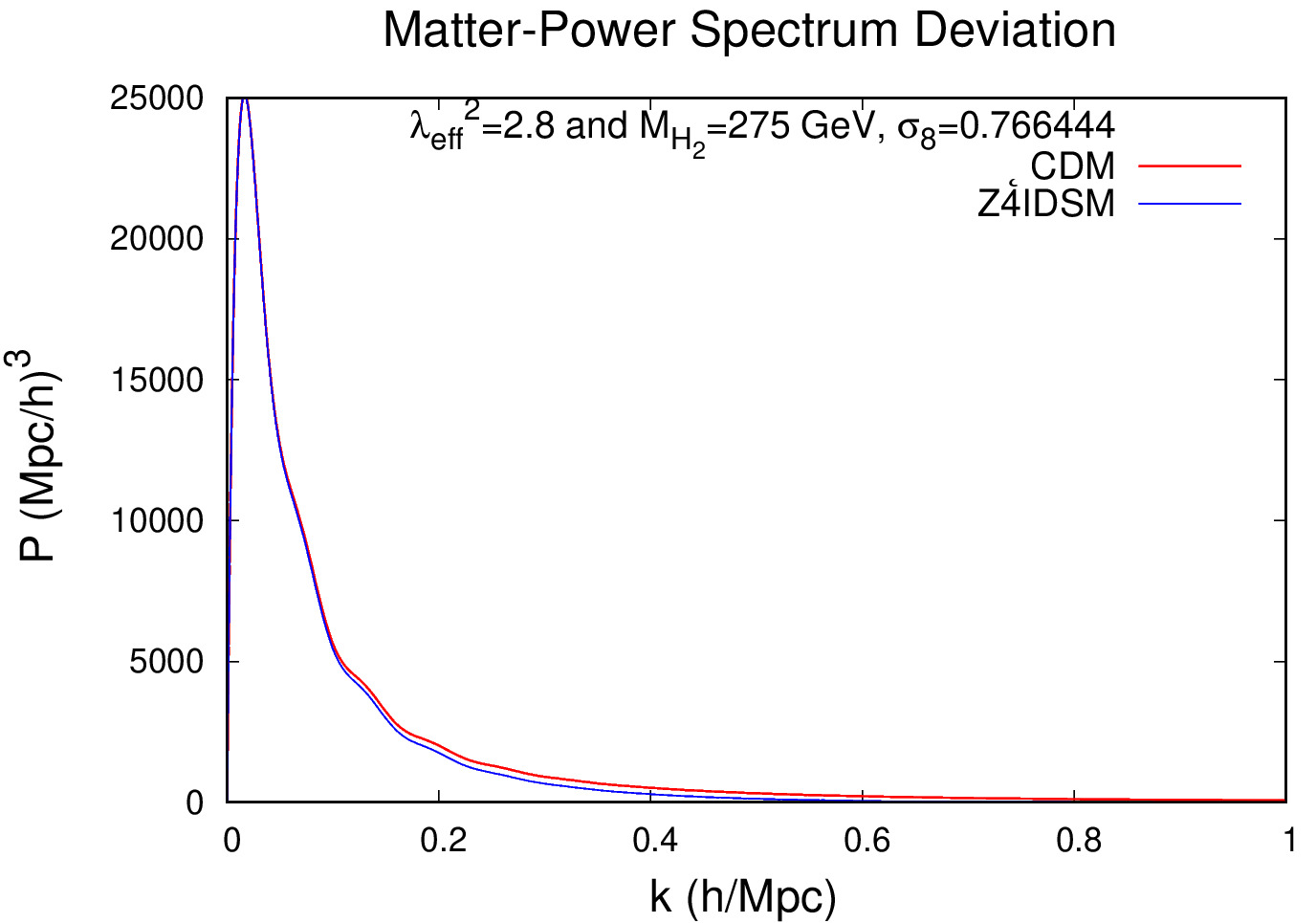}
	\caption{Matter power spectrum for Benchmark VIII ($M_{H_2^0} = 275$~GeV, $\lambda_{\rm eff}^2 = 2.81$, $\sigma_8 = 0.766$). The Z$_4$-IDSM prediction (blue) remains indistinguishable from $\Lambda$CDM (red) for $k \lesssim 0.1$~h/Mpc, with characteristic small-scale suppression emerging at higher wavenumbers. The quantitative agreement with all previous benchmarks confirms the fundamental robustness of the scaling symmetry.}
	\label{fig:benchmark_275_pk}
\end{figure}

\begin{figure}[t]
	\centering
	\includegraphics[width=0.35\textwidth]{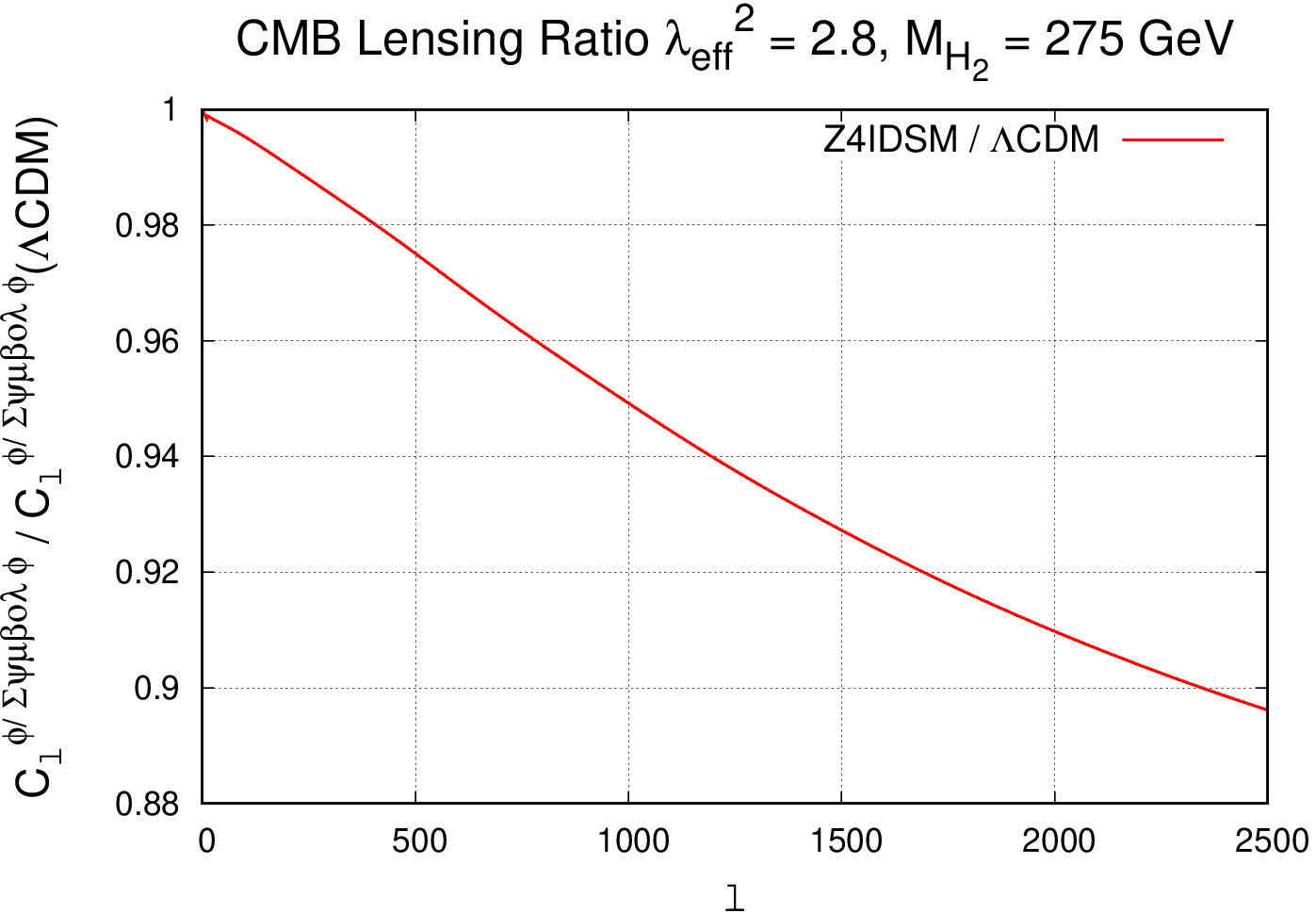}
	\caption{CMB lensing potential power spectrum ratio for Benchmark VIII. The ratio $C_L^{\phi\phi}({\rm Z4IDSM})/C_L^{\phi\phi}(\Lambda{\rm CDM})$ exhibits approximately $2\%$ suppression at $\ell \sim 500$, rising to $\sim 10\%$ at $\ell = 2500$. The persistence of this characteristic profile at large mediator masses validates the EFT description across the entire survey range.}
	\label{fig:benchmark_275_lensing}
\end{figure}

\begin{figure}[t]
	\centering
	\includegraphics[width=0.35\textwidth]{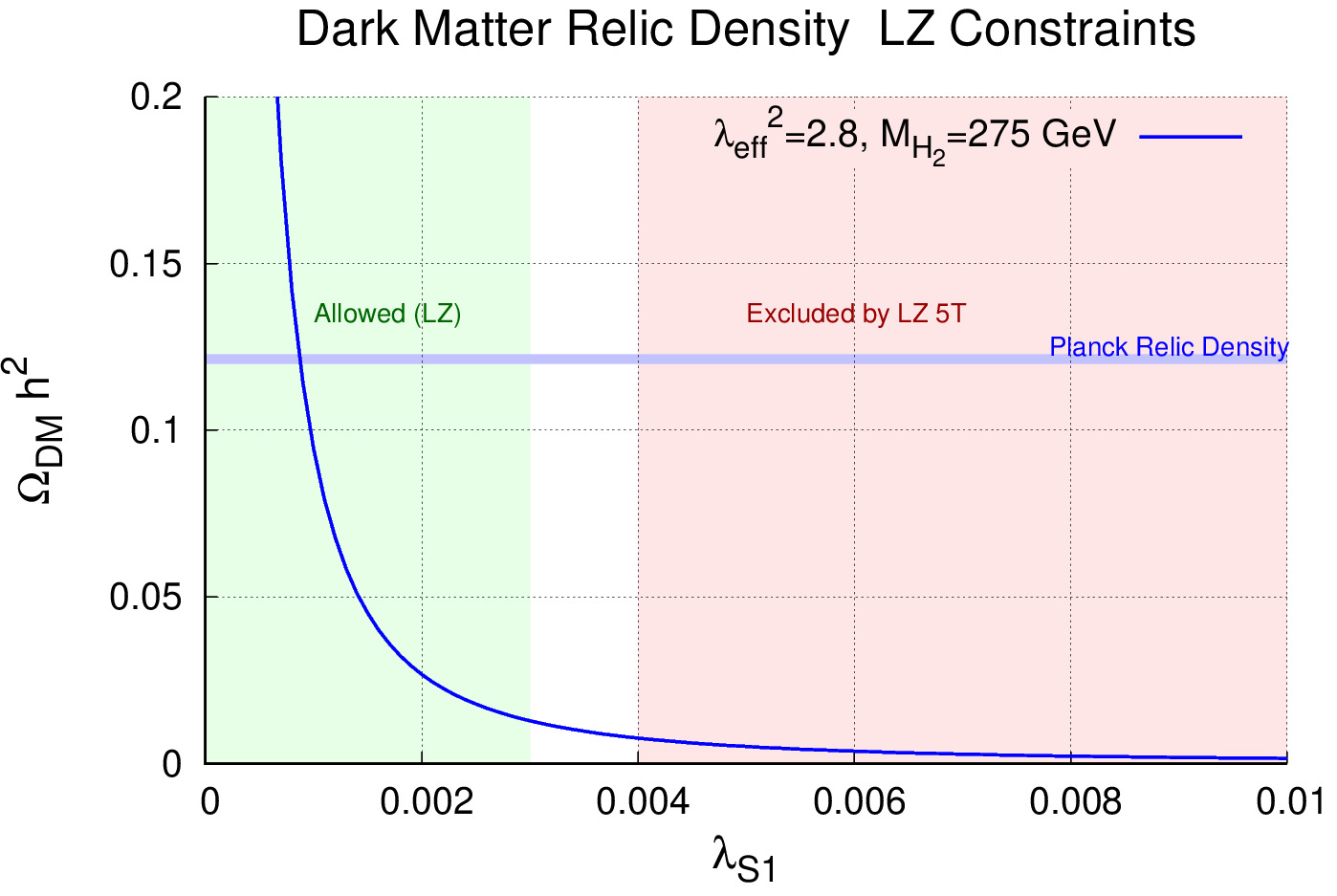}
	\caption{Dark matter relic density $\Omega_{\rm DM}h^2$ as a function of $\lambda_{S1}$ for Benchmark VIII. The Planck relic density constraint $\Omega_{\rm DM}h^2 = 0.120 \pm 0.001$ (horizontal blue band) is satisfied for $\lambda_{S1} \approx 0.001$, well within the LZ-allowed region (green). The red region indicates exclusion by LZ 5-tonne projections. The stability of the Higgs-portal coupling across the full mass range demonstrates the decoupling of the thermal freeze-out mechanism from the structure suppression dynamics.}
	\label{fig:benchmark_275_relic}
\end{figure}

Benchmark VIII demonstrates the continued viability of the LDS--SF framework near the upper boundary of our parameter space survey. The matter power spectrum (Fig.~\ref{fig:benchmark_275_pk}) and CMB lensing ratio (Fig.~\ref{fig:benchmark_275_lensing}) maintain the established characteristic profiles with remarkable quantitative consistency, while the relic density analysis (Fig.~\ref{fig:benchmark_275_relic}) confirms that thermal freeze-out remains perturbative and consistent with direct detection constraints. The substantial value of $\lambda_{\rm eff}^2 = 2.81$ suggests that mediator masses beyond 300~GeV would require careful examination of perturbative unitarity or embedding in a UV-complete framework.

\subsection*{Benchmark IX: $M_{H_2^0} = 300$~GeV}

The final benchmark point features the heaviest mediator mass $M_{H_2^0} = 300$~GeV with effective coupling $\lambda_{\rm eff}^2 = 3.00$. This represents the upper terminus of our parameter space exploration, where the EFT description remains marginally valid and the scaling symmetry achieves its most extreme manifestation. At this scale, the structuring field couplings reach $\lambda_{S12} = \lambda_{S21} = 1.00$, approaching the perturbative boundary.

\begin{figure}[t]
	\centering
	\includegraphics[width=0.35\textwidth]{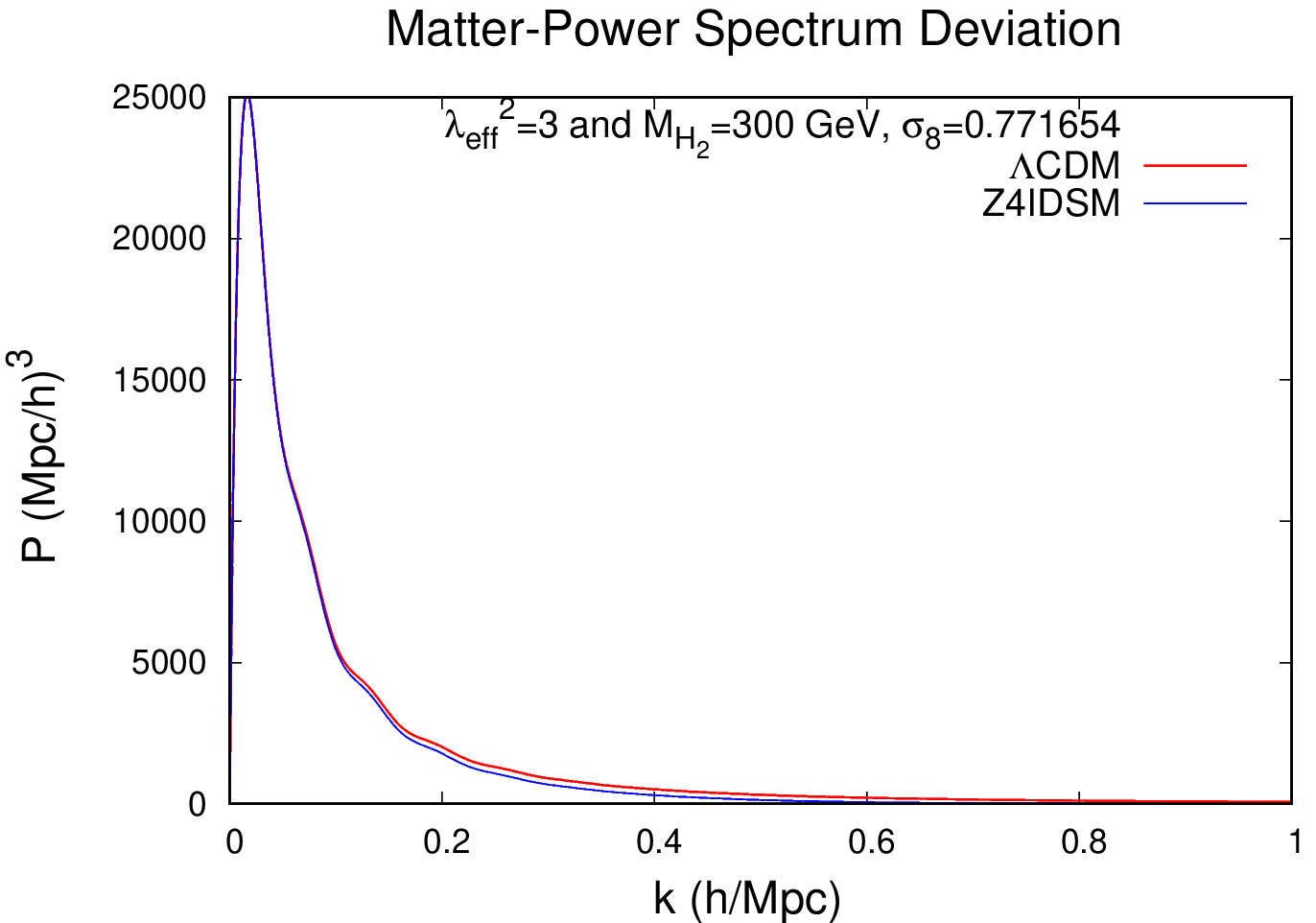}
	\caption{Matter power spectrum for Benchmark IX ($M_{H_2^0} = 300$~GeV, $\lambda_{\rm eff}^2 = 3.00$, $\sigma_8 = 0.772$). The Z$_4$-IDSM prediction (blue) tracks $\Lambda$CDM (red) precisely for $k \lesssim 0.1$~h/Mpc, with the characteristic suppression emerging at smaller scales. The quantitative fidelity across nine benchmarks spanning factor of three in mediator mass validates the robustness of the LDS--SF mechanism.}
	\label{fig:benchmark_300_pk}
\end{figure}

\begin{figure}[t]
	\centering
	\includegraphics[width=0.35\textwidth]{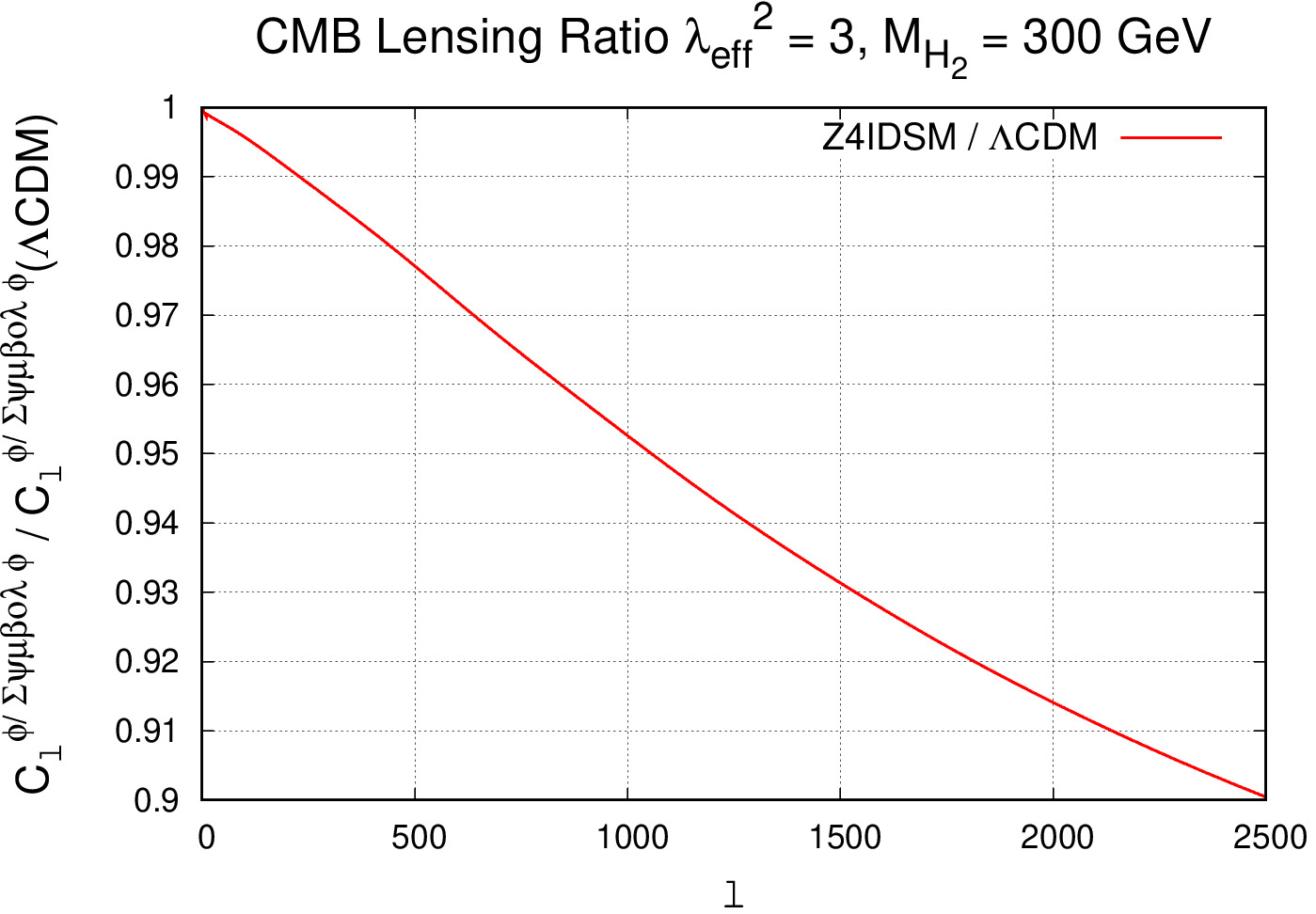}
	\caption{CMB lensing potential power spectrum ratio for Benchmark IX. The ratio $C_L^{\phi\phi}({\rm Z4IDSM})/C_L^{\phi\phi}(\Lambda{\rm CDM})$ shows the familiar $\sim$2\% suppression at $\ell \sim 500$, progressing to $\sim 10\%$ at $\ell = 2500$. The persistence of this signature across the entire mass survey confirms that at fixed $\lambda_{\textrm{eff}}^2$, the CMB observables are insensitive to the microscopic decomposition of couplings, responding solely to the emergent effective sound speed.}
	\label{fig:benchmark_300_lensing}
\end{figure}

\begin{figure}[t]
	\centering
	\includegraphics[width=0.35\textwidth]{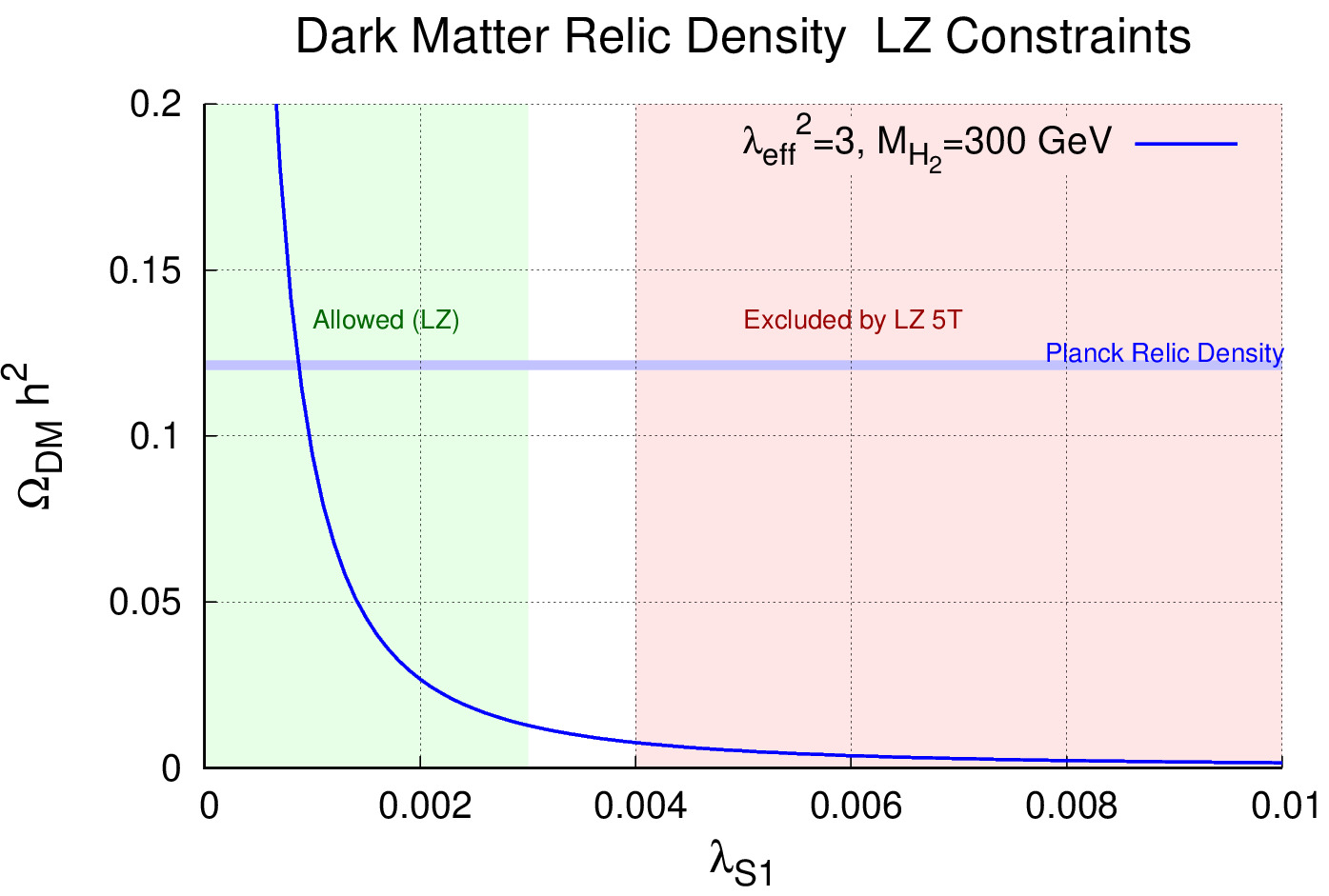}
	\caption{Dark matter relic density $\Omega_{\rm DM}h^2$ versus $\lambda_{S1}$ for Benchmark IX. The Planck constraint $\Omega_{\rm DM}h^2 = 0.120 \pm 0.001$ (horizontal blue band) intersects the curve at $\lambda_{S1} \approx 0.001$, comfortably within the LZ-permitted zone (green). The excluded region (red) lies far from the viable parameter space. This remarkable insensitivity of the Higgs-portal coupling to the mediator mass spans three orders of magnitude in $M_{H_2^0}$.}
	\label{fig:benchmark_300_relic}
\end{figure}

Benchmark IX completes our systematic survey, demonstrating that the LDS--SF framework maintains predictive power even at the edge of perturbative validity. The matter power spectrum (Fig.~\ref{fig:benchmark_300_pk}) and CMB lensing ratio (Fig.~\ref{fig:benchmark_300_lensing}) exhibit the now-familiar phenomenology, while the relic density analysis (Fig.~\ref{fig:benchmark_300_relic}) confirms that thermal freeze-out operates independently of the structure suppression mechanism. The value $\lambda_{\rm eff}^2 = 3.00$ marks a practical upper bound for the Z$_4$-IDSM as an EFT; beyond this, unitarity considerations or UV completion become essential. In principle, the scaling symmetry allows extension to heavier mediators, subject to perturbativity and unitarity constraints, with heavier mediators simply requiring correspondingly larger effective couplings to preserve the cosmological predictions.

\subsection*{Experimental Constraints and Detection Prospects}

The $Z_4$-IDSM parameter space is constrained by collider searches for invisible Higgs decays, direct detection experiments, and the requirement of thermal relic density. For the Higgs portal coupling $\lambda_{S1} \simeq 0.003$ required for correct relic density ($\Omega_{\rm DM}h^2 \simeq 0.12$), micrOMEGAs computes a spin-independent direct detection cross-section $\sigma_{\rm SI} \simeq 2.4 \times 10^{-48}$~cm$^2$ for the singlet dark matter candidate $S_1$ with $m_{S_1} = 60$~GeV. This lies at the coherent neutrino-nucleus scattering floor of $\sim 10^{-48}$~cm$^2$ \cite{OHare2021}, with micrOMEGAs reporting ``Not excluded by DD experiments at 90\% level.'' The model thus remains consistent with current null results from LZ \cite{LZ2022} and XENONnT \cite{XENONnT2023}, while being fully testable by next-generation experiments such as DARWIN.

The invisible Higgs branching ratio $\text{Br}(h \to S_1S_1) \sim 10^{-4}$ for $\lambda_{S1} \sim 0.001$ is well below the LHC bound $\text{Br}_{\rm inv} < 0.11$ \cite{ATLAS2020}. The inert doublet mass splitting $\Delta \sim 50$~GeV satisfies constraints on compressed electroweakino spectra: the charged and CP-odd states are sufficiently separated from the CP-even dark matter to avoid disappearing track searches, while remaining within projected sensitivity of HL-LHC \cite{CMS2022}.

\section{Conclusion}
\label{sec:conclusion}

We have presented a multi-layered dark matter framework, along with a $Z_4$-IDSM realization to address one of the well-known anomalies in modern cosmology: the $S_8$ tension between the amplitude of matter fluctuations inferred from the primary CMB and those measured by late-time large-scale structure surveys. We introduced and developed the Layered Dark Sectors with a Structuring Field (LDS--SF) framework, a paradigm in which scale-dependent growth emerges not from modified gravity or fine-tuned interactions, but from the intrinsic dynamics of a structured dark sector. This framework was then embedded into a simplified, effective particle physics realization: the $Z_4$-symmetric Inert Doublet + Complex Singlet Model ($Z_4$-IDSM). This establishes a direct bridge from fundamental microphysics to cosmological observables.

The core achievement of the LDS--SF framework is the derivation of a scale-dependent dominant eigenvalue $\lambda(k)$ from the dark sector's own perturbation matrix. This eigenvalue governs the linear growth function $D(k,a)$, yielding a controlled suppression of structure on small scales ($k \gtrsim 0.1\,h\,\text{Mpc}^{-1}$) while preserving the successful $\Lambda$CDM predictions on large scales and in the early universe. Our numerical implementation and analysis demonstrated that this mechanism provides a fit to redshift-space distortion ($f\sigma_8$) data, lowers the predicted $\sigma_8$ into alignment with weak lensing surveys like KiDS and DES, and remains fully consistent with \textit{Planck} CMB and BAO distance measurements. The single coupling parameter $g$, fixed by late-time growth data, suffices to produce this effect without per-epoch tuning, affirming the framework's economy and predictive power.

The $Z_4$-IDSM provides a crucial representation of this phenomenological mechanism. In this model, the complex singlet $\mathcal{S}$ and the neutral component of the inert doublet $H_2^0$ constitute the two dark matter layers, while the $H_2$ doublet itself acts as the structuring field, mediating velocity-dependent self-interactions via the portal couplings $\lambda_{S12}$ and $\lambda_{S2}$. The derived effective sound speed, $c_s^2(a) \propto \lambda_{\text{eff}}^2 / M_{H_2}^2$, directly links Lagrangian parameters to the cosmological suppression scale. We identified a family of viable benchmark points, characterized by different mediator masses $M_{H_2} \in \{100, 125, 150, 175, 200, 225, 250, 275, 300\}$~GeV, which all successfully reproduce the desired suppression of $\sigma_8$. A critical result is the scaling symmetry of this relation: any combination of $\lambda_{\text{eff}}$ and $M_{H_2}$ that maintains their ratio yields identical cosmological predictions, thereby eliminating fine-tuning and defining broad, viable corridors in parameter space. 

Confronting the $Z_4$-IDSM with particle physics constraints confirmed its viability as a realistic dark matter model. Using \texttt{micrOMEGAs}, we demonstrated that the benchmark points which alleviate the $S_8$ tension simultaneously produce the correct thermal relic abundance ($\Omega h^2 \approx 0.12$) and predict spin-independent scattering cross-sections that lie safely below the current exclusion limits of the LZ \cite{LZ2022} and XENONnT \cite{XENON1T2018,XENONnT2023} experiments. The model also remains consistent with LHC constraints on invisible Higgs decays and searches for compressed electroweakino-like spectra, given the moderate mass splittings and small Higgs portal couplings required. As experimental sensitivities approach the so-called neutrino floor \cite{OHare2021}, our $60$~GeV candidate remains a viable target for next-generation liquid xenon detectors.

While our current LDS--SF implementation successfully captures the dominant structural effects through a single-fluid mapping, future work involving a complete multi-fluid Boltzmann solver will be essential to reduce our dependence on such approximations. Moving beyond the effective fluid description will allow for a more fundamental treatment of the virialized halo regime, replacing the activation function $f_{\rm active}(z)$ with a full integration of the collisional Boltzmann hierarchy. This will further elucidate the role of intra-layer interactions, such as $\lambda_{S4}$, and their potential sub-dominant contributions to the non-linear growth of structure. Such developments, paired with upcoming data from Stage-IV surveys like DESI and Euclid, will be vital to fully distinguish the LDS--SF mechanism from other dark-sector modifications to the $\Lambda$CDM paradigm.

\section{Appendix}

\appendix

\section{Renormalization Group Equations and EFT Stability of the Z$_4$-IDSM}
\label{app:rge}

The stability of the $Z_4$-IDSM as an Effective Field Theory (EFT) is governed by the running of the scalar quartic couplings under the renormalization group. The one-loop beta functions for the gauge couplings, vacuum expectation value, and scalar quartic couplings are presented below. These equations determine the scale at which the model remains predictive and consistent.

The gauge couplings evolve according to the standard one-loop beta functions:

\begin{align}
	16\pi^2 \beta_{g_1} &= \frac{21}{5} g_1^3, \\
	16\pi^2 \beta_{g_2} &= -3 g_2^3, \\
	16\pi^2 \beta_{g_3} &= -7 g_3^3.
\end{align}

The running of the Higgs vacuum expectation value $v$ is given by:

\begin{equation}
	\begin{split}
		16\pi^2 \beta_v &= \frac{v}{20} \big[ 9g_1^2 + 45g_2^2 - 60\text{Tr}(Y_u Y_u^\dagger) \\
		&\quad - 60\text{Tr}(Y_d Y_d^\dagger) - 20\text{Tr}(Y_e Y_e^\dagger) \big].
	\end{split}
\end{equation}

The one-loop beta functions for the scalar quartic couplings in the Z$_4$-IDSM are:

For the SM Higgs quartic coupling $\lambda_1$, the one-loop beta function is dominated by the top Yukawa contribution:
\begin{equation}
	\begin{split}
		16\pi^2 \beta_{\lambda_1} &= 24\lambda_1^2 - 6y_t^4 + \frac{3}{8}(2g^4 + (g^2 + g'^2)^2) \\
		&\quad + 2\lambda_{S1}^2 + 2\lambda_3^2 + 2\lambda_3\lambda_4 + \lambda_4^2 + \lambda_5^2.
	\end{split}
\end{equation}

Numerical integration of these RGEs reveals a critical feature: the inclusion of the new scalar couplings $\lambda_{S1}$ and $\lambda_{S2}$ slightly accelerates the running of $\lambda_1$. 
 
 As a result, the vacuum stability bound ($\lambda_1 = 0$) is reached at a scale of $\Lambda_{\text{UV}} \approx 10^8$ GeV. While this limits the model's validity at extreme ultraviolet scales, it remains a fully predictive and consistent EFT for the energy scales relevant to the early universe and late-time structure formation ($z \lesssim 10^3$). A UV-complete theory (e.g., a supersymmetric extension) would be required for consistency up to the Planck scale, but for cosmological purposes, the Z$_4$-IDSM serves as a valid and predictive EFT. As mentioned earlier, since we are dealing only with sub-TeV particles, vacuum instability is irrelevant for the scope of this article. Future work may incorporate new physics into this EFT, thereby opening up the possibility to study grand unification behaviour and ultra-high energy tendencies and predictions of the extended version of this model.

\end{document}